\documentclass[prb,superscriptaddress,longbibliography,twocolumn]{revtex4-1}
\pdfoutput=1
\usepackage{geometry}
\usepackage{verbatim}
\usepackage{amsmath}
\usepackage{amssymb}
\usepackage{graphicx,picture,calc}
\usepackage{hyperref}
\usepackage{braket}
\usepackage{bm}
\usepackage{epstopdf}
\usepackage[normalem]{ulem}
\usepackage{xcolor}
\usepackage[multidot]{grffile}
\usepackage{overpic}
\usepackage{dsfont}
\newcommand{\be}{\begin{equation}}
\newcommand{\ee}{\end{equation}}
\newcommand{\bk}{{{\bf{k}}}}

\newcommand{\bq}{{\bf{q}}}

\newcommand{\bea}{\begin{eqnarray}}
\newcommand{\eea}{\end{eqnarray}}
\newcommand{\beal}{\begin{align}}
\newcommand{\eeal}{\end{align}}
\renewcommand{\i}{{\mathrm{i}}}
\newcommand{\ra}{\rangle}
\newcommand{\la}{\langle}

\newcommand{\dg}{{\dagger}}
\newcommand{\pdg}{{\phantom\dagger}}

\begin{document}

\title{Anisotropic Magnetoresistance in Multiband Systems: \\ 
2DEGs and Polar Metals at Oxide Interfaces}

\author{Nazim Boudjada}
\affiliation{Department of Physics, University of Toronto, Toronto, Ontario M5S1A7, Canada.}
\author{Ilia Khait}
\affiliation{Department of Physics, University of Toronto, Toronto, Ontario M5S1A7, Canada.}
\author{Arun Paramekanti}
\email{arunp@physics.utoronto.ca}
\affiliation{Department of Physics, University of Toronto, Toronto, Ontario M5S1A7, Canada.}

\date{\today}

\begin{abstract}
Low density two-dimensional electron gases (2DEGs) with spin-orbit coupling are highly sensitive to an in-plane magnetic field,
which impacts their Fermi surfaces and transport properties. Such 2DEGs, formed at transition metal oxide surfaces or interfaces,
can also undergo surface phase transitions leading to polar metals that exhibit electronic nematicity.
Motivated by experiments on such systems, we theoretically study 
magnetotransport in $t_{2g}$ orbital systems, using Hamiltonians that include atomic spin-orbit coupling and 
broken inversion symmetry, for both square symmetry (001) and hexagonal symmetry (111) 2DEGs. Using a numerical solution to the 
full multiband matrix-Boltzmann equation, together with insights gleaned from the impurity scattering overlap matrix, we explore the
anisotropic magnetoresistance (AMR) in the presence of impurities which favor small momentum scattering. We find that transport in
the (001) 2DEG is dominated by a single pair of bands, weakly coupled by impurity scattering, one of which has a larger Fermi velocity
while the other provides an efficient current-relaxation mechanism. This leads to strong angle-dependent current damping and a large AMR with many angular 
harmonics. In contrast, AMR in the (111) 2DEG typically features a single $\cos(2\vartheta)$ harmonic,
with the angle-averaged magnetoresistance being highly tunable by a symmetry-allowed trigonal distortion.
We also explore how the (111) 2DEG Fermi sufaces are impacted by electronic
nematicity via  a surface phase transition into a 2D polar metal for which we discuss a Landau theory, and we show that this leads to distinct symmetry components
and higher angular harmonics in the AMR.
Our results are in qualitative agreement with experiments from various groups for 2DEGs at the SrTiO$_3$ surface and the LaAlO$_3$-SrTiO$_3$ interface.
\end{abstract}

\pacs{Valid PACS appear here}
\maketitle

\section{Introduction}

The ability to control the layer-by-layer growth of transition metal oxide heterostructures has led to the discovery and exploration of
two-dimensional electron gases (2DEGs) formed at carefully engineered oxide surfaces and interfaces \cite{Review2014} .
Such 2DEGs, formed by a combination of a polarization catastrophe \cite{nakagawa2006some} and oxygen vacancies,
can combine the multiple functionalities of the two bulk quantum materials forming the interface or potentially host
new low-dimensional phases of matter. 
This has led to significant focus on magnetism and superconductivity at the (001) LaAlO$_3$-SrTiO$_3$ (LAO-STO) interface.
\cite{ohtomo2004high,thiel2006tunable,nakagawa2006some,Millis2006,reyren2007superconducting,caviglia2008electric,caviglia2010tunable,Ariando2011,bert2011direct,li2011coexistence, macdonald_001,mehta2012evidence,held_001,Michaeli2012,Fidkowski2013,fischer_spinorbit_2013, kim2013origin, banerjee2013ferromagnetic, chen2013,Park2013,ruhman_competition_2014,caviglia_001,Taraphder2016,tolsma_orbital_2016,atkinson2017influence}
Oxide surfaces and interfaces also offer a novel setting to study the role of spin-charge coupling and 
magnetoelectric effects in 2DEGs \cite{burkov2004}. In particular, it has been shown that an electric field can tune the strength of Rashba spin-orbit coupling (SOC) which
arises from broken inversion symmetry, permitting control of the Fermi surface (FS) spin-texture and spin-to-charge conversion \cite{caviglia2010tunable,Bibes_NMat2016,Han_Science2017}. Similarly, an in-plane magnetic field
is found to have a significant impact on charge transport, leading to a large negative magnetoresistance
in (001) 2DEGs over a range of densities \cite{caviglia_001,dagan2009,ariando2011mr}.
More recent experiments have begun to create and probe 2DEGs at oxide (111)
surfaces and interfaces
\cite{rodel_orientational_2014,mckeown_walker_control_2014,Dudy, miao_anisotropic_2016,
rout_six-fold_2017,ADMI:ADMI201600830,monteiro_two-dimensional_2017, davis_superconductivity_2017,Caviglia2018,Goldstein2019}
which have been proposed to host topological phases \cite{Xiao_NComm2011,Ruegg_PRB2011,Cook_PRL2014,Okamoto_PRB2014,Fiete_SciRep2015,Baidya_PRB2016,Held_2016,Kee_NPJQM2017}.
Such (111) interfaces typically lead to more tightly confined 2DEGs \cite{baidya2015}.

Transport experiments \cite{rout_six-fold_2017,ruhman_competition_2014,miao_anisotropic_2016,Venkatesan2017,ADMI:ADMI201600830,davis_superconductivity_2017} on such 2DEGs have studied the {\it anisotropic magnetoresistance} (AMR):
the change in the diagonal resistivity when the angle between the current direction and the in-plane magnetic field is varied. AMR has also been studied theoretically in the context of spin-dependent impurity scattering on 2DEG samples with finite magnetization \cite{AMR2Band,AMRMagnetic} and in the context of ferromagnetic semiconductors \cite{mcguire1975anisotropic,AMRgaas,De_Ranieri_2008}
A useful analysis
of the symmetry constraints on the AMR is presented in Ref. ~\onlinecite{rout_six-fold_2017}.
These experiments
raise the issues of what controls magnetoresistance in multiband systems such as oxide 2DEGs,
why their resistance drops when the magnetic fields is rotated from being aligned parallel to the
current to being perpendicular to the current direction, and
to what extent higher angular harmonics in the AMR directly reflect FS symmetries (e.g., four-fold versus six-fold rotational symmetry
of the FS).

Motivated by these questions, we present here a theoretical study of the AMR 
using a 
full solution to the semiclassical matrix-Boltzmann equation for the multi-band 2DEG with SOC, showing that it captures
experimental observations in both (001) and (111) oxide 2DEGs.
Focusing on the momentum-dependence of the relevant impurity scattering overlap-matrix in the
transport equation provides useful insights that are broadly applicable to other multiband systems.
Our results also suggest that the AMR harmonics do not directly reflect the underlying lattice
rotational symmetries of the FS.

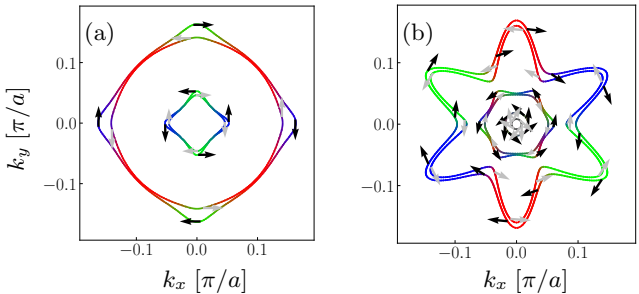
\begin{figure}[t]
	\centering
	\begin{overpic}[width=.49\linewidth]{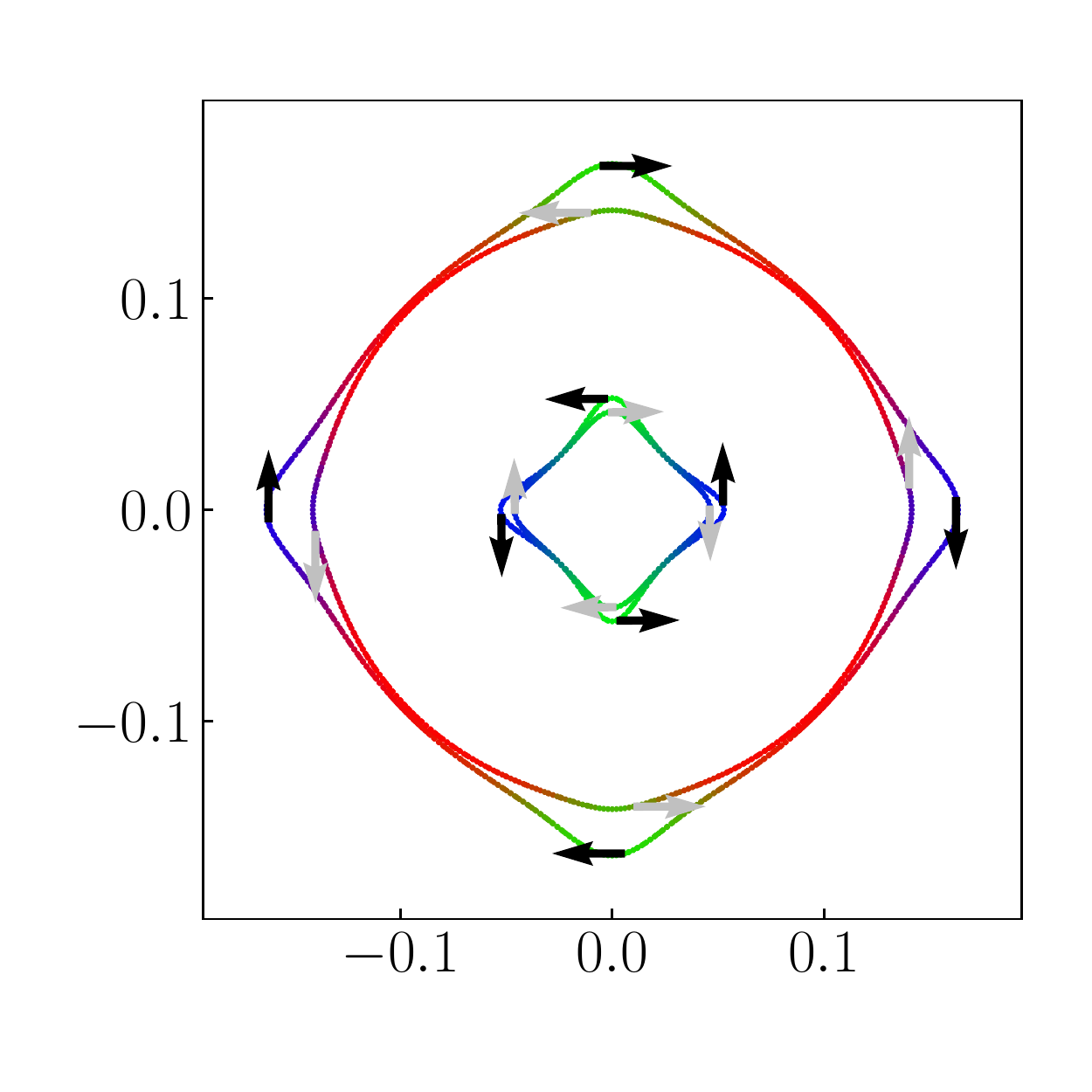} \put(20,80){(a)}\put(45,1){$k_x\;[\pi/a]$}\put(-5,38){\rotatebox{90}{$k_y\;[\pi/a]$}} \end{overpic}
	\begin{overpic}[width=0.49\linewidth]{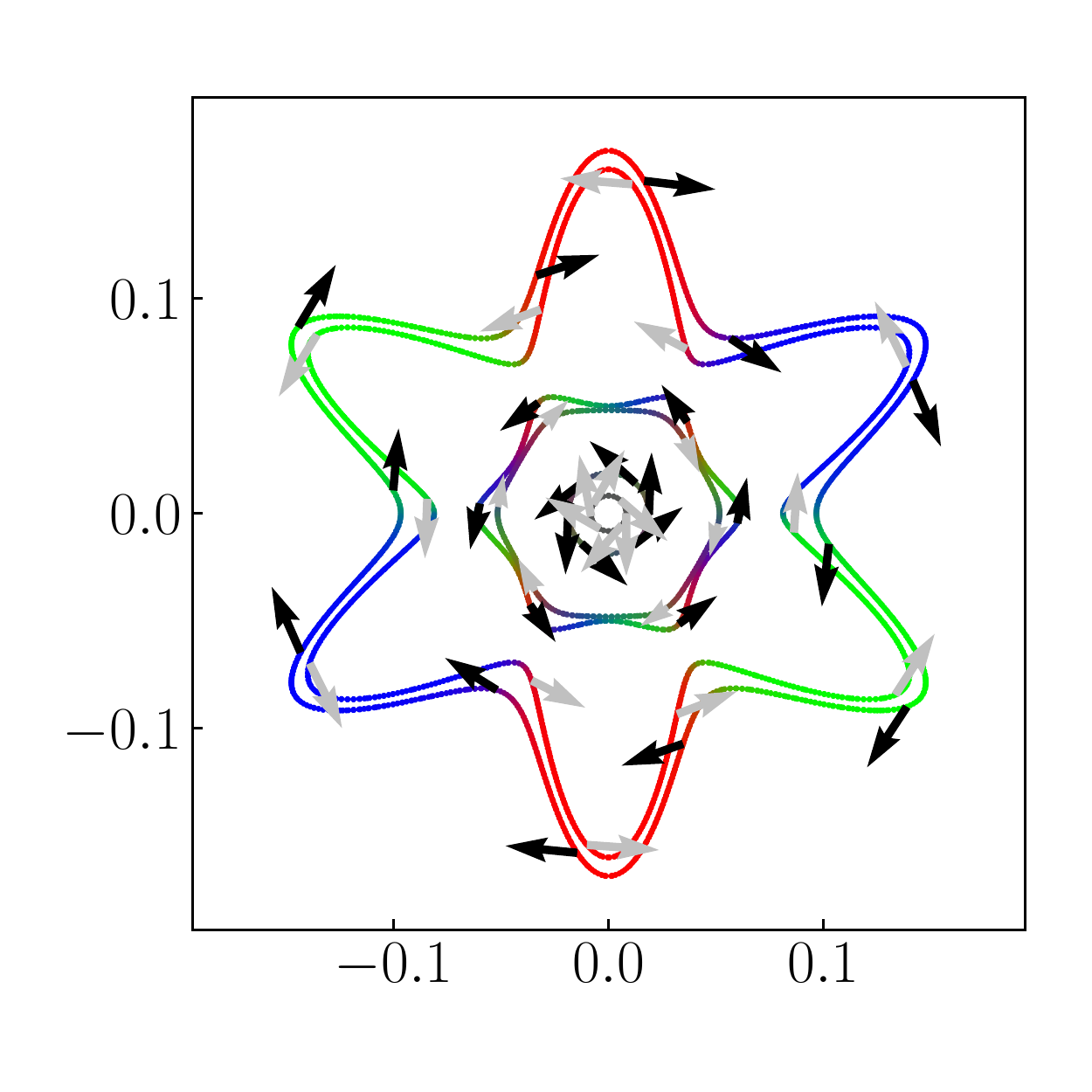} \put(19,80){(b)}\put(45,1){$k_x\;[\pi/a]$} \end{overpic}
	\caption{Fermi surfaces for low density spin-orbit coupled 2DEGs at zero magnetic field, with colors indicating orbital content ($yz$-blue, $zx$-green, $xy$-red), 
	and Rashba spin texture indicated by black/grey arrows for opposite chiralities. (a) (001) 2DEG at a density $n\! =\! 0.035$e/Ti
	showing outer `circular' bands with significant $xy$ orbital content (except along $k_x \! =\! 0$ and $k_y \!=\! 0$)
	and inner `propeller' bands with dominant $yz$-$zx$ character. (b) (111) 2DEG with $n\! =\! 0.05$e/Ti showing outer `flower-like' bands and 
	inner `hexagon' bands. Each band has equal (momentum-dependent) admixture of all orbitals.} 
	\label{fig:FSB0}
\end{figure}

For the (001) 2DEG, our work builds upon previous important numerical
studies \cite{caviglia_001,diez2016}. Here, we further suggest simple physical mechanisms for the observations. 
At zero magnetic field and low density, Fig.~\ref{fig:FSB0}(a) shows the FSs and the
corresponding Rashba spin texture. We find that an effective Rashba SOC enables weak scattering between a `circular' $xy$-orbital 
dominated band, and an inner $(zx,yz)$-orbital dominated `propeller'-band.
For small SOC, this interband scattering gets suppressed by an in-plane magnetic field
leading to large negative magnetoresistance.
Changing the angle of the in-plane magnetic field, we find strong angle-dependent 
current damping and a large AMR, with many angular harmonics at higher fields, as observed experimentally \cite{dagan2009} and in previous
theoretical work \cite{caviglia_001,diez2016}.

In striking contrast, transport in the (111) 
2DEG is dominated by the outermost 
pair of `flower-like' bands, which arise from strong hybridization of {\it all} three $t_{2g}$ orbitals. 
The corresponding FSs are shown in Fig.~\ref{fig:FSB0}(b) along with the Rashba spin texture.
While the overall resistivity has some
contribution from scattering to an inner small `hexagon' band, the angle-dependence of the AMR is dominated by the two large bands.
The AMR in this case typically reveals a single $\cos(2\vartheta)$ angular harmonic; although higher harmonics are symmetry-allowed \cite{rout_six-fold_2017}
we find they have essentially vanishing weight.
Our results are in qualitative agreement with experimental data \cite{rout_six-fold_2017}.

Finally, we explore the impact of directional symmetry-breaking on the (111) 2DEG which leads to a 2D polar metal phase.
Such electronic nematicity has been reported in transport experiments \cite{Chandrasekhar2018,Venkatesan2017,ADMI:ADMI201600830}.
Polar order in the 2DEG could arise as a result of a bulk structural transition or strong electronic correlations.
We argue, within Landau theory, that a polar 2DEG could also arise via 
a surface phase transition, when the insulating bulk is a paraelectric close to a ferroelectric quantum critical point; this may be of 
potential relevance to SrTiO$_3$ interfaces. Ferroelectricity in epitaxially strained [111]-oriented SrTiO$_3$ has in fact recently been studied using
first-principles calculations \cite{Neaton2019}. Independent of its microscopic origin, we show that 
incorporating the nematicity associated with such a polar metal
leads to the readily visible violation of some basic symmetry constraints obeyed by the AMR of
symmetry-unbroken phases, and also generates higher angular harmonics in the AMR signal.
Such simultaneous violations of symmetry constraints
and generation of higher harmonics may have been observed in recent experiments at the LAO-STO (111) interface and STO surface.

This paper is organized as follows. We begin with a quick review of the Boltzmann equation
for multiband materials, relegating technical details of the computation to Appendix \ref{boltz}. We then 
discuss AMR in the (001) 2DEG and a simplified picture for its origin.
We next turn to analogous results for the (111) 2DEG, incorporating the effect of a symmetry-allowed trigonal distortion. 
Finally, we discuss the impact of directional symmetry breaking on the AMR in (111) 2DEGs in light
of recent experimental observations.

\section{Boltzmann equation}
For a weak electric field $\vec{E}$, the semiclassical Botzmann equation for a multiband system is given by
\begin{eqnarray}
- \frac{\partial f_{n,\bk}}{\partial \varepsilon_{n,\bk}}  e  E^i v^i_{n,\bk} =\mathcal{N} \sum_m \! \int\frac{\mathrm{d}^2\bk'}{(2\pi)^2}(g_{m,\bk'}-g_{n,\bk})\nonumber\\
\times|\langle n\bk|\hat{U}|m\bk'\rangle|^2\delta(\varepsilon_{n,\bk}-\varepsilon_{m,\bk'}),
\label{eqn:boltzmann}
\end{eqnarray}
where $f_{n,\bk}$ and $g_{n,\bk}$ correspond respectively to equilibrium distribution function and its perturbed non-equilibrium
part, labeled by band $n$ and momentum $\bk$.
We use $i$ to denote component indices ($i=x,y$), with implied summation for repeated indices.
$\mathcal{N}$ is a constant (proportional to the impurity concentration) which drops out of transport ratio coefficients. The
spin-orbit coupled band eigenfunctions, energies, and velocities, are denoted by $|n\bk\rangle$, $\varepsilon_{n,\bk}$ and $v^i_{n,\bk}$
respectively; these depend on the magnitude and direction of the applied in-plane magnetic field $\mathcal{\vec{B}}$.

The matrix elements for elastic impurity scattering $\langle n\bk|\hat{U}|m\bk'\rangle$ are obtained using a scalar scattering potential
\begin{equation}
\hat{U}=\sum_{\bk,\bk',\ell\sigma} V(\bk-\bk')c^\dagger_{\ell\sigma}(\bk')c^\pdg_{\ell\sigma}(\bk),
\end{equation}
where $V(\bq)=V_0\text{e}^{-\Lambda^2|\bq|^2/4}$. Setting $\Lambda=0$ corresponds to scattering off a point-like impurity, while 
$\Lambda \gg a$ (with $a$ being the lattice constant) corresponds to small momentum transfer scattering as appropriate for a smooth 
real space impurity potential. For SrTiO$_3$, the lattice constant $a \approx 3.9\rm\AA$ is the nearest neighbor Ti-Ti distance in the 
bulk cubic crystal.

The conductivity tensor within the Boltzmann formalism is then
\be
\sigma_{ij} = e^2 \sum_n \int \frac{\rm {d}^2 \bk}{(2\pi)^2} v_{n,\bk}^i \frac{\partial g_{n,\bk}}{\partial E^j}.
\ee
Here, $\partial g_{n,\bk}/\partial E^j$ is computed by taking a derivative of the Boltzmann equation with respect to the electric field, 
and solving the resulting equations iteratively on a finely discretized momentum mesh in a temperature-dependent window around the
FS. The technical details of this approach, including issues related to choice of the momentum mesh and 
convergence of the solution, are discussed in Appendix \ref{boltz}.

The angle-dependent resistivity tensor is calculated by inverting the conductivity tensor ${[\rho(\vartheta)]_{ij}=[\sigma(\vartheta)^{-1}]_{ij}}$, where $\vartheta$ is the angle between the in-plane magnetic field and the current direction. The AMR
and its Fourier amplitudes $C_m$ are defined via
\begin{eqnarray}
	\text{AMR}(\vartheta)&=&\frac{\rho_{xx}(\vartheta)-\rho_{xx}(|\mathcal{\vec{B}}|=0)}{\rho_{xx}(|\mathcal{\vec{B}}|=0)},\\
	C_m&=&\int_{0}^{2\pi} \frac{\rm{d}{\vartheta}}{2\pi}\;\mathrm{e}^{-\i m \vartheta}  \text{AMR}(\vartheta).
\end{eqnarray}
The invariance of the resistivity under $\vartheta\rightarrow\vartheta+\pi$ (which is equivalent to flipping the direction of the current)
leads to the vanishing of all odd harmonics $C_{2m+1}$, while (mirror) symmetry under $\vartheta\rightarrow-\vartheta$ yields purely real Fourier coefficients $C_m$. 
This latter symmetry is broken for the $(111)$ surface when we include nematic order (see Section \ref{nem}).

\section{(001) 2DEG}

There have been extensive density functional theory studies of the (001) 2DEG \cite{held_001}.
Experiments have realized the (001) 2DEG at the LaAlO$_3$-SrTiO$_3$ interface, and at SrTiO$_3$ surfaces via photo-doping \cite{rodel_orientational_2014} or Ar ion bombardment \cite{mckeown_walker_control_2014, miao_anisotropic_2016}. In this
last setting, angle-resolved photoemission spectroscopy (ARPES) provides an experimental guide to the FSs of the $t_{2g}$-orbital derived bands.
Based on these, and previous work on magneto-transport \cite{ruhman_competition_2014,caviglia_001,diez2016}, we consider the following model.

\subsection{Model Hamiltonian}

We begin with the 2DEG square lattice Hamiltonian in the absence of SOC and a magnetic field:
\be
{H^{001}_0=\sum_{\bk\sigma\ell\ell'} c^\dg_{\ell \sigma}(\bk) {\cal M}^{001}_{\ell\ell'}(\bk) c^\pdg_{\ell' \sigma}(\bk)}.
\ee
Working in the $\{yz,zx,xy\}$ basis, and using abbreviated notation $s_i \equiv \sin (k_i)$, $c_i \equiv \cos (k_i)$ (with $i=x,y$)
we have
\begin{eqnarray}
\small
{\cal M}^{001} =
\begin{pmatrix} \varepsilon_{yz} & \delta_{xy} & \i \zeta_x \\
\delta_{xy} & \varepsilon_{zx} & \i \zeta_y \\
-\i \zeta_x & -\i \zeta_y & \varepsilon_{xy} - \Delta_T
\end{pmatrix},
\end{eqnarray}
where the orbital dispersions are given by
\bea
\varepsilon_{yz} &=& 2 t_1 (1-c_y) + 2 t_2 (1-c_x), \\
\varepsilon_{zx} &=& 2 t_1 (1-c_x) + 2 t_2 (1-c_y), \\
\varepsilon_{xy} &=& 2 t_1 (2-c_x - c_y).
\eea

Here, $t_1$ and $t_2$ are the strong and weak nearest-neighbour intra-orbital hoppings respectively and
${\delta_{xy} \equiv 4t_3 s_x s_y}$ is the inter-orbital next-neighbour hybridization. The tetragonal splitting $\Delta_T$ arises from 2D
confinement which lowers the $xy$-orbital energy, while the odd-in-momentum
inter-orbital hopping $\zeta_i=2\zeta s_i$ represents the impact of surface inversion breaking.

Finally, the two important terms for the AMR are the atomic SOC and the coupling to the in-plane magnetic field. 
SOC is captured by the additional term
\be
H_{\rm SOC} =  
\i \frac{\lambda}{2} \sum_{\bk} \epsilon^\pdg_{\ell m n} c^\dg_{\ell \sigma}(\bk) \tau^n_{\sigma\sigma'} c^\pdg_{m \sigma'}(\bk),
\ee
(with sums over repeated indices) where $\epsilon_{\ell m n}$ is the Levi-Civita symbol and $\tau^n$ are Pauli matrices.
Atomic SOC together with the inversion breaking inter-orbital hopping $\zeta$ leads to an effective Rashba SOC.
The in-plane magnetic field $\vec{\cal B}$ leads to the term
\be
H_{\cal B} = (g_\ell\vec{L}+g_s\vec{S})\cdot\vec{{\cal B}},
\ee
with orbital and spin g-factors $g_\ell, g_s$. Here, the angular momentum components are ${L_n = 
\i \sum_{\bk} \epsilon^\pdg_{n \ell m} c^\dg_{\ell \sigma}(\bk) c^\pdg_{m \sigma}(\bk)}$.

For a quantitative study of the 2DEG, we follow Ref. ~\onlinecite{caviglia_001} and fix ${(t_1,t_2,t_3,\Delta_T,\zeta,\lambda) \equiv (340,13,6,60,8,5)}$meV. We set the orbital
$g$-factor to be $g_\ell=1$, and the spin $g$-factor to be $g_s=5$, with $g_s > 2$ chosen to
mimic enhanced ferromagnetic correlations observed in certain such 2DEGs. The transport properties for this system are calculated
as a function of the angle $\vartheta$ between the in-plane $\vec{\cal B}$ field and the current (which we assume to run along the $[100]$ crystal axis).

\begin{figure}[t]
	\centering
	\includegraphics[width=\linewidth]{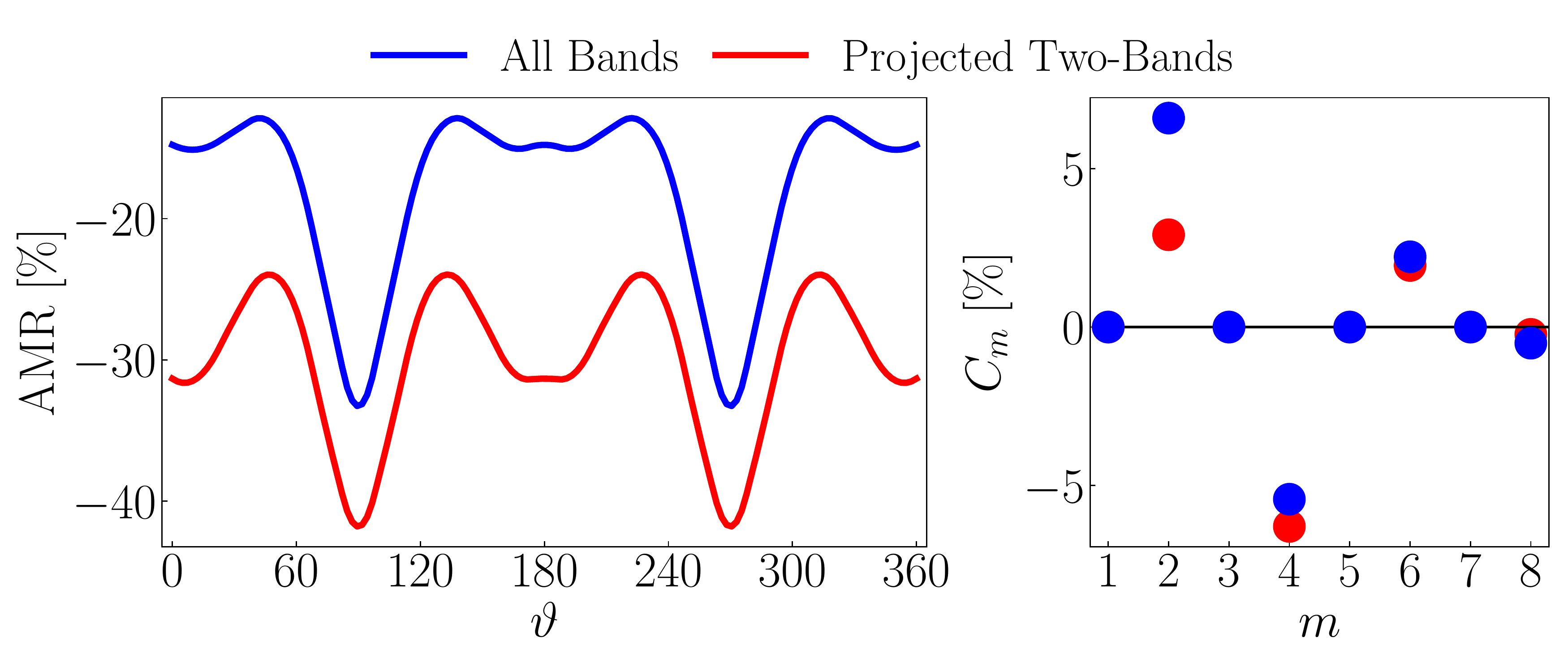} 
	\caption{AMR (left panel) and its Fourier modes (right panel) for the $(001)$ surface computed from the Boltzmann equation. The electronic density is 
	$n=0.035$e/Ti (which is $2.2\times10^{13}/\text{cm}^{2}$) and the temperature is $T=10$K. We fix the magnetic field
	strength $|\vec{\mathcal{B}}|=20$T and vary its angle $\vartheta$ with respect to the current which is along the $[100]$ crystal direction. 
	The impurity scattering length is fixed at $\Lambda=5a$.
	The blue curve corresponds to the full solution including all four partially filled bands at the Fermi level, as shown in Fig.~\ref{fig:FSB0}(a),
	while the red curve is for the `projected' Boltzmann calculation in which we only retain bands $2$ and $3$.}
	\label{fig:AMR001n0035}
\end{figure}

\begin{figure*}[t]
	\centering
	\begin{overpic}[width=0.32\textwidth]{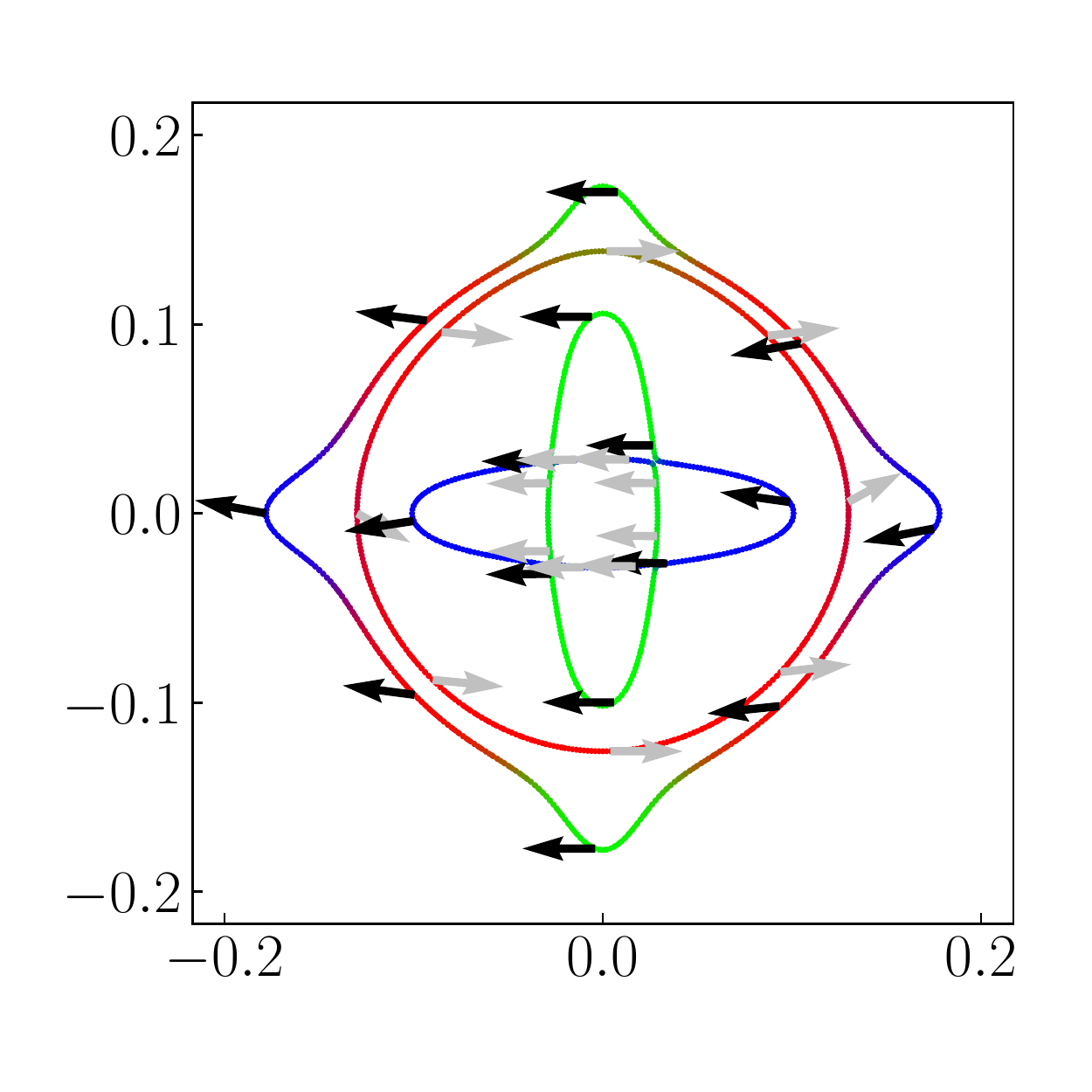}\put(75,82){$\vartheta=0$}\put(47,5){$k_x\;[\pi/a]$}\put(0,42){\rotatebox{90}{$k_y\;[\pi/a]$}} \end{overpic}
	\begin{overpic}[width=0.32\textwidth]{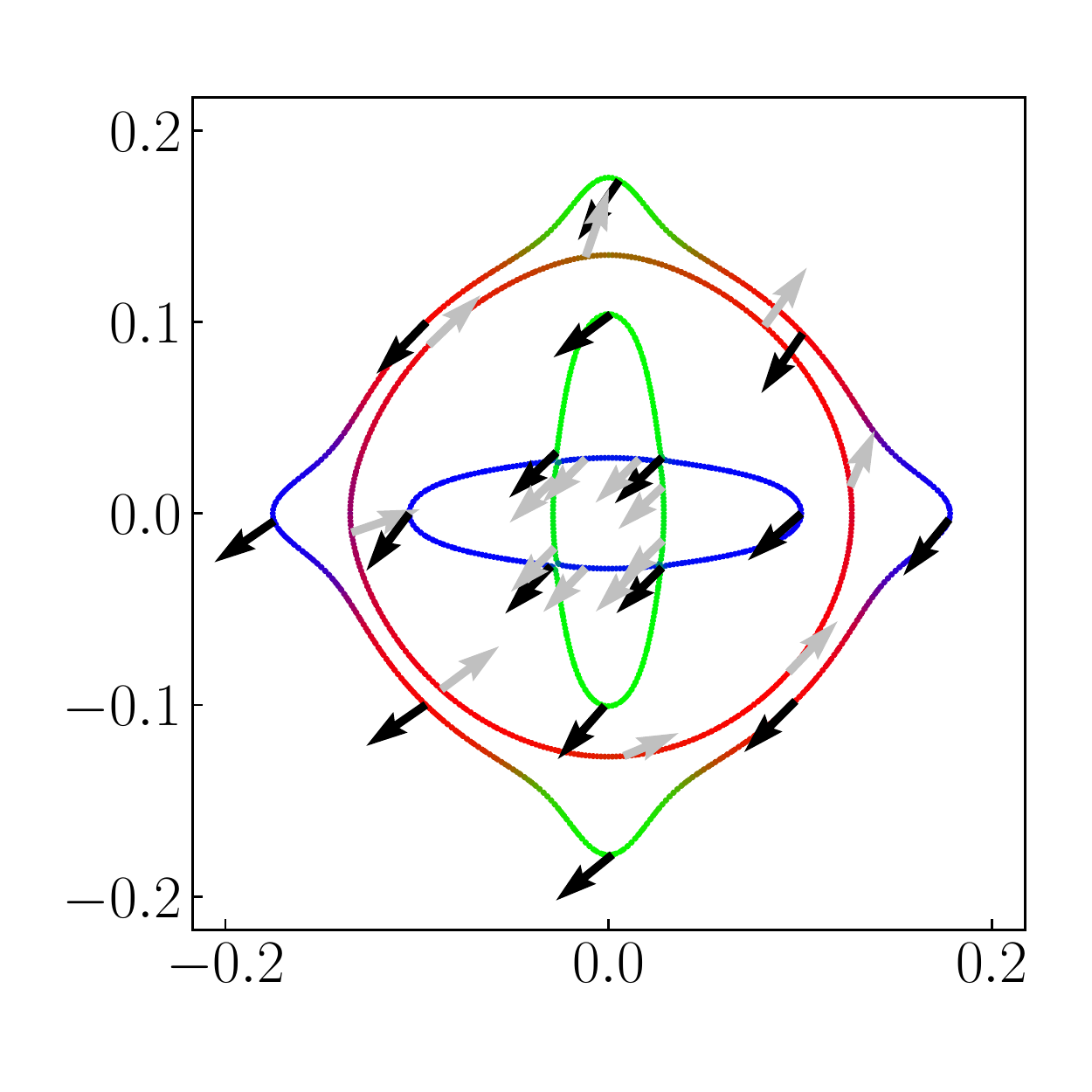}\put(75,82){$\vartheta=\frac{\pi}{4}$}\put(47,5){$k_x\;[\pi/a]$} \end{overpic}	\begin{overpic}[width=0.32\textwidth]{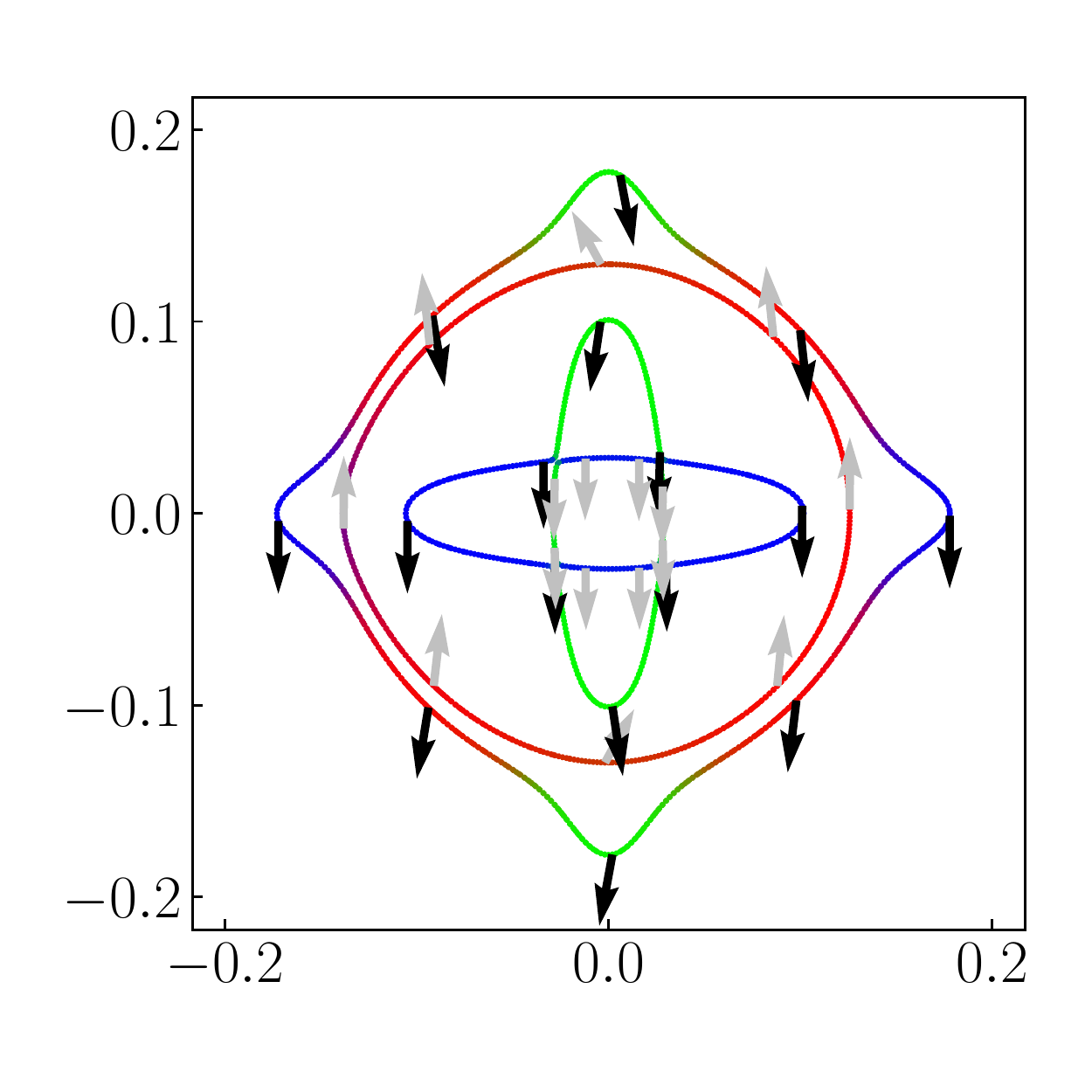}\put(75,82){$\vartheta=\frac{\pi}{2}$}\put(47,5){$k_x\;[\pi/a]$} \end{overpic}
	\begin{overpic}[width=0.32\textwidth]{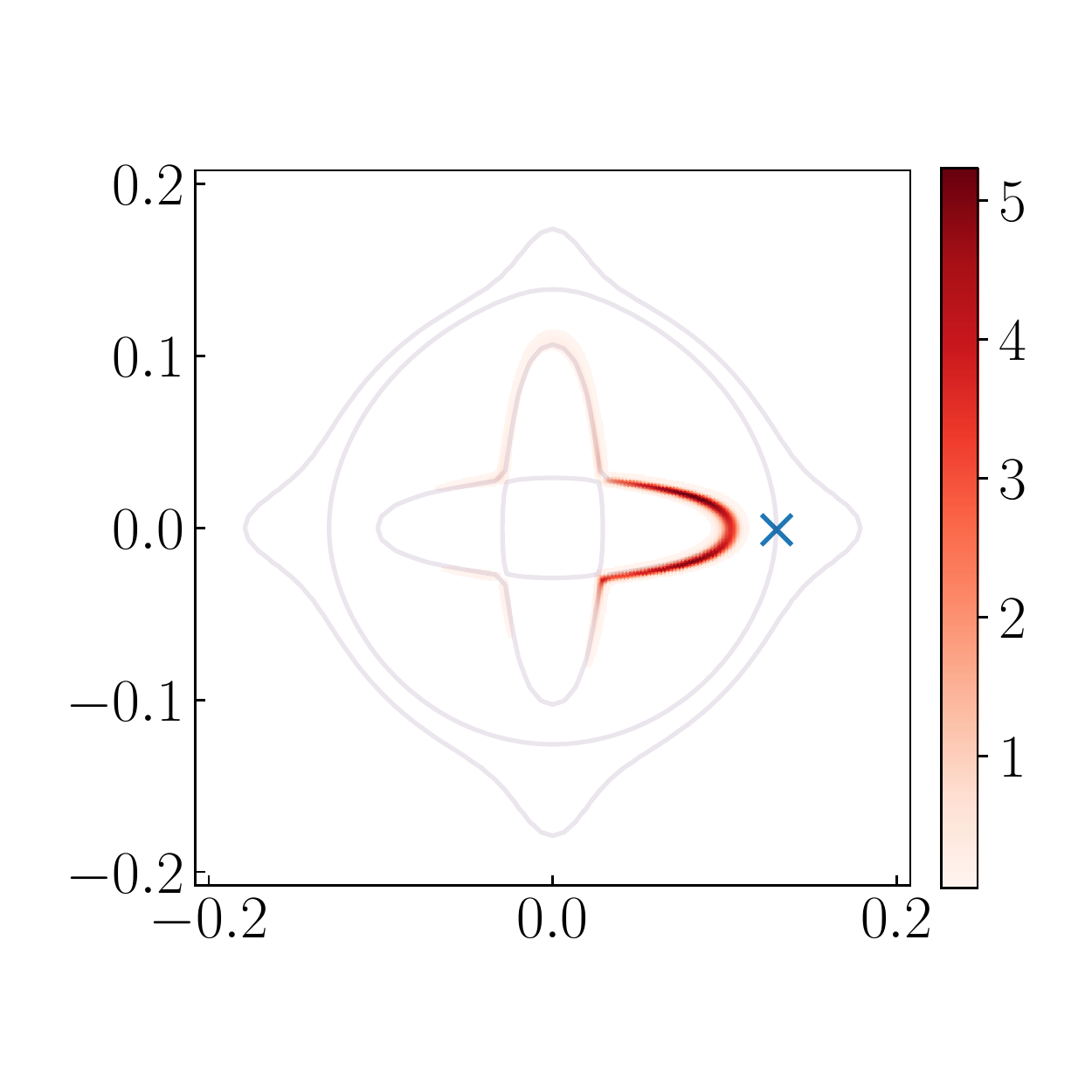}\put(66,73){$\vartheta=0$}\put(42,8){$k_x\;[\pi/a]$}\put(0,38){\rotatebox{90}{$k_y\;[\pi/a]$}}\end{overpic}
	\begin{overpic}[width=0.32\textwidth]{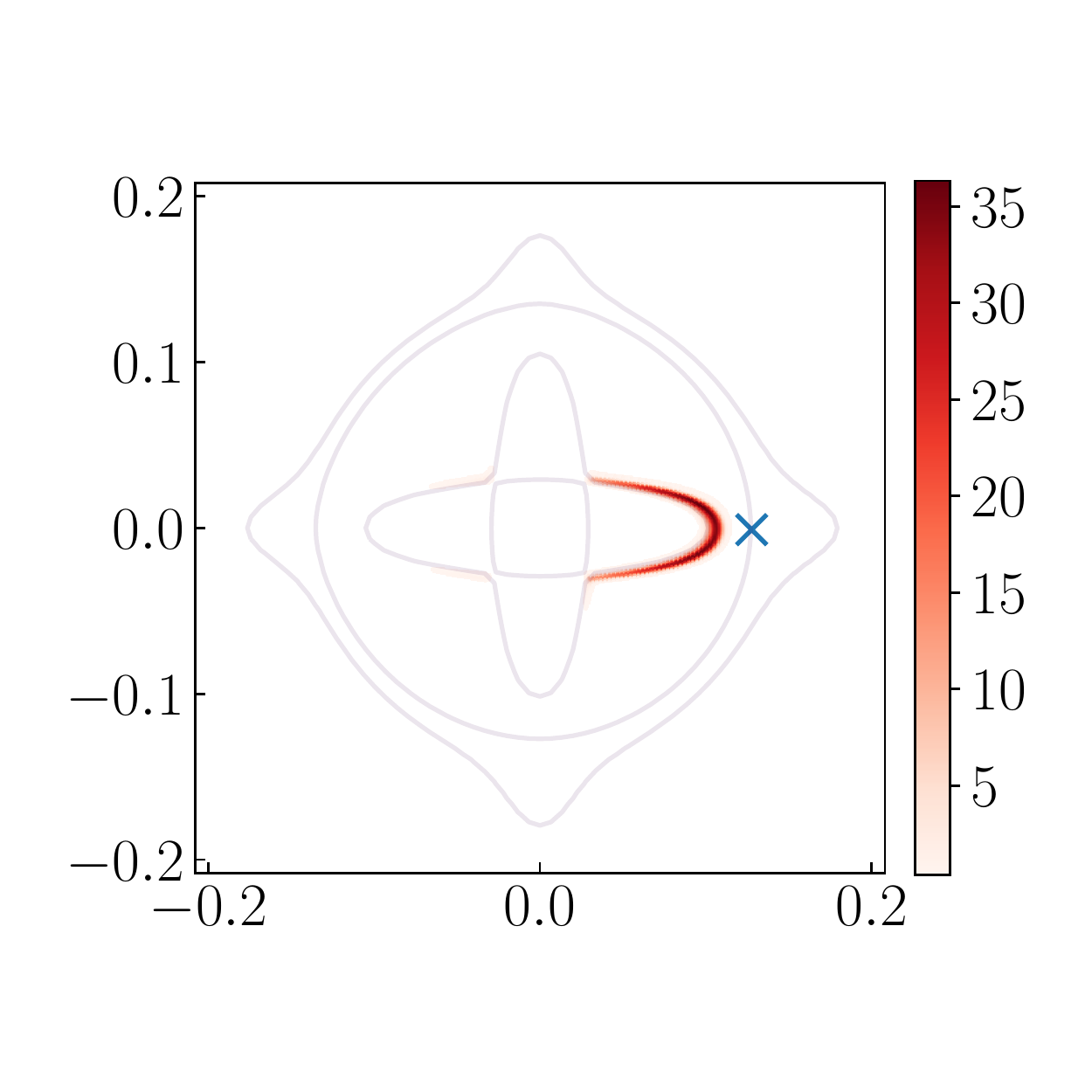}\put(62,73){$\vartheta=\frac{\pi}{4}$}\put(41,9){$k_x\;[\pi/a]$}\end{overpic}
	\begin{overpic}[width=0.32\textwidth]{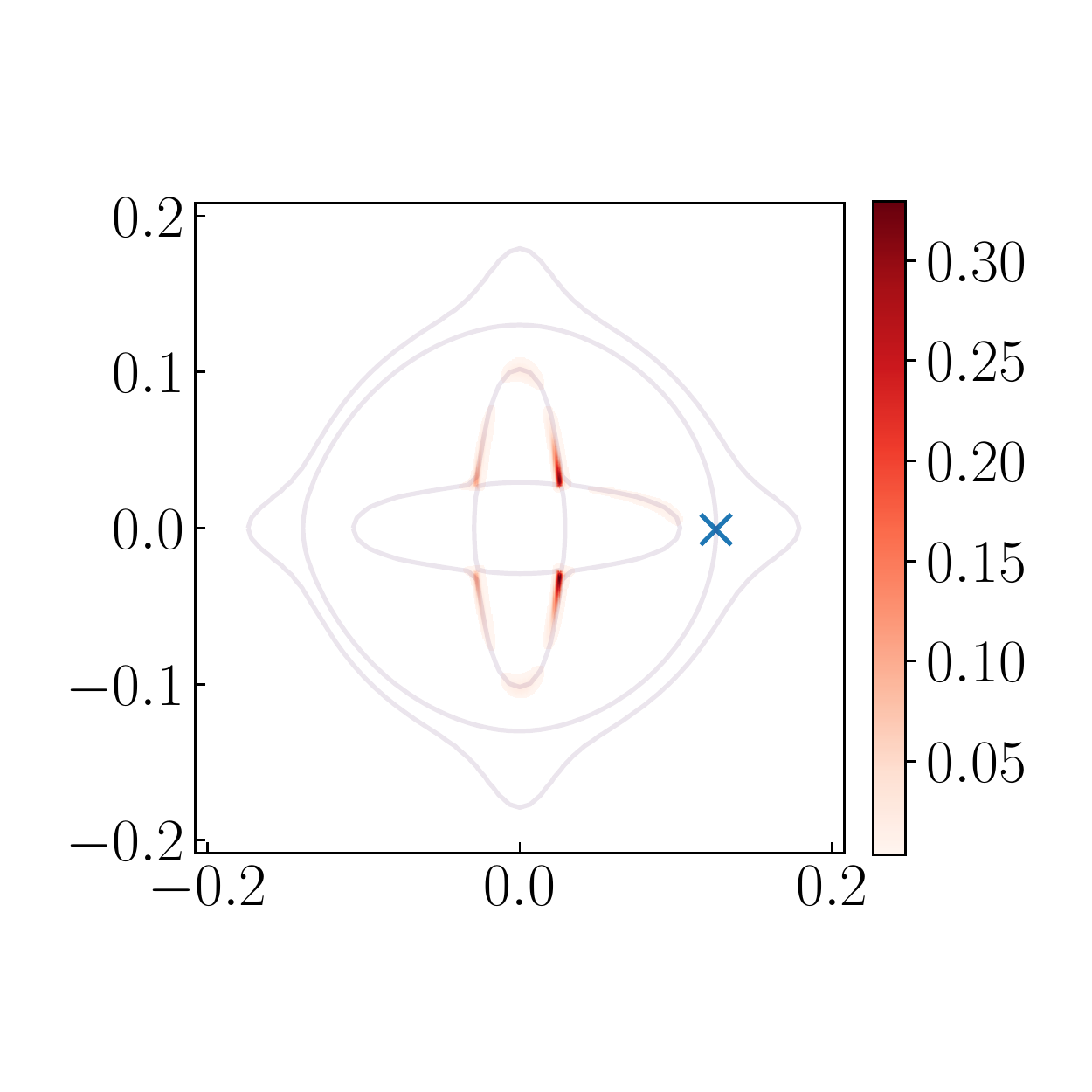}\put(59,73){$\vartheta=\frac{\pi}{2}$}\put(39,11){$k_x\;[\pi/a]$} \end{overpic}
	\caption{{\bf Top panels:} (001) Fermi surfaces and spin textures for fixed magnetic field strength $|\vec{\mathcal{B}}|=20$T and various indicated field angles $\vartheta$. 
	{\bf Bottom panels:}  Colorscale plot of impurity scattering overlap matrix, i.e., integrand in the Boltzmann equation,
	$|\langle n\bk|\hat{U}|m\bk'\rangle|^2\delta(\varepsilon_{n,\bk}-\varepsilon_{m,\bk'})$, with $\Lambda=5a$ for scattering from band $m=2$ and fixed $\bk' = k_F\hat{x}$ 
	(marked by blue cross) to band $n=3$ as a function of $\bk$ for field angles corresponding to top panels.} \label{fig:FS001B20}
\end{figure*}
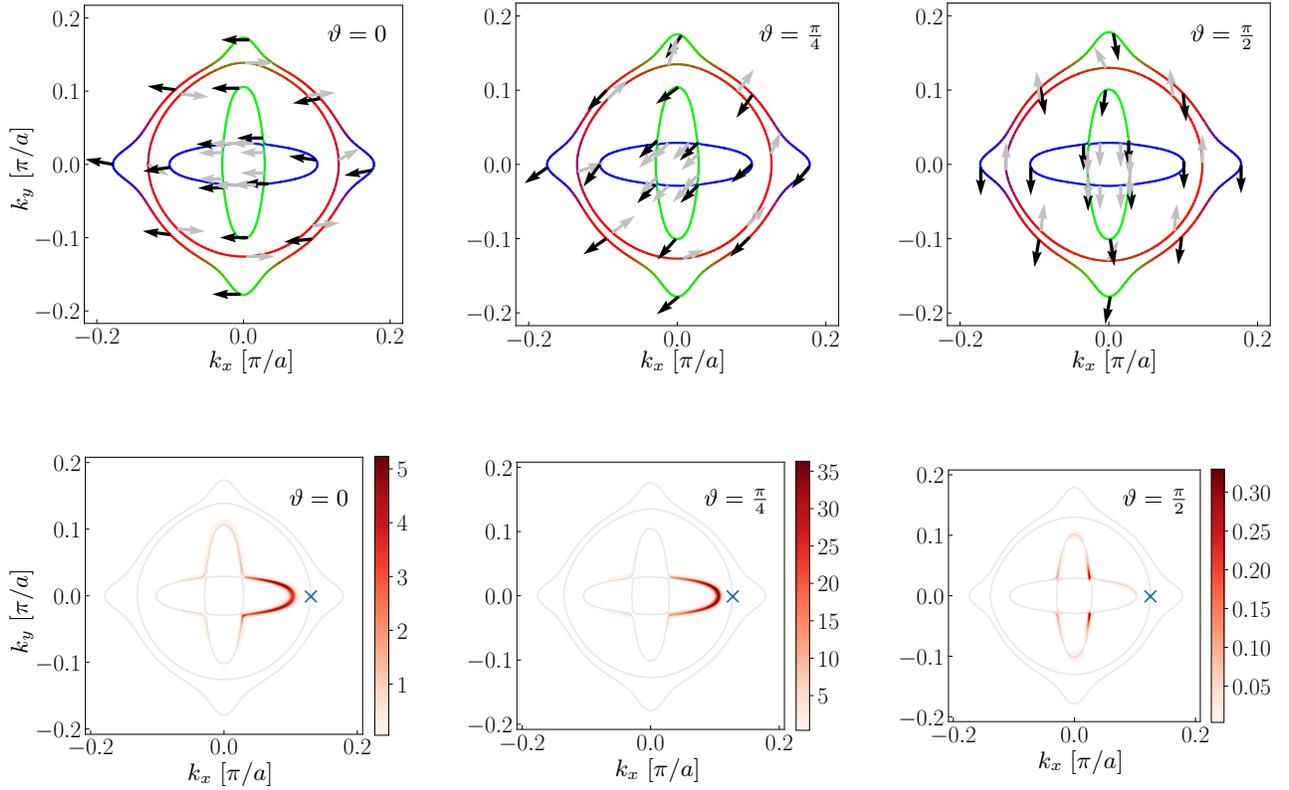

\subsection{Magnetotransport}

The zero field FS for the Hamiltonian $H^{001}_0+ H_{\rm SOC}$ is shown in 
Fig.~\ref{fig:FSB0}(a) for $n=0.035$ electrons per Ti, with colors indicating
the orbital content. We also show the corresponding spin texture on each band (with a different color arrow for spin chirality),
from which the opposite chirality on pairs of effective Rashba bands is clearly visible.
We observe four FSs; the outer two FSs have dominant $xy$-character, with some $zx$-$yz$ character
near the $k_x=0$ and $k_y=0$ directions, while the inner two FSs have dominant $zx$-$yz$ character.

Fig.~\ref{fig:AMR001n0035}(a) shows the calculated AMR signal (blue curve) as a function of magnetic field angle 
$\vartheta$, with current along the [100] direction, for ${\vec{\cal |B|}} \! =\! 20$T, temperature $T\! =\! 10$K, and 
impurity scattering length $\Lambda=5a$.
We highlight a few observations from these results: (i)
The angle averaged value of the AMR signal indicates a large negative MR as observed 
in the experiments. 
(ii) We find that resistivity is significantly lower for fields perpendicular to the current. (iii) From the Fourier modes 
plotted in Fig.~\ref{fig:AMR001n0035}(b), we find several nonzero harmonics reflecting the complexity of the AMR signal.

In order to obtain a simple understanding of these results, we have carried out a detailed analysis to see which bands dominate the AMR.
Labelling the four bands as $n=1,2,3,4$ from the largest to the smallest, we find that the AMR mostly arises from a {\it single 
pair} of bands: the `circular' band-$2$ and the `propeller' band-$3$. We show the AMR and its harmonic decomposition
calculated by projecting to just these two bands in red in Fig.~\ref{fig:AMR001n0035}(a) and (b). In this simplified
calculation, we diagonalize the full Hamiltonian for the band wavefunctions and energies, but retain just the two 
coupled bands in solving the Boltzmann equation. We see that this two-band system semi-quantitatively captures the AMR and its harmonics.
To see why this simplified two-band scenario works well, we observe that the outer two bands, derived largely from the $xy$ orbitals,
are expected to dominate the diagonal conductivity of the system since they have a larger Fermi surface and a higher Fermi velocity 
compared with the quasi-1D inner bands. For impurity scattering with small momentum transfer, we also see that intra-band scattering
will not efficiently degrade the current carried by these bands since it will mostly lead to forward scattering. Thus, the transport
lifetime of these bands is limited by inter-band scattering. Indeed, scattering to the inner `propeller bands' provides an extremely 
efficient way to degrade current since these bands have a smaller Fermi velocity, and a FS shape which 
results in their Fermi velocity mostly pointing away from $\hat{x}$. 

We emphasize that keeping only two bands has a large impact on the 
angle-averaged MR ($C_0$) since the conductivity involves a sum over all bands, and throwing out
the largest FS is expected to significantly alter $\sigma_{xx}$. However, this procedure correctly captures the angular behavior of the signal. Furthermore, we observe that a more negative AMR is obtained when we restrict the calculation to bands 2 and 3. This is because this procedure involves explicitly switching off the scattering channel between bands 1 and 2 which is expected to lead to very little current decay since the group velocity is very large for both of these bands.

Fig.~\ref{fig:FS001B20}(a)-(c) show the Fermi surface for ${\vartheta=0,\pi/4,\pi/2}$. From the spin textures on the FSs, it is
clear that the outermost band-$1$ can continue to scatter strongly to the inner `propeller' band-$3$ whose shape permits efficient current decay. Thus,
band-$1$ does not have its conduction significantly altered by the magnetic field. However,
at this magnetic field, band-$2$ has its spin nearly antiparallel to the inner band-3 at any field angle while the spins were parallel at zero field (Fig. \ref{fig:FSB0}(a)), so any impurity-induced interband 
scattering to the inner band is strongly suppressed by applying a field. This leads to a magnetic field induced enhancement of transport lifetime for band-$2$, 
and a concomitant large negative magnetoresistance.

To further analyze the angle-dependence of the AMR, we plot in Fig.~\ref{fig:FS001B20}(a)-(c) the impurity scattering overlap matrix 
$|\langle n\bk|\hat{U}|m\bk'\rangle|^2\delta(\varepsilon_{n,\bk}-\varepsilon_{m,\bk'})$, keeping 
a fixed momentum $\bk$ on band-$2$ indicated by the blue cross (which is the current-carrying region), and for varying momentum 
$\bk'$ on band $m=3$. For $\vartheta=0$ as in Fig.~\ref{fig:FS001B20}(a), we see that the dominant scattering occurs to the forward elongated ellipse part of band-$3$, which has
its Fermi velocity typically directed away from $\hat{x}$, so the resulting transport lifetime will be short. Carriers preferably scatter to this region of the FS because it is closer to the initial momentum which satisfies the small-momentum transfering potential, and the spins are not fully antiparallel with the spin at $\bk$ (marked with the blue $\times$). However, when $\vartheta=\pi/2$ as in Fig.~\ref{fig:FS001B20}(c), the
scattering occurs into the vertical ellipse part of band-$3$, which has its Fermi velocity along $\hat{x}$, leading to less efficient current
decay, resulting in a longer transport lifetime. This portion of Fermi momenta is preferred because the previously favored region in the horizontal ellipse now has a fully antiparallel spin with the spin at $\bk$. Since the momentum transfer between $\bk$ and the vertical ellipse is large, the scattering potential suppresses this scattering channel which explains the small overlap matrix as seen from the colorbar. Thus, we expect the resistivity to be much lower for $\vartheta=\pi/2$. The reason the spins are not fully parallel or antiparallel stems from the competition between the Rashba energy scale which becomes important for large Fermi momenta and the magnetic field energy scale. A simple model that captures this behavior is provided in Appendix \ref{2band}. If the 
$\vartheta$-dependence of this overlap pattern was smooth, we would expect a single $\cos(2\vartheta)$ harmonic in the AMR; however,
the pattern at $\vartheta=\pi/4$ (see Fig.~\ref{fig:FS001B20}(b)) is nearly the same as for $\vartheta=0$, so the pattern changes abruptly with angle 
for $\vartheta > \pi/4$, resulting
in many harmonics $\cos  (2 m \vartheta)$ in the AMR. In particular, the AMR has $\cos (6\vartheta)$ components, which are symmetry-allowed harmonics 
although the FS itself does not have six-fold symmetry. We do not, at this point, have a simple intuitive understanding of the abrupt change in the overlap
matrix for $\vartheta > \pi/4$.

\section{(111) 2DEG}
The (111) 2DEG at oxide surfaces and interfaces has also been studied using ARPES \cite{rodel_orientational_2014,Dudy}. Based on such experiments, we can write a general $6\times6$ Hamiltonian and fit hopping parameters to qualitatively reproduce the shape of the Fermi surfaces seen in experiments.

\subsection{Model Hamiltonian}
We begin with the zero field Hamiltonian 
\be
{H^{111}_0=\sum_{\bk\sigma\ell\ell'} c^\dg_{\ell \sigma}(\bk) {\cal M}^{111}_{\ell\ell'}(\bk) c^\pdg_{\ell' \sigma}(\bk)},
\ee
defined on a triangular lattice, where
\bea
\!\!\!\!\!\! {\cal M}^{111} &=&
\begin{pmatrix}\!\!
\! \varepsilon_{yz} & \gamma_{a} \!+\! \i \zeta_{bc} \!-\! \frac{\Delta}{3} &  \gamma_{b} \!-\! \i \zeta_{ac} \!-\! \frac{\Delta}{3} \! \\
\! \gamma_{a} \!-\! \i \zeta_{bc} \!-\! \frac{\Delta}{3}   & \varepsilon_{zx} & \gamma_{c} \!+\!  \i \zeta_{ab} \!-\! \frac{\Delta}{3} \!  \\
\! \gamma_{b} \!+\! \i \zeta_{ac}  \!-\! \frac{\Delta}{3}   &  \gamma_{c} \!-\! \i \zeta_{ab} \!-\! \frac{\Delta}{3}  & \varepsilon_{xy} \!
\end{pmatrix}
\eea
Similar to the (001) case, we use the abbreviated notation $c_i,s_i$, with $\hat{i}=\hat{a},\hat{b},\hat{c}$, where
$\hat{a} \!=\! \hat{x}$, ${\hat{b} \!=\! -\hat{x}/2-\hat{y} \sqrt{3}/2}$, and $\hat{c} \!=\! -\hat{x}/2+\hat{y} \sqrt{3}/2$.
With this notation, the intra-orbital dispersions are
\bea
\varepsilon_{yz} &=& 2 t_1 (1-c_c) + 2 t_2 (2-c_a-c_b), \\
\varepsilon_{zx} &=& 2 t_1 (1-c_b) + 2 t_2 (2-c_c-c_a), \\
\varepsilon_{xy} &=& 2 t_1 (1-c_a) + 2 t_2 (2-c_b-c_c),
\eea
while the inter-orbital hybridization
$\gamma_i \equiv -2 t_3 c_i$, the odd-in-momentum term $\zeta_{ij} = 2\zeta (s_i + s_j)$ represents hopping
permitted by broken inversion symmetry at the interface, and $\Delta$ is a symmetry-allowed trigonal distortion term.
This Hamiltonian is again supplemented with $H_{\rm SOC} + H_{\cal B}$
as for the (001) 2DEG.

The ARPES data on the STO (111) surface \cite{rodel_orientational_2014}
are reasonably fit by choosing $(t_1, t_2, t_3) = (320,13,-13)$meV. In addition, for simplicity, we pick $(\zeta,\lambda) \equiv (8,5)$meV 
as for the (001) 2DEG. We also use the same $g$-factors $g_\ell=1$ and $g_s=5$ for coupling to the in-plane $\vec{\cal{B}}$
field, with the field angle $\vartheta$ being defined with respect to the $[\bar{1}10]$ crystal axis.

We have explored several values of the symmetry-allowed trigonal distortion scale $\Delta$ as shown in Fig. \ref{fig:Delta}. We find that the AMR is reasonably
described using values of ${\Delta \approx 50-80}$meV, which are similar to the scale of ${\Delta_T = 60}$meV in the (001) 2DEG.
It is important to note that this local distortion energy scale is different from the band degeneracy splitting at the $\Gamma$-point
of the BZ; this is given by 
$6 t_3 + \Delta$ for the (111) 2DEG, unlike the (001) case where it is just $\Delta_T$. Furthermore, it has been recently pointed out
that this scale can be density-dependent due to renormalization by electron-electron interactions \cite{Caviglia2018,Goldstein2019}. Here, for simplicity, we focus on
a single density $n=0.05$e/Ti, and we view $\Delta$ as the renormalized distortion.

\subsection{Magnetotransport}
\begin{figure}[t]
	\centering
	\includegraphics[width=\linewidth]{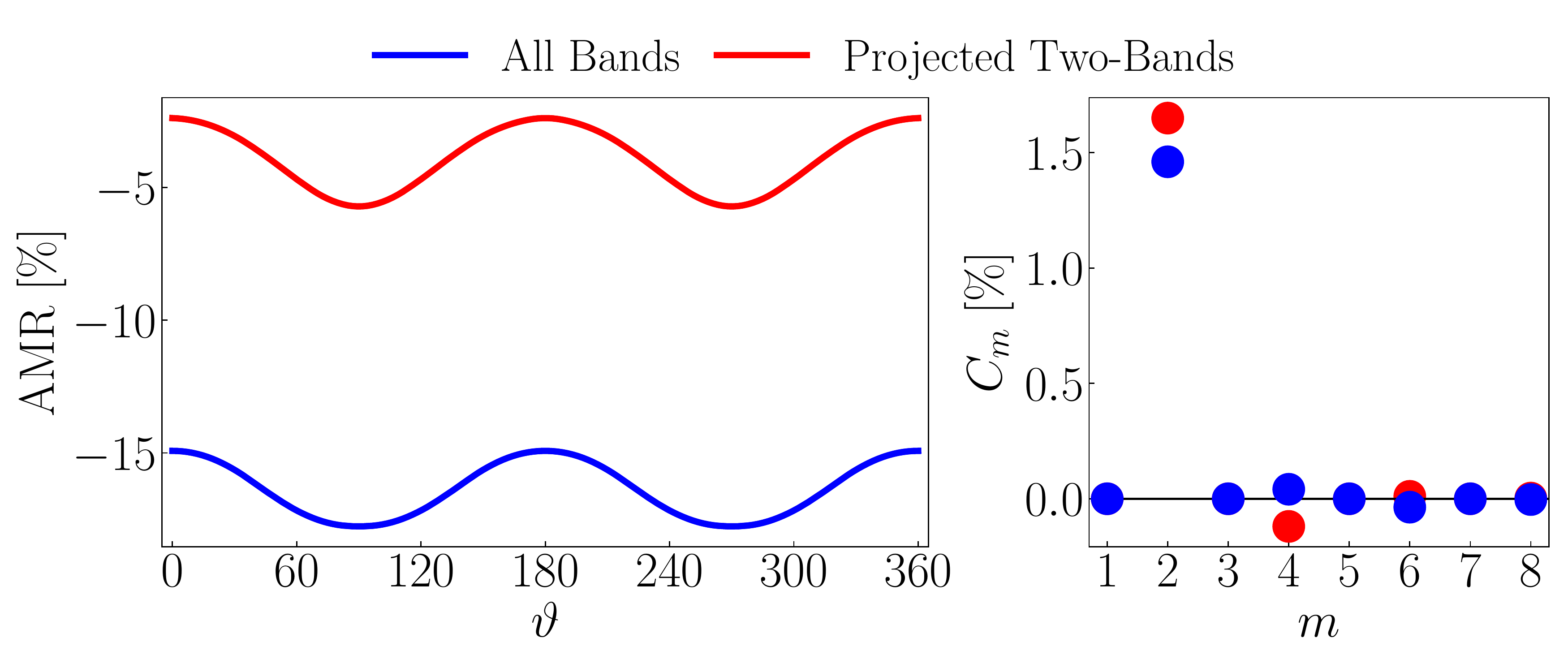} 
	\caption{AMR (left) and its Fourier modes (right) for the $(111)$ surface for electronic density $n=0.05$e/Ti (corresponding to $1.6\times10^{13}\text{cm}^{2}$) 
	and at $T=5$K. The strength of the magnetic field is kept constant at $|\vec{\mathcal{B}}|=20$T and its angle $\vartheta$ is varied with respect to the 
	current which is along the $[\bar{1}10]$ crystal direction. 
	The impurity scattering length is taken to be $\Lambda=5a$. Similar to Fig. \ref{fig:AMR001n0035}, the blue curve is the solution obtained 
	when keeping all Fermi surfaces shown in Fig.\ref{fig:FSB0}(b), while the red curve is obtained when only keeping the two outermost `flower-like' bands $1$ and $2$.}
	\label{fig:AMR111n005}
\end{figure}

\begin{figure}[t]
\centering
	\includegraphics[width=\linewidth]{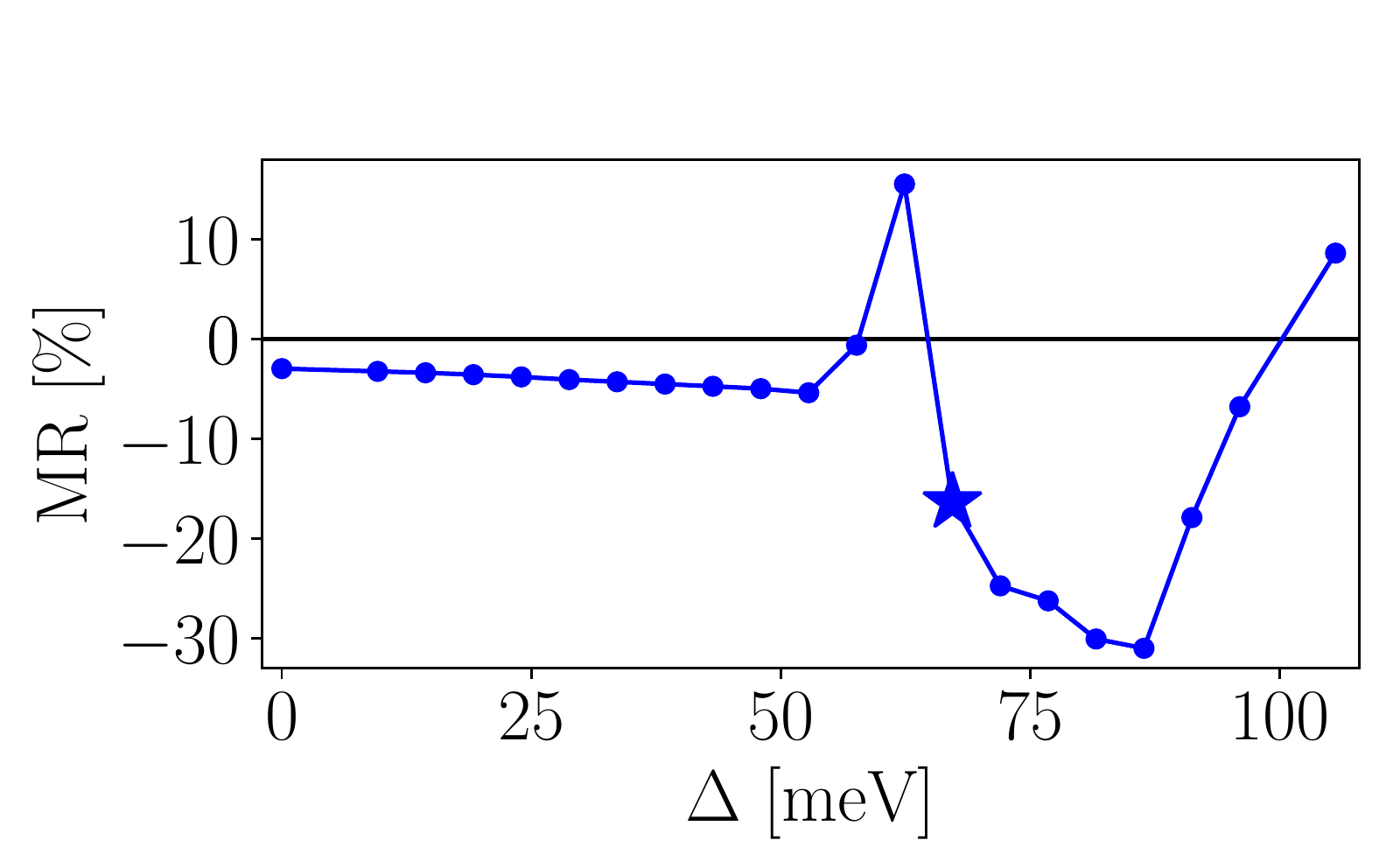} 
	\caption{Effect of trigonal distortion $\Delta$ on the (111) MR at a fixed density $n=0.05$e/Ti and a temperature $T=5$K. 
	The point marked $\star$ corresponds to $\Delta=70$meV, which is used to compute the full AMR signal. }
	\label{fig:Delta}
\end{figure}

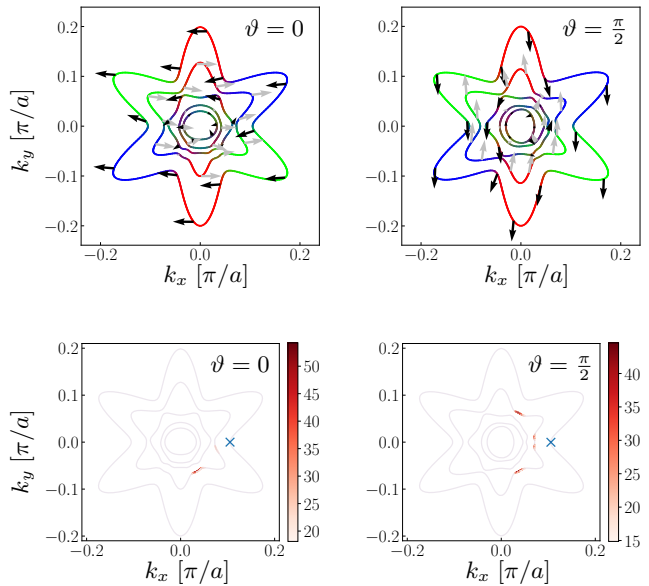
\begin{figure}[t]
	\centering
	\begin{overpic}[width=0.49\linewidth]{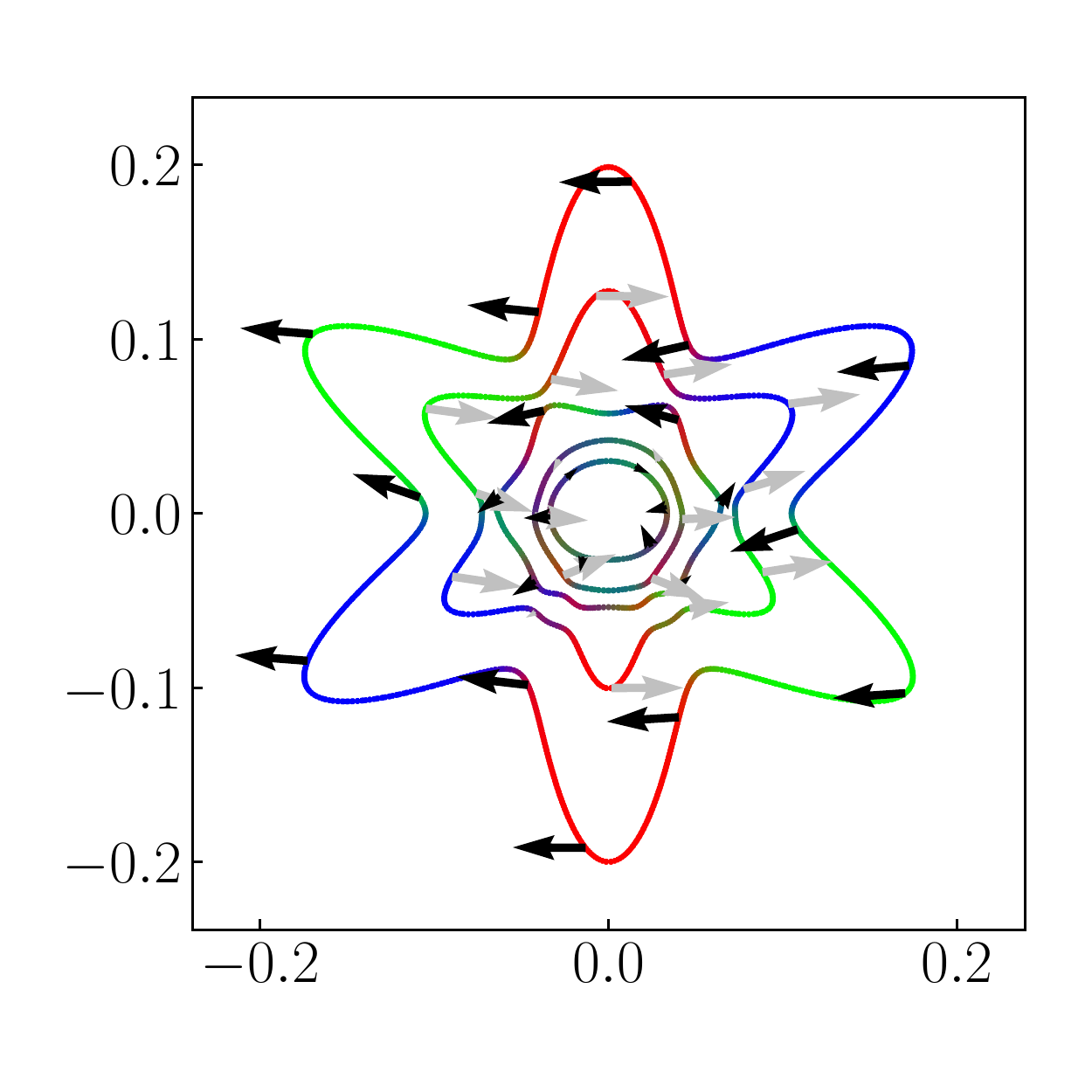}\put(70,82){$\vartheta=0$}\put(44,3){$k_x\;[\pi/a]$}\put(-5,38){\rotatebox{90}{$k_y\;[\pi/a]$}}\end{overpic}
	\begin{overpic}[width=0.49\linewidth]{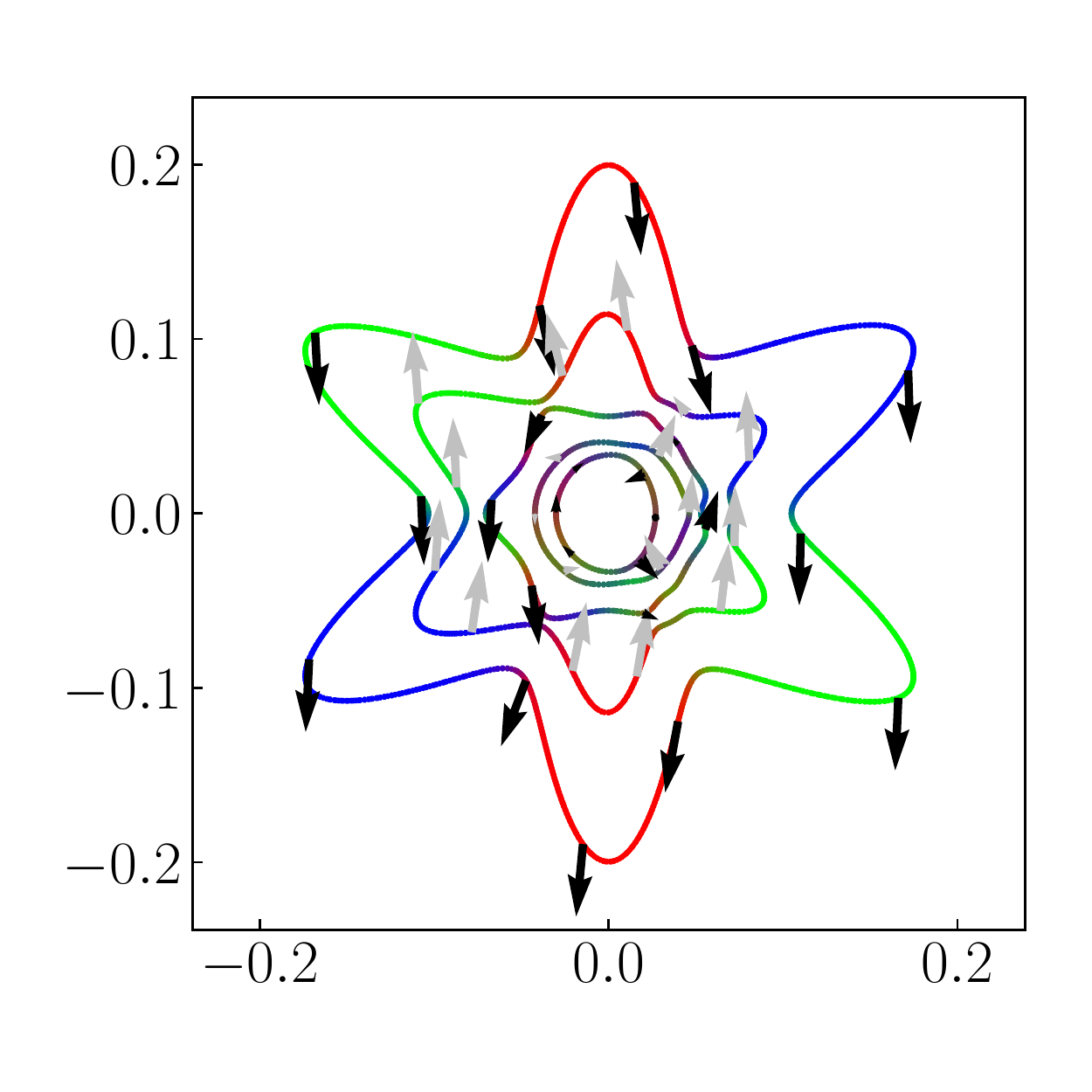}\put(69,82){$\vartheta=\frac{\pi}{2}$}\put(44,3){$k_x\;[\pi/a]$}\end{overpic} 
	\begin{overpic}[width=0.49\linewidth]{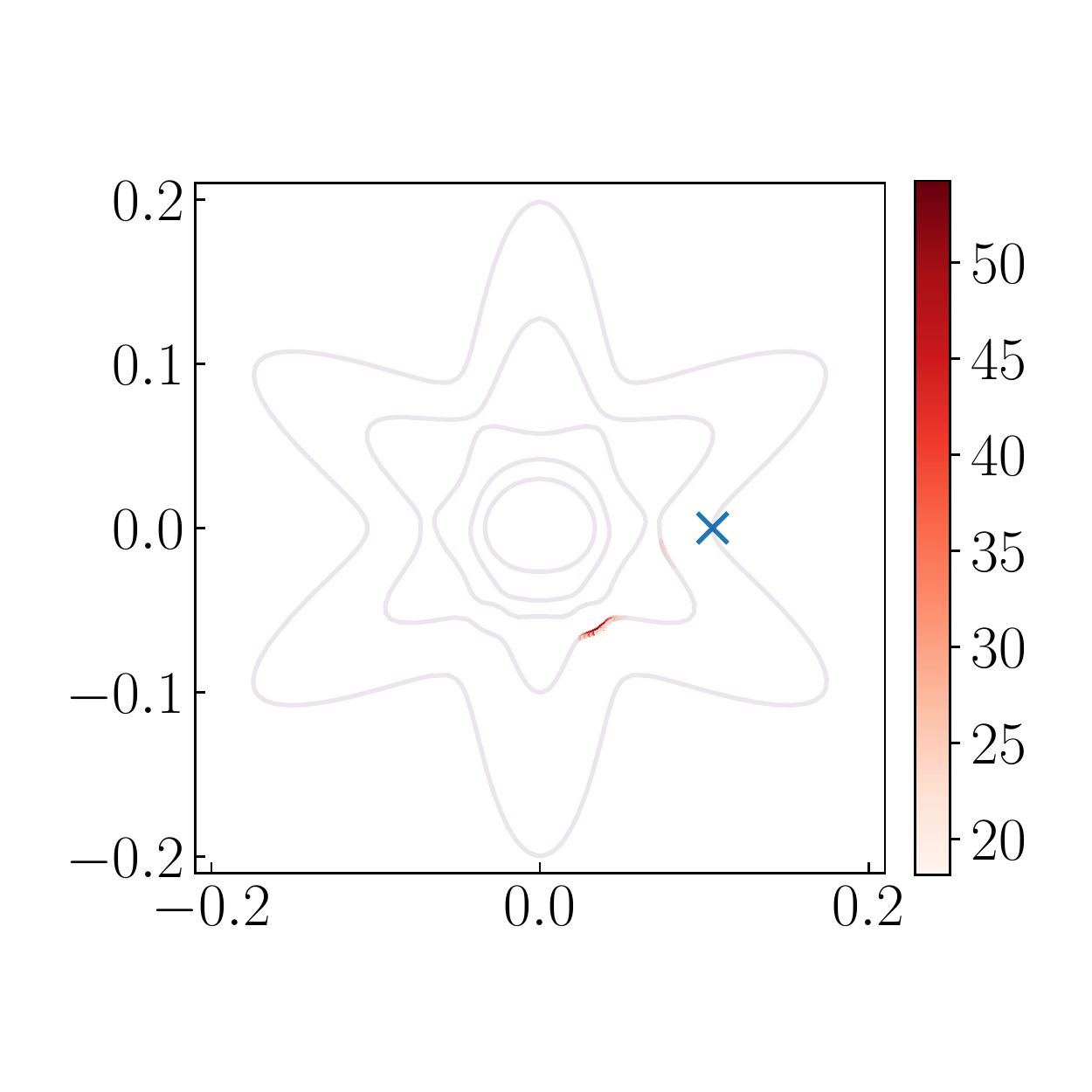}\put(59,74){$\vartheta=0$}\put(38,8){$k_x\;[\pi/a]$}\put(-5,35){\rotatebox{90}{$k_y\;[\pi/a]$}}\end{overpic}
	\begin{overpic}[width=0.49\linewidth]{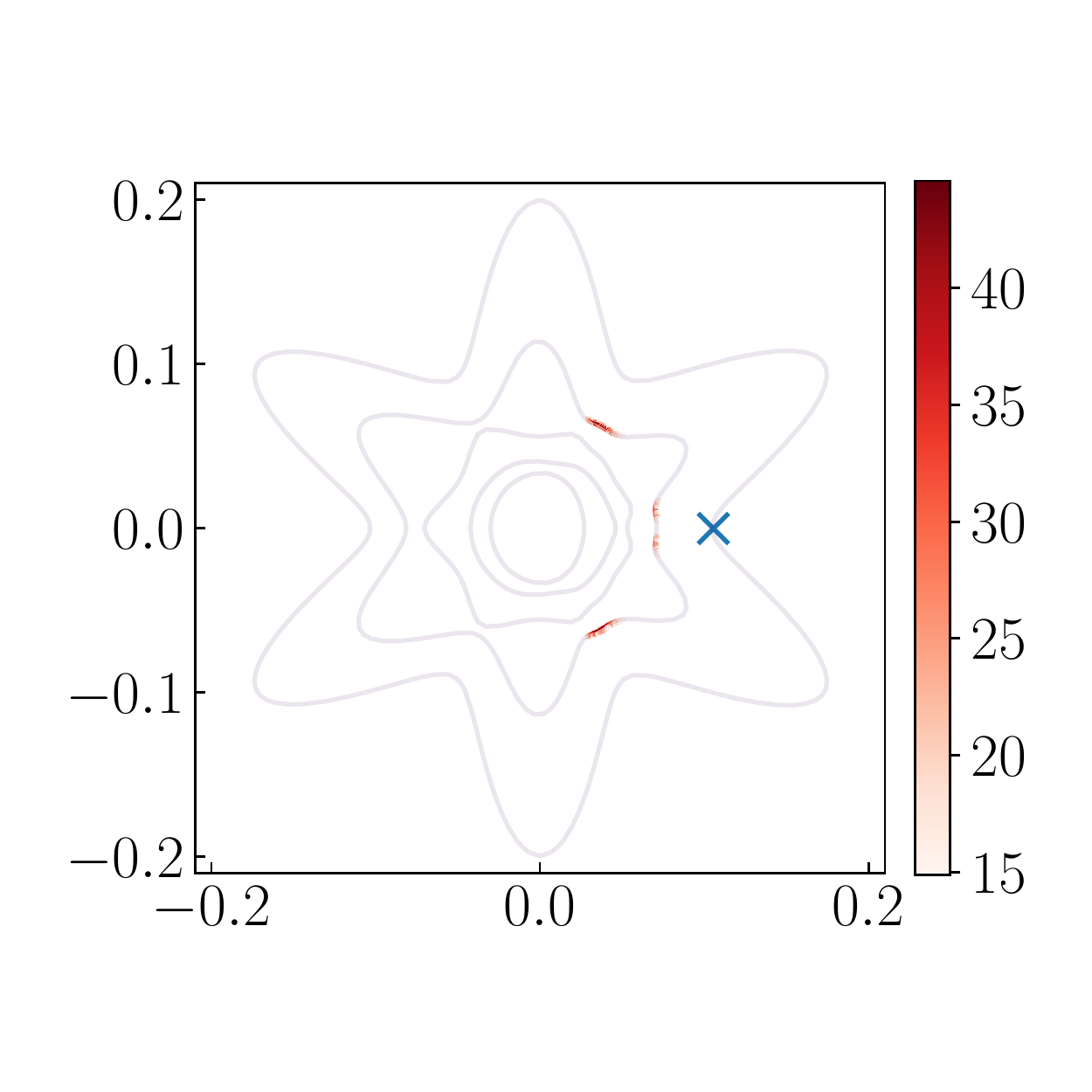}\put(58,74){$\vartheta=\frac{\pi}{2}$}\put(38,8){$k_x\;[\pi/a]$}\end{overpic}
	\caption{\textbf{Top panel:} (111) Fermi surfaces and spin textures at nonzero magnetic field $|\vec{\mathcal{B}}|=20$T for
	$\vartheta=0,\pi/2$; black/grey arrows indicate spin textures on bands with opposite Rashba spin chiralities at zero field. 
	\textbf{Bottom panel:} Integrand of the Boltzmann equation ${|\langle n\bk|\hat{U}|m\bk'\rangle|^2\delta(\varepsilon_{n,\bk}-\varepsilon_{m,\bk'})}$ with $\Lambda=5a$ for scattering
	from fixed band $m=1$ and $\bk'=k_F\hat{x}$ (marked with a blue x) to band $n=2$ as a function of final momentum $\bk$.}
	\label{fig:FS111B20}
\end{figure}

For $n \! =\! 0.05$e/Ti and $\Delta \! =\! 70$meV, the (111) 2DEG exhibits four Fermi surfaces with Rashba-like spin textures
as shown in Fig. \ref{fig:FSB0}(b). The calculated AMR signal for this case is shown as the blue curve 
in Fig.~\ref{fig:AMR111n005}. We find
that the magnitude of the angle-averaged MR is smaller than for the (001) case, but it is tunable by changing the
trigonal distortion $\Delta$ as seen from Fig.~\ref{fig:Delta}.
In all cases, we find that the AMR has $C_2$ character, with no sign
of higher harmonics. This is qualitatively consistent with experimental observations.
We have found that this $C_2$ dominant response, with almost no sign of higher harmonics, holds true even upto much higher 
densities $n \sim 0.4$e/Ti. 

To understand the average MR, we plot the spin texture on the FSs in Fig.~\ref{fig:FS111B20} at nonzero ${ \cal \vec B}$, which shows that there is
a significant contrast with the (001) 2DEG --- namely,
the outer pair of bands features opposite spin polarizations, and so does the inner pair of bands. As a result,
each outer band (band-$1$ or band-$2$) has a corresponding inner band (band-$3$ or band-$4$) into which it can scatter
even when ${|\vec{\mathcal{B}}|}\neq0$; this leads to a 
reduced MR, since no scattering channel is ``switched off'' by the field unlike in the (001) 2DEG. The tunability
of the MR with $\Delta$ arises due to changes in the FS spin textures, such that an inner band flips its 
spin polarization; the mechanism for large negative MR then parallels that of the (001) 2DEG. 

We observe that 
the projected two-band calculation leads to a more positive angle-averaged AMR, unlike in the (001) case.
This is specific to the choice of $\Delta$ presented here, and is not a universal feature. Indeed, since the (111) surface
is mainly governed by the outermost FSs which have large velocities, no efficient current-degrading scattering channel is 
switched-off when the calculation is projected to these two outer bands. For this reason, whether the projected or the full
calculation leads to the most negative $C_0$ coefficient is more difficult to predict and depends on the choice of parameters.

The sign of the $C_2$ harmonic in the AMR may be understood from the overlap matrix plots in Fig.~\ref{fig:FS111B20}. It is clear that
both $\vartheta=0$ and $\vartheta=\pi/2$ have scattering from the marked blue cross on band-$1$ to momenta on band-$2$
where the Fermi velocity points away from $\hat{x}$, leading to current dissipation.
However, for $\vartheta=\pi/2$, the magnitude of this overlap is smaller (see the color scale), and furthermore
it has some scattering into band-$2$ where the Fermi velocity is still along $\hat{x}$,
which suppresses the resistivity for this field angle.

We attribute the absence of higher harmonics in the AMR
signal to the fact that each band in the (111) 2DEG has an equal (momentum dependent)
mixture of all three orbitals $yz,zx,xy$, which independently have strong directional character. This is different from the (001)
2DEG where the more unidirectional $yz,zx$ orbitals are split off from the symmetric $xy$ orbitals.

\section{Impact of symmetry breaking in the (111) 2DEG: A 2D Polar Metal}\label{nem}

We turn next to the question of how directional symmetry breaking at the (111) interface, which leads to a 2D polar metal phase,
might impact the AMR in the 2DEG.
Our work is partly motivated by experimental reports of nematicity in transport measurements in such 2DEGs \cite{miao_anisotropic_2016,rout_six-fold_2017,Chandrasekhar2018,Goldstein2019,ADMI:ADMI201600830}.
Such symmetry breaking might have its origin in the bulk 3D structure; e.g., bulk SrTiO$_3$ has a structural cubic-to-tetragonal transition 
upon lowering temperature \cite{aharony_trigonal--tetragonal_1977} $T \lesssim 100$K. If the tetragonal domains are aligned, it will impose a nematic distortion for 2DEGs at the 
surface or interface of such a crystal. In high density 2DEGs, symmetry breaking may also be driven by electron interactions \cite{paramekanti2018}, which
can lead to orbital-ordering or cause a Pomeranchuk instability of the FS. Finally, as discussed in the Introduction,
the surface or interface hosting the 2DEG might undergo a
surface phase transition, leading to a polar metal breaking discrete rotational and mirror symmetries of the 2DEG.

\subsection{Landau theory and coupling to electrons}

\begin{figure}[t]
	\begin{overpic}[width=0.49\linewidth]{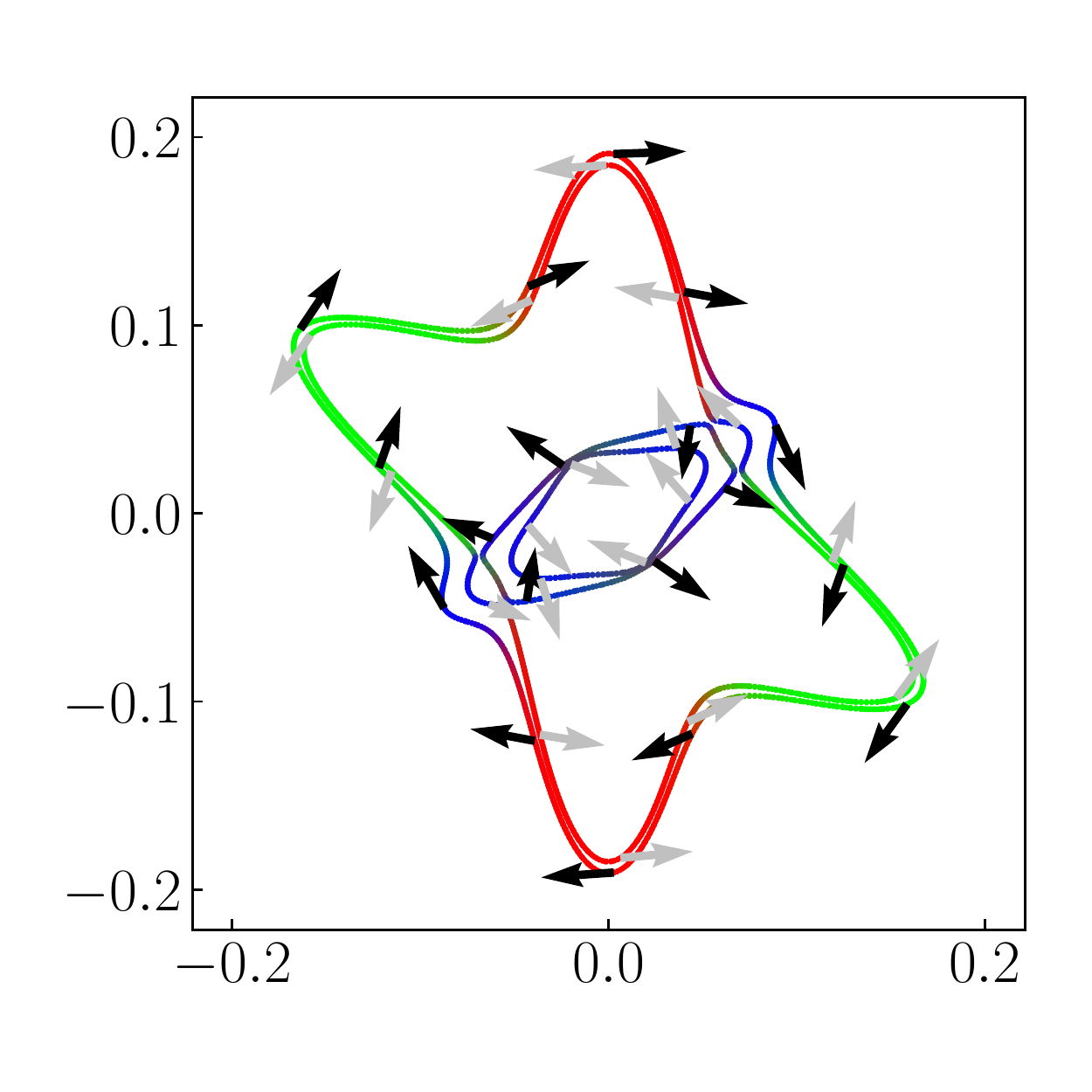}\put(20,80){(a)}\put(45,3){$k_x\;[\pi/a]$}\put(-5,38){\rotatebox{90}{$k_y\;[\pi/a]$}}
	\end{overpic}
	\begin{overpic}[width=0.49\linewidth]{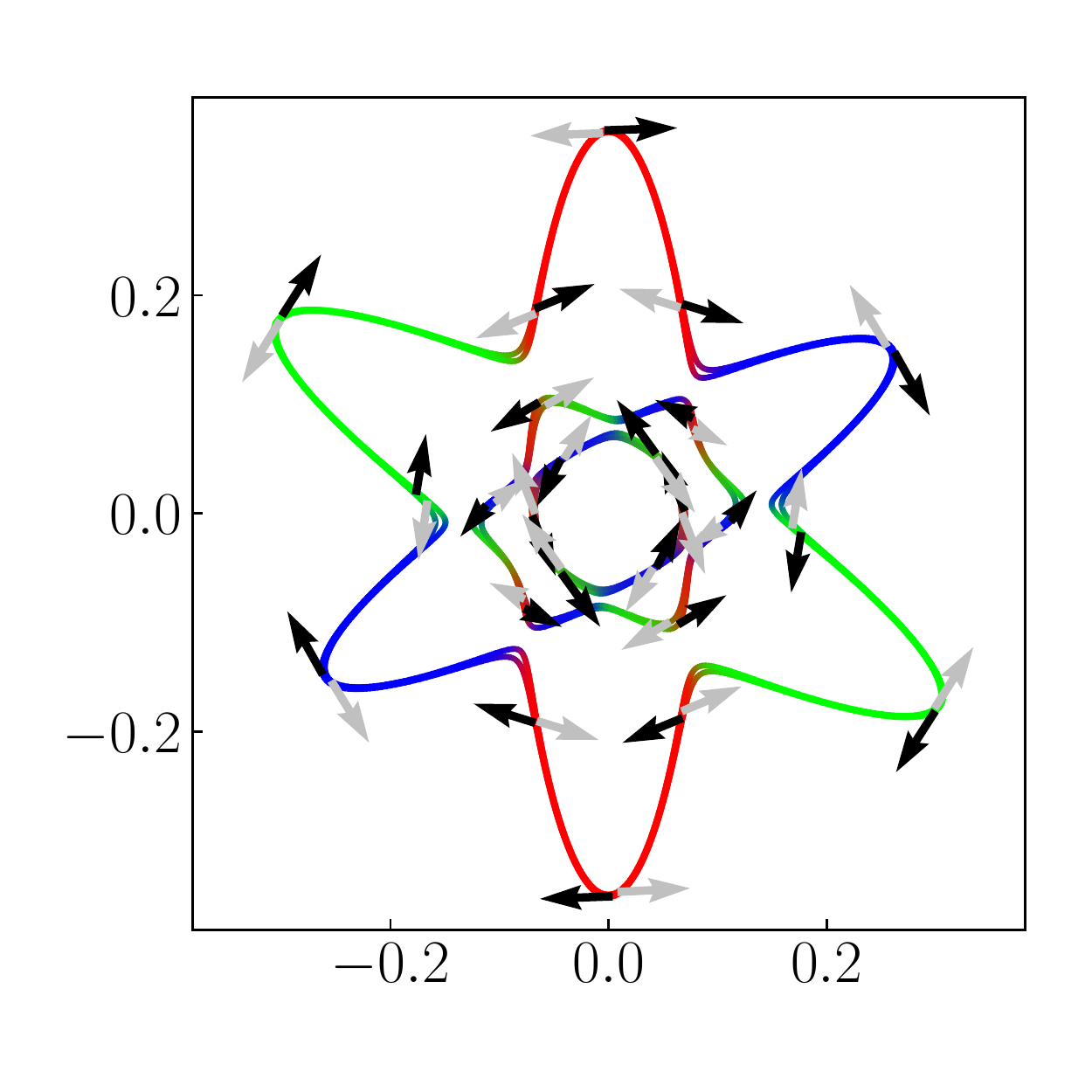}\put(20,80){(b)}\put(45,3){$k_x\;[\pi/a]$}
	\end{overpic}
	\begin{overpic}[width=0.49\linewidth]{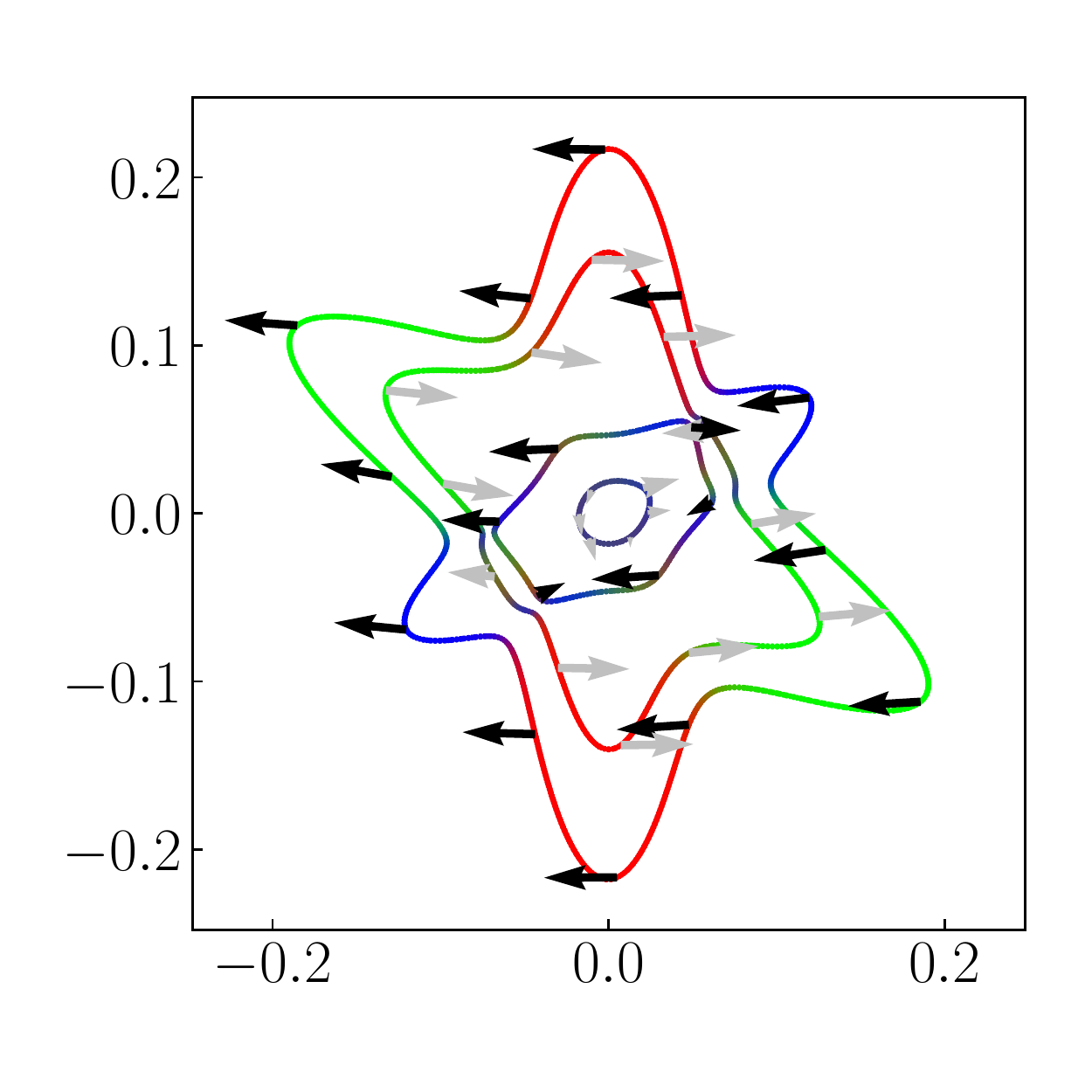}\put(20,80){(c)}\put(45,3){$k_x\;[\pi/a]$}\put(-5,38){\rotatebox{90}{$k_y\;[\pi/a]$}}
	\end{overpic}
	\begin{overpic}[width=0.49\linewidth]{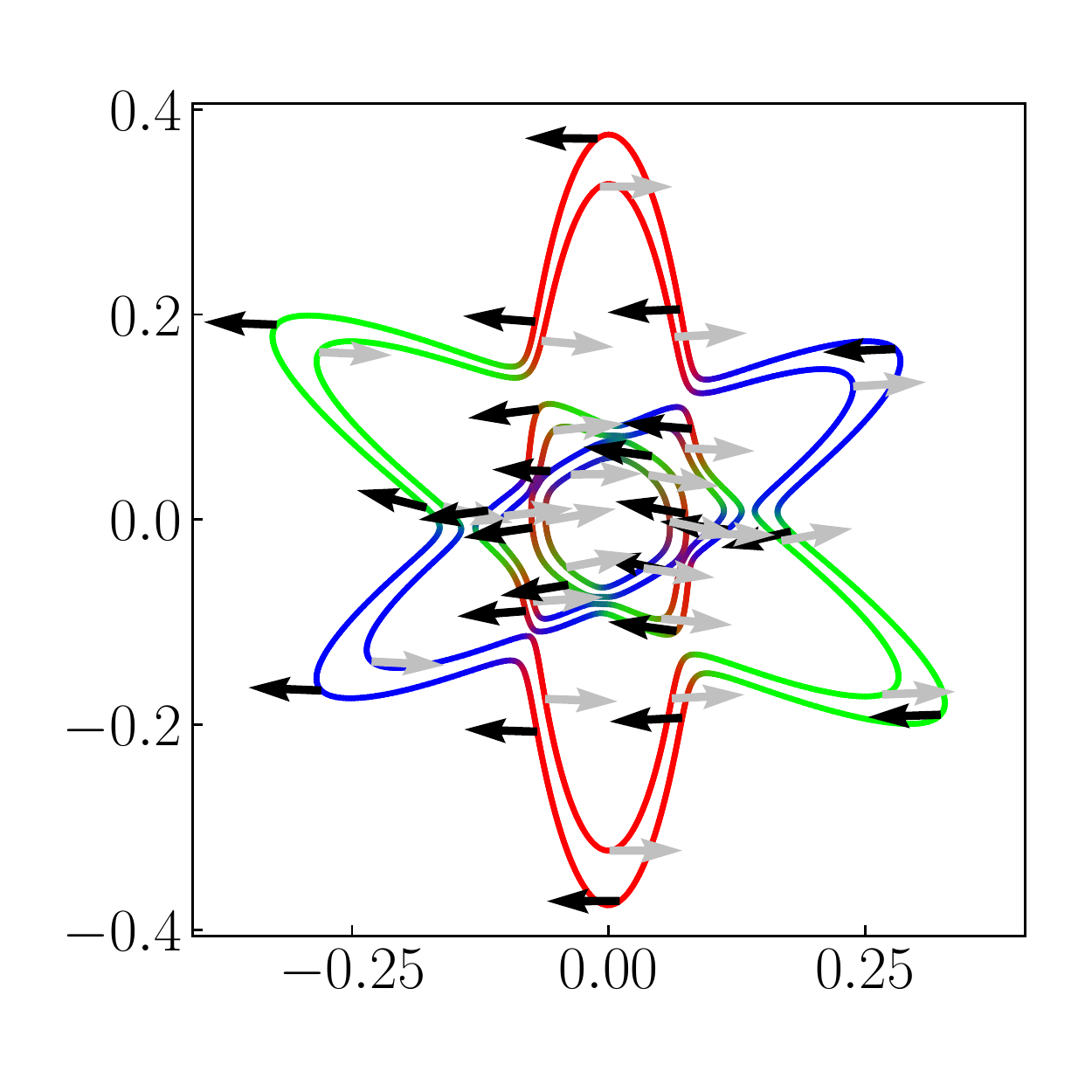}\put(20,80){(d)}\put(45,3){$k_x\;[\pi/a]$}\end{overpic}
	\caption{FSs and spin textures in the nematic 2DEG (polar metal), with couplings
	$v_1=v_2=-11$meV and phase $\chi=\pi$ in Eq.~\ref{Hsymmbroken}, at two different densities: (a) $n=0.05$e/Ti and (b) $0.2$e/Ti. 
	The impact of $\psi\neq0$ is more visually pronounced at smaller densities. Panels (c) and (d) show corresponding FSs at nonzero field ${|\vec{\mathcal{B}}|}=20$T. 
	In panels (a) and (b), arrows of different colors (black/grey) belong to FSs that are almost degenerate (and hence difficult to resolve visually) 
	but which come with opposite chiralities.}
	\label{fig:FS111B0sb}
\end{figure}
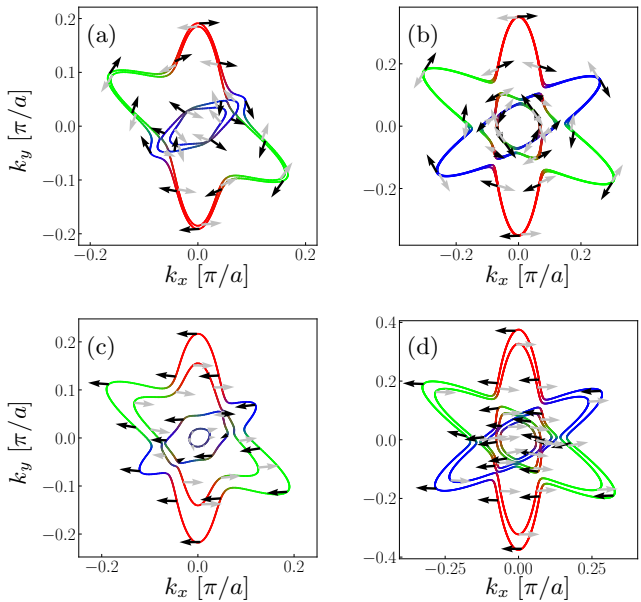

Independent of its microscopic origin, the order parameter for such a nematic in (111) 2DEGs is a complex scalar $\psi$, with the
Landau theory taking the form
\bea
{\cal F}_{111} = r |\psi|^2 + w (\psi^3 + \psi^{*3}) + u |\psi|^4 + \ldots,
\eea
where cubic terms lead to an effective $Z_3$ clock model.
 A similar model was studied long ago for bulk SrTiO$_3$ in
the presence of a stress applied along
the (111) direction \cite{muller_indication_1991}. We present a heuristic derivation of this in Appendix \ref{landau}, discussing its relation
to the proximity of SrTiO$_3$ to a paraelectric-ferroelectric quantum phase transition.

\newcommand{\Lxy}{\overline{L_x L_y}}
\newcommand{\Lyz}{\overline{L_y L_z}}
\newcommand{\Lzx}{\overline{L_z L_x}}
\newcommand{\Lxz}{\overline{L_x L_z}}

The impact of this
symmetry breaking is captured by a local linear-in-$\psi$ coupling to the orbitals via
\bea
\!\!\!\! H_{\psi} &=& - \frac{1}{2} \tilde{v}_1 \left[(L_x^2 \!+\!  \omega L^2_y \!+\!  \omega^2 L^2_z)\;\psi \!+\!  {\rm h.c.} \right] \nonumber\\
\!\!\!\!\!\!&-& \frac{1}{2} \tilde{v}_2 \left[(\Lyz \!+\! \omega \Lzx \!+\! \omega^2 \Lxy)\;\psi \!+\! {\rm h.c.}\right], 
\eea
where $\omega = \mathrm{e}^{\i 2\pi/3}$, and we have defined the symmetrized product
$\overline{L_i L_j} = L_i L_j + L_j L_i$. Explicitly, this leads to an orbital Hamiltonian
\bea
\!\!\!\!\!\! H_{\psi} \! &\!=\! &\! \begin{pmatrix} \!\!\!\! v_1 \! \cos(\chi) & v_2\! \cos(\chi \!+\! \frac{4\pi}{3})  & v_2 \! \cos(\chi \!+\! \frac{2\pi}{3}) \\ v_2\! \cos(\chi \!+\! \frac{4\pi}{3}) & 
v_1\! \cos(\chi \!+\! \frac{2\pi}{3}) 
& v_2\! \cos(\chi) \\
 v_2 \! \cos(\chi \!+\! \frac{2\pi}{3}) & v_2 \! \cos(\chi) & v_1\! \cos(\chi \!+\! \frac{4\pi}{3}) \!\!
 \label{Hsymmbroken}
\end{pmatrix}
\eea
where we have set $\psi = |\psi| \mathrm{e}^{\i\chi}$, and
absorbed the amplitude of the order parameter into redefined coefficients
$v_{1,2} = |\psi| \tilde{v}_{1,2}$ which have dimensions of energy.

The resulting distorted FSs and their spin textures are shown in Fig.~\ref{fig:FS111B0sb} at two different densities $n=0.05$e/Ti and
$n=0.2$e/Ti. For illustrative purposes, we have chosen ${v_1=v_2=-11}$meV and $\chi=\pi$. 
We find that the observed distortion of the FS is harder to resolve at higher densities where ARPES studies \cite{rodel_orientational_2014,Dudy} have been
carried out on the (111) surface 2DEG of
SrTiO$_3$.

\begin{figure}[t]
	\centering
	\begin{overpic}[width=\linewidth]{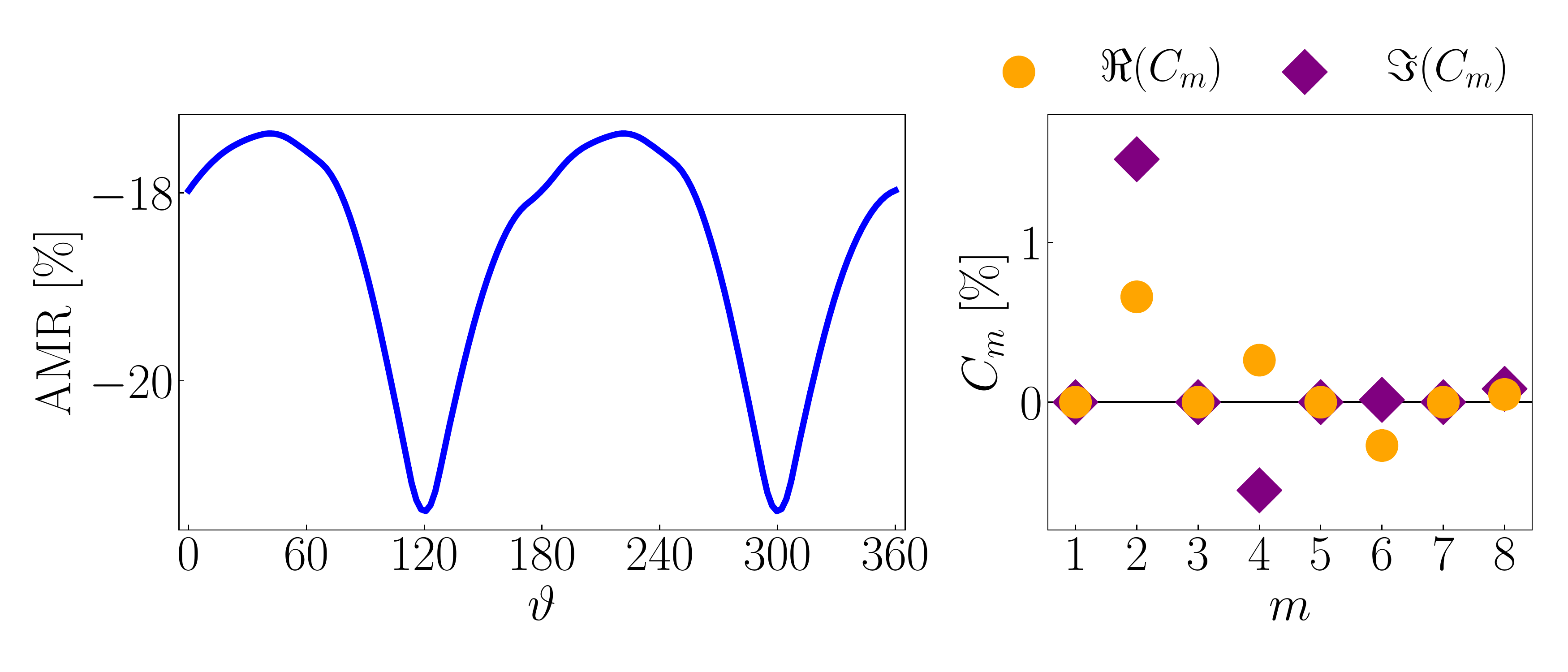}\put(25,37){All Bands}\end{overpic}\vskip 0.5cm
	\begin{overpic}[width=\linewidth]{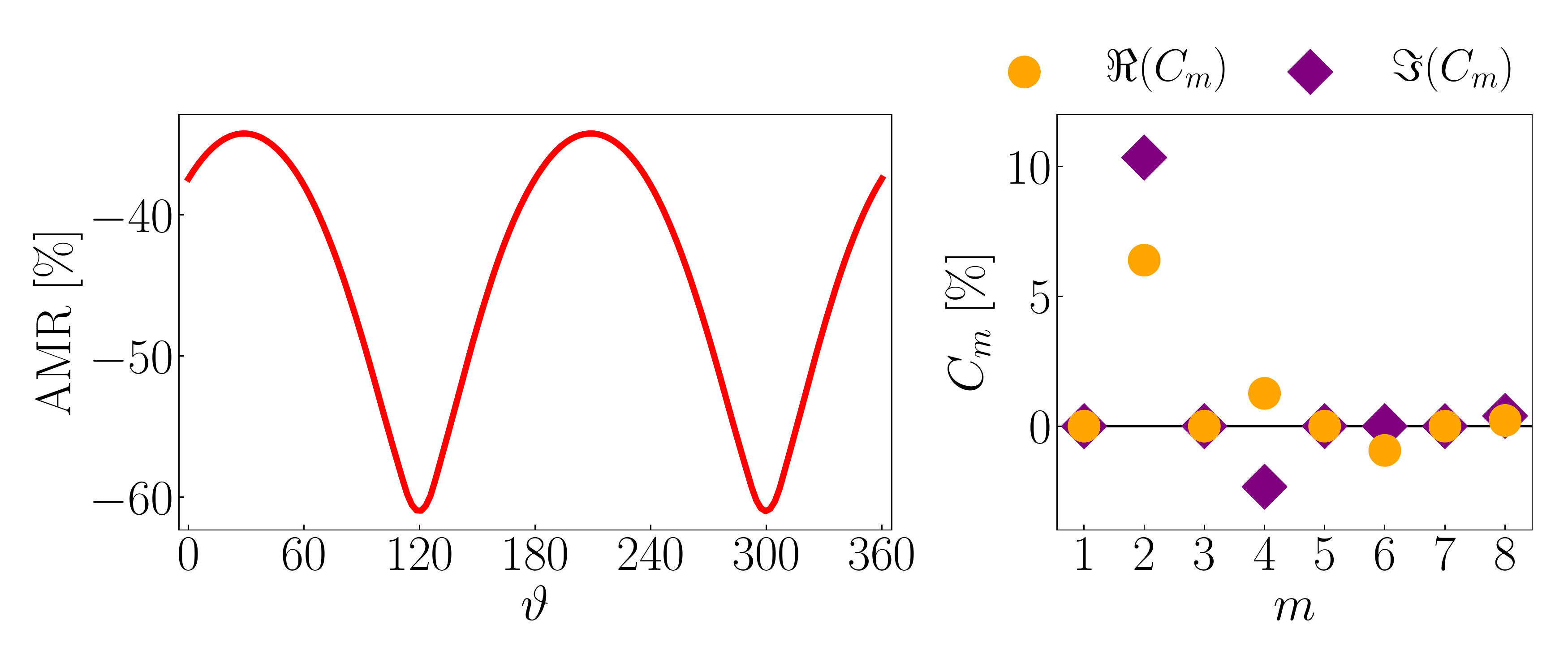} \put(18,37){Projected Two-Band}\end{overpic}
	\caption{AMR (left) and its Fourier modes (right) for the $(111)$ surface for electronic density $n=0.05$e/Ti including the symmetry breaking order parameter $\psi$ with ${v_1=v_2=-11}$meV and $\chi=\pi$. Top panels shows the AMR signal including all bands while the bottom panels are results from the projected calculation using
	band-$1$ and band-$4$. In the Fourier modes panels, circles and diamonds represent the real and imaginary parts of the different Fourier components $C_m$;
	the imaginary part is permitted in this case due to the symmetry breaking field $\psi$ which results in breaking $\vartheta\rightarrow-\vartheta$ symmetry of the AMR.
	Note that the maxima of the AMR signal are shifted away from $\vartheta=0,\pi$.}
	\label{fig:AMR111SB}
\end{figure}

\subsection{Magnetotransport}

The AMR in the presence of this symmetry breaking field, with ${v_1=v_2=-11}$meV and $\chi=\pi$,
is shown in Fig.~\ref{fig:AMR111SB}. We find that the AMR exhibits several new aspects absent
in the symmetric 2DEG:
(i) The AMR is no longer symmetric under
$\vartheta \to -\vartheta$, which allows for sinusoidal components in the Fourier decomposition. (ii) We find higher angular harmonics being generated
by the symmetry breaking. These observations appear consistent with experiments \cite{rout_six-fold_2017}, where the onset of higher $C_6$ harmonics in the AMR appears to
coincide with the breaking of the $\vartheta \to -\vartheta$ symmetry of the signal, suggesting that these two phenomena go hand in hand, and might be tied to
directional symmetry breaking in the 2DEG.
As shown in Fig.~\ref{fig:AMR111SB}, all of these results including the correct sign
of the Fourier modes, can be qualitatively reproduced by keeping just a single pair of bands in solving the Boltzmann equation. Unlike the non-symmetry broken phase discussed previously, the relevant bands are now bands 1 and 4 which have different shapes and orbital content -- the outermost band is mainly composed of $xy,zx$ orbitals while the innermost band has a predominant $yz$ character. This situation is similar to the (001) calculation where the relevant bands 2 and 3 had different geometries/orbital contents leading us to believe that this orbital imbalance and different geometry could be responsible for the generation of higher harmonics.

\section{Summary and Discussion}

We have considered model 2D Hamiltonians for both (001) and (111) oxide interface 2DEGs, and we showed that their AMR is
in reasonable agreement with experimental results. Such simplified model Hamiltonians are thus useful to describe the transport properties of these 2DEGs.
Our results, obtained by solving the full matrix Boltzmann equation, may be rationalized in terms of simplified two-band models and from
the momentum dependence of the scattering overlap matrix. We summarize below some of the
key results and discuss some speculative ideas: 

(i) We have shown that the MR and its angle dependence in the oxide 2DEGs appear to be governed by the field-dependent tuning of interband 
scattering.  

(ii) Octahedral distortions, such as the trigonal distortion at the (111) interface, can significantly affect the FSs and MR, suggesting that 
engineering or tuning such distortions may provide a viable route to controlling transport properties of such 2DEGs. 

(iii) We have argued, and
provided a Landau 
theory reasoning, for why (111) oxide interfaces might stabilize polar metal phases, particularly those involving SrTiO$_3$ which are proximate to a bulk
ferroelectric critical point. Our work suggests that the AMR and its symmetries may be used to indirectly detect such symmetry breaking. Furthermore,
the nematic resistivity observed \cite{Chandrasekhar2018,Venkatesan2017,ADMI:ADMI201600830} in insulating ultra-low density (111) 2DEGs may be viewed as arising 
from an ``Anderson-localized polar metal'' which exhibits anisotropic variable-range hopping. It would be interesting to further explore this regime.

(iv) While the dominant AMR mode is the
uniform $C_2$ angular component, we have shown that higher harmonics can emerge in symmetric (001) 2DEGs and in nematic/polar (111) 2DEGs. 
In particular, the higher AMR harmonics appear not to be directly related to FS symmetries. However, a common emerging picture,
based on comparing  the (001) and (111) 2DEGs, is that these harmonics may arise from a shape mismatch between the two bands that govern the AMR.

(v) Various experiments on the metallic (111) 2DEGs report a change in 
the AMR when cooling below $T^* \lesssim 20$-$25$K, which depends on density and is consistent with underlying polar symmetry breaking. This may reflect the 
actual symmetry breaking temperature scale of the 2DEG, or it might reflect a temperature at which the 2DEG effectively ``approaches'' the
symmetry-broken interface, with the polar order onset already occurring at a higher transition temperature. Future studies of this issue would
be valuable.

(vi) The concavity or convexity of the 2DEG FS is known to strongly impact Hall transport in perpendicular magnetic fields, 
\cite{AMRMagnetic,De_Ranieri_2008}. Here, we have focused on the diagonal resistivity under an in-plane magnetic field, and it 
is unclear how much the AMR is impacted under these circumstances by the FS curvature. We find that multiband scattering plays a more important role.

(vii) Finally, we have considered the simple case of scattering off a scalar impurity (independent of orbital and spin) in this paper. A future direction would be to study
orbital-dependent impurity scattering and the impact of spin-orbit randomness,\cite{Sherman2012,Bovenzi2017} 
using the scattering overlap matrix to provide insights.

\acknowledgements

We thank Venkat Chandrasekhar, Yoram Dagan, and Bill Atkinson, for useful discussions about oxide 2DEGs, and Ehud Altman for insightful conversations
about the possible connection between nematic/polar order of the 2DEG and the proximity of SrTiO$_3$ to a ferroelectric phase. This work was 
supported by NSERC of Canada, the CIfAR Quantum Materials Programme and FRQNT of Quebec. This research was enabled in part by support provided by Compute Ontario, Westgrid and Compute Canada. Computations were performed on the Niagara supercomputer at the SciNet HPC Consortium. SciNet is funded by: the Canada Foundation for Innovation; the Government of Ontario; Ontario Research Fund - Research Excellence; and the University of Toronto.

\bibliography{AMRSTO}

\begin{thebibliography}{64}%
\makeatletter
\providecommand \@ifxundefined [1]{%
 \@ifx{#1\undefined}
}%
\providecommand \@ifnum [1]{%
 \ifnum #1\expandafter \@firstoftwo
 \else \expandafter \@secondoftwo
 \fi
}%
\providecommand \@ifx [1]{%
 \ifx #1\expandafter \@firstoftwo
 \else \expandafter \@secondoftwo
 \fi
}%
\providecommand \natexlab [1]{#1}%
\providecommand \enquote  [1]{``#1''}%
\providecommand \bibnamefont  [1]{#1}%
\providecommand \bibfnamefont [1]{#1}%
\providecommand \citenamefont [1]{#1}%
\providecommand \href@noop [0]{\@secondoftwo}%
\providecommand \href [0]{\begingroup \@sanitize@url \@href}%
\providecommand \@href[1]{\@@startlink{#1}\@@href}%
\providecommand \@@href[1]{\endgroup#1\@@endlink}%
\providecommand \@sanitize@url [0]{\catcode `\\12\catcode `\$12\catcode
  `\&12\catcode `\#12\catcode `\^12\catcode `\_12\catcode `\%12\relax}%
\providecommand \@@startlink[1]{}%
\providecommand \@@endlink[0]{}%
\providecommand \url  [0]{\begingroup\@sanitize@url \@url }%
\providecommand \@url [1]{\endgroup\@href {#1}{\urlprefix }}%
\providecommand \urlprefix  [0]{URL }%
\providecommand \Eprint [0]{\href }%
\providecommand \doibase [0]{http://dx.doi.org/}%
\providecommand \selectlanguage [0]{\@gobble}%
\providecommand \bibinfo  [0]{\@secondoftwo}%
\providecommand \bibfield  [0]{\@secondoftwo}%
\providecommand \translation [1]{[#1]}%
\providecommand \BibitemOpen [0]{}%
\providecommand \bibitemStop [0]{}%
\providecommand \bibitemNoStop [0]{.\EOS\space}%
\providecommand \EOS [0]{\spacefactor3000\relax}%
\providecommand \BibitemShut  [1]{\csname bibitem#1\endcsname}%
\let\auto@bib@innerbib\@empty
\bibitem [{\citenamefont {Chakhalian}\ \emph {et~al.}(2014)\citenamefont
  {Chakhalian}, \citenamefont {Freeland}, \citenamefont {Millis}, \citenamefont
  {Panagopoulos},\ and\ \citenamefont {Rondinelli}}]{Review2014}%
  \BibitemOpen
  \bibfield  {author} {\bibinfo {author} {\bibfnamefont {Jak}\ \bibnamefont
  {Chakhalian}}, \bibinfo {author} {\bibfnamefont {John~W.}\ \bibnamefont
  {Freeland}}, \bibinfo {author} {\bibfnamefont {Andrew~J.}\ \bibnamefont
  {Millis}}, \bibinfo {author} {\bibfnamefont {Christos}\ \bibnamefont
  {Panagopoulos}}, \ and\ \bibinfo {author} {\bibfnamefont {James~M.}\
  \bibnamefont {Rondinelli}},\ }\bibfield  {title} {\enquote {\bibinfo {title}
  {Colloquium},}\ }\href {\doibase 10.1103/RevModPhys.86.1189} {\bibfield
  {journal} {\bibinfo  {journal} {Rev. Mod. Phys.}\ }\textbf {\bibinfo {volume}
  {86}},\ \bibinfo {pages} {1189--1202} (\bibinfo {year} {2014})}\BibitemShut
  {NoStop}%
\bibitem [{\citenamefont {Nakagawa}\ \emph {et~al.}(2006)\citenamefont
  {Nakagawa}, \citenamefont {Hwang},\ and\ \citenamefont
  {Muller}}]{nakagawa2006some}%
  \BibitemOpen
  \bibfield  {author} {\bibinfo {author} {\bibfnamefont {Naoyuki}\ \bibnamefont
  {Nakagawa}}, \bibinfo {author} {\bibfnamefont {Harold~Y}\ \bibnamefont
  {Hwang}}, \ and\ \bibinfo {author} {\bibfnamefont {David~A}\ \bibnamefont
  {Muller}},\ }\bibfield  {title} {\enquote {\bibinfo {title} {Why some
  interfaces cannot be sharp},}\ }\href {\doibase 10.1038/nmat1569} {\bibfield
  {journal} {\bibinfo  {journal} {Nature materials}\ }\textbf {\bibinfo
  {volume} {5}},\ \bibinfo {pages} {204--209} (\bibinfo {year}
  {2006})}\BibitemShut {NoStop}%
\bibitem [{\citenamefont {Ohtomo}\ and\ \citenamefont
  {Hwang}(2004)}]{ohtomo2004high}%
  \BibitemOpen
  \bibfield  {author} {\bibinfo {author} {\bibfnamefont {A.}~\bibnamefont
  {Ohtomo}}\ and\ \bibinfo {author} {\bibfnamefont {H.Y.}\ \bibnamefont
  {Hwang}},\ }\bibfield  {title} {\enquote {\bibinfo {title} {A high-mobility
  electron gas at the {LaAlO}$_3$/{SrTiO}$_3$ heterointerface},}\ }\href
  {\doibase 10.1038/nature02308} {\bibfield  {journal} {\bibinfo  {journal}
  {Nature (London)}\ }\textbf {\bibinfo {volume} {427}},\ \bibinfo {pages}
  {423--426} (\bibinfo {year} {2004})}\BibitemShut {NoStop}%
\bibitem [{\citenamefont {Thiel}\ \emph {et~al.}(2006)\citenamefont {Thiel},
  \citenamefont {Hammerl}, \citenamefont {Schmehl}, \citenamefont {Schneider},\
  and\ \citenamefont {Mannhart}}]{thiel2006tunable}%
  \BibitemOpen
  \bibfield  {author} {\bibinfo {author} {\bibfnamefont {S.}~\bibnamefont
  {Thiel}}, \bibinfo {author} {\bibfnamefont {G.}~\bibnamefont {Hammerl}},
  \bibinfo {author} {\bibfnamefont {A.}~\bibnamefont {Schmehl}}, \bibinfo
  {author} {\bibfnamefont {C.~W.}\ \bibnamefont {Schneider}}, \ and\ \bibinfo
  {author} {\bibfnamefont {J.}~\bibnamefont {Mannhart}},\ }\bibfield  {title}
  {\enquote {\bibinfo {title} {Tunable {Quasi}-{Two}-{Dimensional} {Electron}
  {Gases} in {Oxide} {Heterostructures}},}\ }\href {\doibase
  10.1126/science.1131091} {\bibfield  {journal} {\bibinfo  {journal}
  {Science}\ }\textbf {\bibinfo {volume} {313}},\ \bibinfo {pages} {1942--1945}
  (\bibinfo {year} {2006})}\BibitemShut {NoStop}%
\bibitem [{\citenamefont {Takizawa}\ \emph {et~al.}(2006)\citenamefont
  {Takizawa}, \citenamefont {Wadati}, \citenamefont {Tanaka}, \citenamefont
  {Hashimoto}, \citenamefont {Yoshida}, \citenamefont {Fujimori}, \citenamefont
  {Chikamatsu}, \citenamefont {Kumigashira}, \citenamefont {Oshima},
  \citenamefont {Shibuya}, \citenamefont {Mihara}, \citenamefont {Ohnishi},
  \citenamefont {Lippmaa}, \citenamefont {Kawasaki}, \citenamefont {Koinuma},
  \citenamefont {Okamoto},\ and\ \citenamefont {Millis}}]{Millis2006}%
  \BibitemOpen
  \bibfield  {author} {\bibinfo {author} {\bibfnamefont {M.}~\bibnamefont
  {Takizawa}}, \bibinfo {author} {\bibfnamefont {H.}~\bibnamefont {Wadati}},
  \bibinfo {author} {\bibfnamefont {K.}~\bibnamefont {Tanaka}}, \bibinfo
  {author} {\bibfnamefont {M.}~\bibnamefont {Hashimoto}}, \bibinfo {author}
  {\bibfnamefont {T.}~\bibnamefont {Yoshida}}, \bibinfo {author} {\bibfnamefont
  {A.}~\bibnamefont {Fujimori}}, \bibinfo {author} {\bibfnamefont
  {A.}~\bibnamefont {Chikamatsu}}, \bibinfo {author} {\bibfnamefont
  {H.}~\bibnamefont {Kumigashira}}, \bibinfo {author} {\bibfnamefont
  {M.}~\bibnamefont {Oshima}}, \bibinfo {author} {\bibfnamefont
  {K.}~\bibnamefont {Shibuya}}, \bibinfo {author} {\bibfnamefont
  {T.}~\bibnamefont {Mihara}}, \bibinfo {author} {\bibfnamefont
  {T.}~\bibnamefont {Ohnishi}}, \bibinfo {author} {\bibfnamefont
  {M.}~\bibnamefont {Lippmaa}}, \bibinfo {author} {\bibfnamefont
  {M.}~\bibnamefont {Kawasaki}}, \bibinfo {author} {\bibfnamefont
  {H.}~\bibnamefont {Koinuma}}, \bibinfo {author} {\bibfnamefont
  {S.}~\bibnamefont {Okamoto}}, \ and\ \bibinfo {author} {\bibfnamefont
  {A.~J.}\ \bibnamefont {Millis}},\ }\bibfield  {title} {\enquote {\bibinfo
  {title} {Photoemission from {Buried} {Interfaces} in
  {LaAlO}$_{3}$/{SrTiO}$_{3}$ {Superlattices}},}\ }\href {\doibase
  10.1103/PhysRevLett.97.057601} {\bibfield  {journal} {\bibinfo  {journal}
  {Phys. Rev. Lett.}\ }\textbf {\bibinfo {volume} {97}},\ \bibinfo {pages}
  {057601} (\bibinfo {year} {2006})}\BibitemShut {NoStop}%
\bibitem [{\citenamefont {Reyren}\ \emph {et~al.}(2007)\citenamefont {Reyren},
  \citenamefont {Thiel}, \citenamefont {Caviglia}, \citenamefont {Kourkoutis},
  \citenamefont {Hammerl}, \citenamefont {Richter}, \citenamefont {Schneider},
  \citenamefont {Kopp}, \citenamefont {R{\"u}etschi}, \citenamefont {Jaccard},
  \citenamefont {Gabay}, \citenamefont {Muller}, \citenamefont {Triscone},\
  and\ \citenamefont {Mannhart}}]{reyren2007superconducting}%
  \BibitemOpen
  \bibfield  {author} {\bibinfo {author} {\bibfnamefont {N.}~\bibnamefont
  {Reyren}}, \bibinfo {author} {\bibfnamefont {S.}~\bibnamefont {Thiel}},
  \bibinfo {author} {\bibfnamefont {A.~D.}\ \bibnamefont {Caviglia}}, \bibinfo
  {author} {\bibfnamefont {L.~Fitting}\ \bibnamefont {Kourkoutis}}, \bibinfo
  {author} {\bibfnamefont {G.}~\bibnamefont {Hammerl}}, \bibinfo {author}
  {\bibfnamefont {C.}~\bibnamefont {Richter}}, \bibinfo {author} {\bibfnamefont
  {C.~W.}\ \bibnamefont {Schneider}}, \bibinfo {author} {\bibfnamefont
  {T.}~\bibnamefont {Kopp}}, \bibinfo {author} {\bibfnamefont {A.-S.}\
  \bibnamefont {R{\"u}etschi}}, \bibinfo {author} {\bibfnamefont
  {D.}~\bibnamefont {Jaccard}}, \bibinfo {author} {\bibfnamefont
  {M.}~\bibnamefont {Gabay}}, \bibinfo {author} {\bibfnamefont {D.~A.}\
  \bibnamefont {Muller}}, \bibinfo {author} {\bibfnamefont {J.-M.}\
  \bibnamefont {Triscone}}, \ and\ \bibinfo {author} {\bibfnamefont
  {J.}~\bibnamefont {Mannhart}},\ }\bibfield  {title} {\enquote {\bibinfo
  {title} {Superconducting interfaces between insulating oxides},}\ }\href
  {\doibase 10.1126/science.1146006} {\bibfield  {journal} {\bibinfo  {journal}
  {Science}\ }\textbf {\bibinfo {volume} {317}},\ \bibinfo {pages} {1196--1199}
  (\bibinfo {year} {2007})}\BibitemShut {NoStop}%
\bibitem [{\citenamefont {Caviglia}\ \emph {et~al.}(2008)\citenamefont
  {Caviglia}, \citenamefont {Gariglio}, \citenamefont {Reyren}, \citenamefont
  {Jaccard}, \citenamefont {Schneider}, \citenamefont {Gabay}, \citenamefont
  {Thiel}, \citenamefont {Hammerl}, \citenamefont {Mannhart},\ and\
  \citenamefont {Triscone}}]{caviglia2008electric}%
  \BibitemOpen
  \bibfield  {author} {\bibinfo {author} {\bibfnamefont {AD}~\bibnamefont
  {Caviglia}}, \bibinfo {author} {\bibfnamefont {Stefano}\ \bibnamefont
  {Gariglio}}, \bibinfo {author} {\bibfnamefont {Nicolas}\ \bibnamefont
  {Reyren}}, \bibinfo {author} {\bibfnamefont {Didier}\ \bibnamefont
  {Jaccard}}, \bibinfo {author} {\bibfnamefont {T}~\bibnamefont {Schneider}},
  \bibinfo {author} {\bibfnamefont {M}~\bibnamefont {Gabay}}, \bibinfo {author}
  {\bibfnamefont {S}~\bibnamefont {Thiel}}, \bibinfo {author} {\bibfnamefont
  {G}~\bibnamefont {Hammerl}}, \bibinfo {author} {\bibfnamefont {Jochen}\
  \bibnamefont {Mannhart}}, \ and\ \bibinfo {author} {\bibfnamefont {J-M}\
  \bibnamefont {Triscone}},\ }\bibfield  {title} {\enquote {\bibinfo {title}
  {Electric field control of the {LaAlO}$_{3}$/{SrTiO}$_{3}$ interface ground
  state},}\ }\href {\doibase 10.1038/nature07576} {\bibfield  {journal}
  {\bibinfo  {journal} {Nature}\ }\textbf {\bibinfo {volume} {456}},\ \bibinfo
  {pages} {624--627} (\bibinfo {year} {2008})}\BibitemShut {NoStop}%
\bibitem [{\citenamefont {Caviglia}\ \emph {et~al.}(2010)\citenamefont
  {Caviglia}, \citenamefont {Gabay}, \citenamefont {Gariglio}, \citenamefont
  {Reyren}, \citenamefont {Cancellieri},\ and\ \citenamefont
  {Triscone}}]{caviglia2010tunable}%
  \BibitemOpen
  \bibfield  {author} {\bibinfo {author} {\bibfnamefont {A.~D.}\ \bibnamefont
  {Caviglia}}, \bibinfo {author} {\bibfnamefont {M.}~\bibnamefont {Gabay}},
  \bibinfo {author} {\bibfnamefont {S.}~\bibnamefont {Gariglio}}, \bibinfo
  {author} {\bibfnamefont {N.}~\bibnamefont {Reyren}}, \bibinfo {author}
  {\bibfnamefont {C.}~\bibnamefont {Cancellieri}}, \ and\ \bibinfo {author}
  {\bibfnamefont {J.-M.}\ \bibnamefont {Triscone}},\ }\bibfield  {title}
  {\enquote {\bibinfo {title} {Tunable {Rashba} {Spin}-{Orbit} {Interaction} at
  {Oxide} {Interfaces}},}\ }\href {\doibase 10.1103/PhysRevLett.104.126803}
  {\bibfield  {journal} {\bibinfo  {journal} {Phys. Rev. Lett.}\ }\textbf
  {\bibinfo {volume} {104}},\ \bibinfo {pages} {126803} (\bibinfo {year}
  {2010})}\BibitemShut {NoStop}%
\bibitem [{\citenamefont {Ariando}\ \emph {et~al.}(2011)\citenamefont
  {Ariando}, \citenamefont {Wang}, \citenamefont {Baskaran}, \citenamefont
  {Liu}, \citenamefont {Huijben}, \citenamefont {Yi}, \citenamefont {Annadi},
  \citenamefont {Barman}, \citenamefont {Rusydi}, \citenamefont {Dhar},
  \citenamefont {Feng}, \citenamefont {Ding}, \citenamefont {Hilgenkamp},\ and\
  \citenamefont {Venkatesan}}]{Ariando2011}%
  \BibitemOpen
  \bibfield  {author} {\bibinfo {author} {\bibfnamefont {T.}~\bibnamefont
  {Ariando}}, \bibinfo {author} {\bibfnamefont {X.}~\bibnamefont {Wang}},
  \bibinfo {author} {\bibfnamefont {G.}~\bibnamefont {Baskaran}}, \bibinfo
  {author} {\bibfnamefont {Z.~Q.}\ \bibnamefont {Liu}}, \bibinfo {author}
  {\bibfnamefont {J.}~\bibnamefont {Huijben}}, \bibinfo {author} {\bibfnamefont
  {J.~B.}\ \bibnamefont {Yi}}, \bibinfo {author} {\bibfnamefont
  {A.}~\bibnamefont {Annadi}}, \bibinfo {author} {\bibfnamefont {A.~Roy}\
  \bibnamefont {Barman}}, \bibinfo {author} {\bibfnamefont {A.}~\bibnamefont
  {Rusydi}}, \bibinfo {author} {\bibfnamefont {S.}~\bibnamefont {Dhar}},
  \bibinfo {author} {\bibfnamefont {Y.~P.}\ \bibnamefont {Feng}}, \bibinfo
  {author} {\bibfnamefont {J.}~\bibnamefont {Ding}}, \bibinfo {author}
  {\bibfnamefont {H.}~\bibnamefont {Hilgenkamp}}, \ and\ \bibinfo {author}
  {\bibfnamefont {T.}~\bibnamefont {Venkatesan}},\ }\bibfield  {title}
  {\enquote {\bibinfo {title} {Electronic phase separation at the
  {LaAlO}$_{3}$/{SrTiO}$_{3}$ interface},}\ }\href
  {http://dx.doi.org/10.1038/ncomms1192} {\bibfield  {journal} {\bibinfo
  {journal} {Nature Communications}\ }\textbf {\bibinfo {volume} {2}},\
  \bibinfo {pages} {188} (\bibinfo {year} {2011})}\BibitemShut {NoStop}%
\bibitem [{\citenamefont {Bert}\ \emph {et~al.}(2011)\citenamefont {Bert},
  \citenamefont {Kalisky}, \citenamefont {Bell}, \citenamefont {Kim},
  \citenamefont {Hikita}, \citenamefont {Hwang},\ and\ \citenamefont
  {Moler}}]{bert2011direct}%
  \BibitemOpen
  \bibfield  {author} {\bibinfo {author} {\bibfnamefont {Julie~A}\ \bibnamefont
  {Bert}}, \bibinfo {author} {\bibfnamefont {Beena}\ \bibnamefont {Kalisky}},
  \bibinfo {author} {\bibfnamefont {Christopher}\ \bibnamefont {Bell}},
  \bibinfo {author} {\bibfnamefont {Minu}\ \bibnamefont {Kim}}, \bibinfo
  {author} {\bibfnamefont {Yasuyuki}\ \bibnamefont {Hikita}}, \bibinfo {author}
  {\bibfnamefont {Harold~Y}\ \bibnamefont {Hwang}}, \ and\ \bibinfo {author}
  {\bibfnamefont {Kathryn~A}\ \bibnamefont {Moler}},\ }\bibfield  {title}
  {\enquote {\bibinfo {title} {Direct imaging of the coexistence of
  ferromagnetism and superconductivity at the {LaAlO}$_{3}$/{SrTiO}$_{3}$
  interface},}\ }\href {\doibase 10.1038/nphys2079} {\bibfield  {journal}
  {\bibinfo  {journal} {Nature physics}\ }\textbf {\bibinfo {volume} {7}},\
  \bibinfo {pages} {767--771} (\bibinfo {year} {2011})}\BibitemShut {NoStop}%
\bibitem [{\citenamefont {Li}\ \emph {et~al.}(2011)\citenamefont {Li},
  \citenamefont {Richter}, \citenamefont {Mannhart},\ and\ \citenamefont
  {Ashoori}}]{li2011coexistence}%
  \BibitemOpen
  \bibfield  {author} {\bibinfo {author} {\bibfnamefont {Lu}~\bibnamefont
  {Li}}, \bibinfo {author} {\bibfnamefont {C}~\bibnamefont {Richter}}, \bibinfo
  {author} {\bibfnamefont {J}~\bibnamefont {Mannhart}}, \ and\ \bibinfo
  {author} {\bibfnamefont {RC}~\bibnamefont {Ashoori}},\ }\bibfield  {title}
  {\enquote {\bibinfo {title} {Coexistence of magnetic order and
  two-dimensional superconductivity at {LaAlO}$_{3}$/{SrTiO}$_{3}$
  interfaces},}\ }\href {\doibase 10.1038/nphys2080} {\bibfield  {journal}
  {\bibinfo  {journal} {Nature physics}\ }\textbf {\bibinfo {volume} {7}},\
  \bibinfo {pages} {762--766} (\bibinfo {year} {2011})}\BibitemShut {NoStop}%
\bibitem [{\citenamefont {Khalsa}\ and\ \citenamefont
  {MacDonald}(2012)}]{macdonald_001}%
  \BibitemOpen
  \bibfield  {author} {\bibinfo {author} {\bibfnamefont {Guru}\ \bibnamefont
  {Khalsa}}\ and\ \bibinfo {author} {\bibfnamefont {A.~H.}\ \bibnamefont
  {MacDonald}},\ }\bibfield  {title} {\enquote {\bibinfo {title} {Theory of the
  {SrTiO}$_{3}$ surface state two-dimensional electron gas},}\ }\href {\doibase
  10.1103/PhysRevB.86.125121} {\bibfield  {journal} {\bibinfo  {journal} {Phys.
  Rev. B}\ }\textbf {\bibinfo {volume} {86}},\ \bibinfo {pages} {125121}
  (\bibinfo {year} {2012})}\BibitemShut {NoStop}%
\bibitem [{\citenamefont {Mehta}\ \emph {et~al.}(2012)\citenamefont {Mehta},
  \citenamefont {Dikin}, \citenamefont {Bark}, \citenamefont {Ryu},
  \citenamefont {Folkman}, \citenamefont {Eom},\ and\ \citenamefont
  {Chandrasekhar}}]{mehta2012evidence}%
  \BibitemOpen
  \bibfield  {author} {\bibinfo {author} {\bibfnamefont {MM}~\bibnamefont
  {Mehta}}, \bibinfo {author} {\bibfnamefont {DA}~\bibnamefont {Dikin}},
  \bibinfo {author} {\bibfnamefont {CW}~\bibnamefont {Bark}}, \bibinfo {author}
  {\bibfnamefont {S}~\bibnamefont {Ryu}}, \bibinfo {author} {\bibfnamefont
  {CM}~\bibnamefont {Folkman}}, \bibinfo {author} {\bibfnamefont
  {CB}~\bibnamefont {Eom}}, \ and\ \bibinfo {author} {\bibfnamefont
  {V}~\bibnamefont {Chandrasekhar}},\ }\bibfield  {title} {\enquote {\bibinfo
  {title} {Evidence for charge-vortex duality at the {LaAlO}$_3$/{SrTiO}$_3$
  interface},}\ }\href@noop {} {\bibfield  {journal} {\bibinfo  {journal}
  {Nature Communications}\ }\textbf {\bibinfo {volume} {3}},\ \bibinfo {pages}
  {955} (\bibinfo {year} {2012})}\BibitemShut {NoStop}%
\bibitem [{\citenamefont {Zhong}\ \emph {et~al.}(2013)\citenamefont {Zhong},
  \citenamefont {T\'oth},\ and\ \citenamefont {Held}}]{held_001}%
  \BibitemOpen
  \bibfield  {author} {\bibinfo {author} {\bibfnamefont {Zhicheng}\
  \bibnamefont {Zhong}}, \bibinfo {author} {\bibfnamefont {Anna}\ \bibnamefont
  {T\'oth}}, \ and\ \bibinfo {author} {\bibfnamefont {Karsten}\ \bibnamefont
  {Held}},\ }\bibfield  {title} {\enquote {\bibinfo {title} {Theory of
  spin-orbit coupling at {LaAlO}$_{3}$/{SrTiO}$_{3}$ interfaces and
  {SrTiO}$_{3}$ surfaces},}\ }\href {\doibase 10.1103/PhysRevB.87.161102}
  {\bibfield  {journal} {\bibinfo  {journal} {Phys. Rev. B}\ }\textbf {\bibinfo
  {volume} {87}},\ \bibinfo {pages} {161102} (\bibinfo {year}
  {2013})}\BibitemShut {NoStop}%
\bibitem [{\citenamefont {Michaeli}\ \emph {et~al.}(2012)\citenamefont
  {Michaeli}, \citenamefont {Potter},\ and\ \citenamefont
  {Lee}}]{Michaeli2012}%
  \BibitemOpen
  \bibfield  {author} {\bibinfo {author} {\bibfnamefont {Karen}\ \bibnamefont
  {Michaeli}}, \bibinfo {author} {\bibfnamefont {Andrew~C.}\ \bibnamefont
  {Potter}}, \ and\ \bibinfo {author} {\bibfnamefont {Patrick~A.}\ \bibnamefont
  {Lee}},\ }\bibfield  {title} {\enquote {\bibinfo {title} {Superconducting and
  ferromagnetic phases in {SrTiO}$_3$/{LaAlO}$_3$ oxide interface structures:
  Possibility of finite momentum pairing},}\ }\href {\doibase
  10.1103/PhysRevLett.108.117003} {\bibfield  {journal} {\bibinfo  {journal}
  {Phys. Rev. Lett.}\ }\textbf {\bibinfo {volume} {108}},\ \bibinfo {pages}
  {117003} (\bibinfo {year} {2012})}\BibitemShut {NoStop}%
\bibitem [{\citenamefont {Fidkowski}\ \emph {et~al.}(2013)\citenamefont
  {Fidkowski}, \citenamefont {Jiang}, \citenamefont {Lutchyn},\ and\
  \citenamefont {Nayak}}]{Fidkowski2013}%
  \BibitemOpen
  \bibfield  {author} {\bibinfo {author} {\bibfnamefont {Lukasz}\ \bibnamefont
  {Fidkowski}}, \bibinfo {author} {\bibfnamefont {Hong-Chen}\ \bibnamefont
  {Jiang}}, \bibinfo {author} {\bibfnamefont {Roman~M.}\ \bibnamefont
  {Lutchyn}}, \ and\ \bibinfo {author} {\bibfnamefont {Chetan}\ \bibnamefont
  {Nayak}},\ }\bibfield  {title} {\enquote {\bibinfo {title} {Magnetic and
  superconducting ordering in one-dimensional nanostructures at the
  {LaAlO}$_3$/{SrTiO}$_3$ interface},}\ }\href {\doibase
  10.1103/PhysRevB.87.014436} {\bibfield  {journal} {\bibinfo  {journal} {Phys.
  Rev. B}\ }\textbf {\bibinfo {volume} {87}},\ \bibinfo {pages} {014436}
  (\bibinfo {year} {2013})}\BibitemShut {NoStop}%
\bibitem [{\citenamefont {Fischer}\ \emph {et~al.}(2013)\citenamefont
  {Fischer}, \citenamefont {Raghu},\ and\ \citenamefont
  {Kim}}]{fischer_spinorbit_2013}%
  \BibitemOpen
  \bibfield  {author} {\bibinfo {author} {\bibfnamefont {Mark~H.}\ \bibnamefont
  {Fischer}}, \bibinfo {author} {\bibfnamefont {Srinivas}\ \bibnamefont
  {Raghu}}, \ and\ \bibinfo {author} {\bibfnamefont {Eun-Ah}\ \bibnamefont
  {Kim}},\ }\bibfield  {title} {\enquote {\bibinfo {title} {Spin-orbit coupling
  in {LaAlO}$_3$/{SrTiO}$_3$ interfaces: magnetism and orbital ordering},}\
  }\href {\doibase 10.1088/1367-2630/15/2/023022} {\bibfield  {journal}
  {\bibinfo  {journal} {New J. Phys.}\ }\textbf {\bibinfo {volume} {15}},\
  \bibinfo {pages} {023022} (\bibinfo {year} {2013})}\BibitemShut {NoStop}%
\bibitem [{\citenamefont {Kim}\ \emph {et~al.}(2013)\citenamefont {Kim},
  \citenamefont {Lutchyn},\ and\ \citenamefont {Nayak}}]{kim2013origin}%
  \BibitemOpen
  \bibfield  {author} {\bibinfo {author} {\bibfnamefont {Younghyun}\
  \bibnamefont {Kim}}, \bibinfo {author} {\bibfnamefont {Roman~M.}\
  \bibnamefont {Lutchyn}}, \ and\ \bibinfo {author} {\bibfnamefont {Chetan}\
  \bibnamefont {Nayak}},\ }\bibfield  {title} {\enquote {\bibinfo {title}
  {Origin and transport signatures of spin-orbit interactions in one- and
  two-dimensional {SrTiO}$_{3}$-based heterostructures},}\ }\href {\doibase
  10.1103/PhysRevB.87.245121} {\bibfield  {journal} {\bibinfo  {journal} {Phys.
  Rev. B}\ }\textbf {\bibinfo {volume} {87}},\ \bibinfo {pages} {245121}
  (\bibinfo {year} {2013})}\BibitemShut {NoStop}%
\bibitem [{\citenamefont {Banerjee}\ \emph {et~al.}(2013)\citenamefont
  {Banerjee}, \citenamefont {Erten},\ and\ \citenamefont
  {Randeria}}]{banerjee2013ferromagnetic}%
  \BibitemOpen
  \bibfield  {author} {\bibinfo {author} {\bibfnamefont {Sumilan}\ \bibnamefont
  {Banerjee}}, \bibinfo {author} {\bibfnamefont {Onur}\ \bibnamefont {Erten}},
  \ and\ \bibinfo {author} {\bibfnamefont {Mohit}\ \bibnamefont {Randeria}},\
  }\bibfield  {title} {\enquote {\bibinfo {title} {Ferromagnetic exchange,
  spin-orbit coupling and spiral magnetism at the {LaAlO}$_{3}$/{SrTiO}$_{3}$
  interface},}\ }\href {\doibase 10.1038/nphys2702} {\bibfield  {journal}
  {\bibinfo  {journal} {Nature physics}\ }\textbf {\bibinfo {volume} {9}},\
  \bibinfo {pages} {626--630} (\bibinfo {year} {2013})}\BibitemShut {NoStop}%
\bibitem [{\citenamefont {Chen}\ and\ \citenamefont
  {Balents}(2013)}]{chen2013}%
  \BibitemOpen
  \bibfield  {author} {\bibinfo {author} {\bibfnamefont {Gang}\ \bibnamefont
  {Chen}}\ and\ \bibinfo {author} {\bibfnamefont {Leon}\ \bibnamefont
  {Balents}},\ }\bibfield  {title} {\enquote {\bibinfo {title} {Ferromagnetism
  in itinerant two-dimensional ${t}_{2g}$ systems},}\ }\href {\doibase
  10.1103/PhysRevLett.110.206401} {\bibfield  {journal} {\bibinfo  {journal}
  {Phys. Rev. Lett.}\ }\textbf {\bibinfo {volume} {110}},\ \bibinfo {pages}
  {206401} (\bibinfo {year} {2013})}\BibitemShut {NoStop}%
\bibitem [{\citenamefont {Park}\ and\ \citenamefont {Millis}(2013)}]{Park2013}%
  \BibitemOpen
  \bibfield  {author} {\bibinfo {author} {\bibfnamefont {Se~Young}\
  \bibnamefont {Park}}\ and\ \bibinfo {author} {\bibfnamefont {Andrew~J.}\
  \bibnamefont {Millis}},\ }\bibfield  {title} {\enquote {\bibinfo {title}
  {Charge density distribution and optical response of the
  {LaAlO}$_{3}$/{SrTiO}$_{3}$ interface},}\ }\href {\doibase
  10.1103/PhysRevB.87.205145} {\bibfield  {journal} {\bibinfo  {journal} {Phys.
  Rev. B}\ }\textbf {\bibinfo {volume} {87}},\ \bibinfo {pages} {205145}
  (\bibinfo {year} {2013})}\BibitemShut {NoStop}%
\bibitem [{\citenamefont {Ruhman}\ \emph {et~al.}(2014)\citenamefont {Ruhman},
  \citenamefont {Joshua}, \citenamefont {Ilani},\ and\ \citenamefont
  {Altman}}]{ruhman_competition_2014}%
  \BibitemOpen
  \bibfield  {author} {\bibinfo {author} {\bibfnamefont {Jonathan}\
  \bibnamefont {Ruhman}}, \bibinfo {author} {\bibfnamefont {Arjun}\
  \bibnamefont {Joshua}}, \bibinfo {author} {\bibfnamefont {Shahal}\
  \bibnamefont {Ilani}}, \ and\ \bibinfo {author} {\bibfnamefont {Ehud}\
  \bibnamefont {Altman}},\ }\bibfield  {title} {\enquote {\bibinfo {title}
  {Competition {Between} {Kondo} {Screening} and {Magnetism} at the
  {LaAlO}$_{3}$/{SrTiO}$_{3}$ {Interface}},}\ }\href {\doibase
  10.1103/PhysRevB.90.125123} {\bibfield  {journal} {\bibinfo  {journal} {Phys.
  Rev. B}\ }\textbf {\bibinfo {volume} {90}},\ \bibinfo {pages} {125123}
  (\bibinfo {year} {2014})}\BibitemShut {NoStop}%
\bibitem [{\citenamefont {Diez}\ \emph {et~al.}(2015)\citenamefont {Diez},
  \citenamefont {Monteiro}, \citenamefont {Mattoni}, \citenamefont {Cobanera},
  \citenamefont {Hyart}, \citenamefont {Mulazimoglu}, \citenamefont {Bovenzi},
  \citenamefont {Beenakker},\ and\ \citenamefont {Caviglia}}]{caviglia_001}%
  \BibitemOpen
  \bibfield  {author} {\bibinfo {author} {\bibfnamefont {M.}~\bibnamefont
  {Diez}}, \bibinfo {author} {\bibfnamefont {A.~M. R. V.~L.}\ \bibnamefont
  {Monteiro}}, \bibinfo {author} {\bibfnamefont {G.}~\bibnamefont {Mattoni}},
  \bibinfo {author} {\bibfnamefont {E.}~\bibnamefont {Cobanera}}, \bibinfo
  {author} {\bibfnamefont {T.}~\bibnamefont {Hyart}}, \bibinfo {author}
  {\bibfnamefont {E.}~\bibnamefont {Mulazimoglu}}, \bibinfo {author}
  {\bibfnamefont {N.}~\bibnamefont {Bovenzi}}, \bibinfo {author} {\bibfnamefont
  {C.~W.~J.}\ \bibnamefont {Beenakker}}, \ and\ \bibinfo {author}
  {\bibfnamefont {A.~D.}\ \bibnamefont {Caviglia}},\ }\bibfield  {title}
  {\enquote {\bibinfo {title} {{Giant} {Negative} {Magnetoresistance} {Driven}
  by {Spin}-{Orbit} {Coupling} at the {LaAlO}$_{3}$/{SrTiO}$_{3}$
  {Interface}},}\ }\href {\doibase 10.1103/PhysRevLett.115.016803} {\bibfield
  {journal} {\bibinfo  {journal} {Phys. Rev. Lett.}\ }\textbf {\bibinfo
  {volume} {115}},\ \bibinfo {pages} {016803} (\bibinfo {year}
  {2015})}\BibitemShut {NoStop}%
\bibitem [{\citenamefont {Nandy}\ \emph {et~al.}(2016)\citenamefont {Nandy},
  \citenamefont {Mohanta}, \citenamefont {Acharya},\ and\ \citenamefont
  {Taraphder}}]{Taraphder2016}%
  \BibitemOpen
  \bibfield  {author} {\bibinfo {author} {\bibfnamefont {S.}~\bibnamefont
  {Nandy}}, \bibinfo {author} {\bibfnamefont {N.}~\bibnamefont {Mohanta}},
  \bibinfo {author} {\bibfnamefont {S.}~\bibnamefont {Acharya}}, \ and\
  \bibinfo {author} {\bibfnamefont {A.}~\bibnamefont {Taraphder}},\ }\bibfield
  {title} {\enquote {\bibinfo {title} {Anomalous transport near the {Lifshitz}
  transition at the {LaAlO}$_{3}$/{SrTiO}$_{3}$ interface},}\ }\href {\doibase
  10.1103/PhysRevB.94.155103} {\bibfield  {journal} {\bibinfo  {journal} {Phys.
  Rev. B}\ }\textbf {\bibinfo {volume} {94}},\ \bibinfo {pages} {155103}
  (\bibinfo {year} {2016})}\BibitemShut {NoStop}%
\bibitem [{\citenamefont {Tolsma}\ \emph {et~al.}(2017)\citenamefont {Tolsma},
  \citenamefont {Polini},\ and\ \citenamefont
  {MacDonald}}]{tolsma_orbital_2016}%
  \BibitemOpen
  \bibfield  {author} {\bibinfo {author} {\bibfnamefont {John~R.}\ \bibnamefont
  {Tolsma}}, \bibinfo {author} {\bibfnamefont {Marco}\ \bibnamefont {Polini}},
  \ and\ \bibinfo {author} {\bibfnamefont {Allan~H.}\ \bibnamefont
  {MacDonald}},\ }\bibfield  {title} {\enquote {\bibinfo {title} {Orbital and
  spin order in oxide two-dimensional electron gases},}\ }\href {\doibase
  10.1103/PhysRevB.95.205101} {\bibfield  {journal} {\bibinfo  {journal} {Phys.
  Rev. B}\ }\textbf {\bibinfo {volume} {95}},\ \bibinfo {pages} {205101}
  (\bibinfo {year} {2017})}\BibitemShut {NoStop}%
\bibitem [{\citenamefont {Atkinson}\ \emph {et~al.}(2017)\citenamefont
  {Atkinson}, \citenamefont {Lafleur},\ and\ \citenamefont
  {Raslan}}]{atkinson2017influence}%
  \BibitemOpen
  \bibfield  {author} {\bibinfo {author} {\bibfnamefont {W.~A.}\ \bibnamefont
  {Atkinson}}, \bibinfo {author} {\bibfnamefont {P.}~\bibnamefont {Lafleur}}, \
  and\ \bibinfo {author} {\bibfnamefont {A.}~\bibnamefont {Raslan}},\
  }\bibfield  {title} {\enquote {\bibinfo {title} {Influence of the
  ferroelectric quantum critical point on {SrTiO}$_{3}$ interfaces},}\ }\href
  {\doibase 10.1103/PhysRevB.95.054107} {\bibfield  {journal} {\bibinfo
  {journal} {Phys. Rev. B}\ }\textbf {\bibinfo {volume} {95}},\ \bibinfo
  {pages} {054107} (\bibinfo {year} {2017})}\BibitemShut {NoStop}%
\bibitem [{\citenamefont {Burkov}\ \emph {et~al.}(2004)\citenamefont {Burkov},
  \citenamefont {N\'u\~nez},\ and\ \citenamefont {MacDonald}}]{burkov2004}%
  \BibitemOpen
  \bibfield  {author} {\bibinfo {author} {\bibfnamefont {A.~A.}\ \bibnamefont
  {Burkov}}, \bibinfo {author} {\bibfnamefont {Alvaro~S.}\ \bibnamefont
  {N\'u\~nez}}, \ and\ \bibinfo {author} {\bibfnamefont {A.~H.}\ \bibnamefont
  {MacDonald}},\ }\bibfield  {title} {\enquote {\bibinfo {title} {Theory of
  spin-charge-coupled transport in a two-dimensional electron gas with {Rashba}
  spin-orbit interactions},}\ }\href {\doibase 10.1103/PhysRevB.70.155308}
  {\bibfield  {journal} {\bibinfo  {journal} {Phys. Rev. B}\ }\textbf {\bibinfo
  {volume} {70}},\ \bibinfo {pages} {155308} (\bibinfo {year}
  {2004})}\BibitemShut {NoStop}%
\bibitem [{\citenamefont {Lesne}\ \emph {et~al.}(2016)\citenamefont {Lesne},
  \citenamefont {Fu}, \citenamefont {Oyarzun}, \citenamefont
  {Rojas-S{\'a}nchez}, \citenamefont {Vaz}, \citenamefont {Naganuma},
  \citenamefont {Sicoli}, \citenamefont {Attan{\'e}}, \citenamefont {Jamet},
  \citenamefont {Jacquet}, \citenamefont {George}, \citenamefont
  {Barth{\'e}l{\'e}my}, \citenamefont {Jaffr{\`e}s}, \citenamefont {Fert},
  \citenamefont {Bibes},\ and\ \citenamefont {Vila}}]{Bibes_NMat2016}%
  \BibitemOpen
  \bibfield  {author} {\bibinfo {author} {\bibfnamefont {E.}~\bibnamefont
  {Lesne}}, \bibinfo {author} {\bibfnamefont {Yu}~\bibnamefont {Fu}}, \bibinfo
  {author} {\bibfnamefont {S.}~\bibnamefont {Oyarzun}}, \bibinfo {author}
  {\bibfnamefont {J.~C.}\ \bibnamefont {Rojas-S{\'a}nchez}}, \bibinfo {author}
  {\bibfnamefont {D.~C.}\ \bibnamefont {Vaz}}, \bibinfo {author} {\bibfnamefont
  {H.}~\bibnamefont {Naganuma}}, \bibinfo {author} {\bibfnamefont
  {G.}~\bibnamefont {Sicoli}}, \bibinfo {author} {\bibfnamefont {J.~P.}\
  \bibnamefont {Attan{\'e}}}, \bibinfo {author} {\bibfnamefont
  {M.}~\bibnamefont {Jamet}}, \bibinfo {author} {\bibfnamefont
  {E.}~\bibnamefont {Jacquet}}, \bibinfo {author} {\bibfnamefont {J.~M.}\
  \bibnamefont {George}}, \bibinfo {author} {\bibfnamefont {A.}~\bibnamefont
  {Barth{\'e}l{\'e}my}}, \bibinfo {author} {\bibfnamefont {H.}~\bibnamefont
  {Jaffr{\`e}s}}, \bibinfo {author} {\bibfnamefont {A.}~\bibnamefont {Fert}},
  \bibinfo {author} {\bibfnamefont {M.}~\bibnamefont {Bibes}}, \ and\ \bibinfo
  {author} {\bibfnamefont {L.}~\bibnamefont {Vila}},\ }\bibfield  {title}
  {\enquote {\bibinfo {title} {Highly efficient and tunable spin-to-charge
  conversion through {Rashba} coupling at oxide interfaces},}\ }\href
  {https://doi.org/10.1038/nmat4726} {\bibfield  {journal} {\bibinfo  {journal}
  {Nature Materials}\ }\textbf {\bibinfo {volume} {15}},\ \bibinfo {pages}
  {1261 EP --} (\bibinfo {year} {2016})}\BibitemShut {NoStop}%
\bibitem [{\citenamefont {Song}\ \emph {et~al.}(2017)\citenamefont {Song},
  \citenamefont {Zhang}, \citenamefont {Su}, \citenamefont {Yuan},
  \citenamefont {Chen}, \citenamefont {Xing}, \citenamefont {Shi},
  \citenamefont {Sun},\ and\ \citenamefont {Han}}]{Han_Science2017}%
  \BibitemOpen
  \bibfield  {author} {\bibinfo {author} {\bibfnamefont {Qi}~\bibnamefont
  {Song}}, \bibinfo {author} {\bibfnamefont {Hongrui}\ \bibnamefont {Zhang}},
  \bibinfo {author} {\bibfnamefont {Tang}\ \bibnamefont {Su}}, \bibinfo
  {author} {\bibfnamefont {Wei}\ \bibnamefont {Yuan}}, \bibinfo {author}
  {\bibfnamefont {Yangyang}\ \bibnamefont {Chen}}, \bibinfo {author}
  {\bibfnamefont {Wenyu}\ \bibnamefont {Xing}}, \bibinfo {author}
  {\bibfnamefont {Jing}\ \bibnamefont {Shi}}, \bibinfo {author} {\bibfnamefont
  {Jirong}\ \bibnamefont {Sun}}, \ and\ \bibinfo {author} {\bibfnamefont {Wei}\
  \bibnamefont {Han}},\ }\bibfield  {title} {\enquote {\bibinfo {title}
  {Observation of inverse {Edelstein} effect in {Rashba}-split {2DEG} between
  {SrTiO}$_{3}$ and {LaAlO}$_{3}$ at room temperature},}\ }\href {\doibase
  10.1126/sciadv.1602312} {\bibfield  {journal} {\bibinfo  {journal} {Science
  Advances}\ }\textbf {\bibinfo {volume} {3}},\ \bibinfo {pages} {1--6}
  (\bibinfo {year} {2017})}\BibitemShut {NoStop}%
\bibitem [{\citenamefont {Ben~Shalom}\ \emph {et~al.}(2009)\citenamefont
  {Ben~Shalom}, \citenamefont {Tai}, \citenamefont {Lereah}, \citenamefont
  {Sachs}, \citenamefont {Levy}, \citenamefont {Rakhmilevitch}, \citenamefont
  {Palevski},\ and\ \citenamefont {Dagan}}]{dagan2009}%
  \BibitemOpen
  \bibfield  {author} {\bibinfo {author} {\bibfnamefont {M.}~\bibnamefont
  {Ben~Shalom}}, \bibinfo {author} {\bibfnamefont {C.~W.}\ \bibnamefont {Tai}},
  \bibinfo {author} {\bibfnamefont {Y.}~\bibnamefont {Lereah}}, \bibinfo
  {author} {\bibfnamefont {M.}~\bibnamefont {Sachs}}, \bibinfo {author}
  {\bibfnamefont {E.}~\bibnamefont {Levy}}, \bibinfo {author} {\bibfnamefont
  {D.}~\bibnamefont {Rakhmilevitch}}, \bibinfo {author} {\bibfnamefont
  {A.}~\bibnamefont {Palevski}}, \ and\ \bibinfo {author} {\bibfnamefont
  {Y.}~\bibnamefont {Dagan}},\ }\bibfield  {title} {\enquote {\bibinfo {title}
  {Anisotropic magnetotransport at the {LaAlO}$_3$/{SrTiO}$_3$ interface},}\
  }\href {\doibase 10.1103/PhysRevB.80.140403} {\bibfield  {journal} {\bibinfo
  {journal} {Phys. Rev. B}\ }\textbf {\bibinfo {volume} {80}},\ \bibinfo
  {pages} {140403} (\bibinfo {year} {2009})}\BibitemShut {NoStop}%
\bibitem [{\citenamefont {Wang}\ \emph {et~al.}(2011)\citenamefont {Wang},
  \citenamefont {L\"u}, \citenamefont {Annadi}, \citenamefont {Liu},
  \citenamefont {Gopinadhan}, \citenamefont {Dhar}, \citenamefont
  {Venkatesan},\ and\ \citenamefont {Ariando}}]{ariando2011mr}%
  \BibitemOpen
  \bibfield  {author} {\bibinfo {author} {\bibfnamefont {X.}~\bibnamefont
  {Wang}}, \bibinfo {author} {\bibfnamefont {W.~M.}\ \bibnamefont {L\"u}},
  \bibinfo {author} {\bibfnamefont {A.}~\bibnamefont {Annadi}}, \bibinfo
  {author} {\bibfnamefont {Z.~Q.}\ \bibnamefont {Liu}}, \bibinfo {author}
  {\bibfnamefont {K.}~\bibnamefont {Gopinadhan}}, \bibinfo {author}
  {\bibfnamefont {S.}~\bibnamefont {Dhar}}, \bibinfo {author} {\bibfnamefont
  {T.}~\bibnamefont {Venkatesan}}, \ and\ \bibinfo {author} {\bibnamefont
  {Ariando}},\ }\bibfield  {title} {\enquote {\bibinfo {title}
  {Magnetoresistance of two-dimensional and three-dimensional electron gas in
  {LaAlO}$_{3}$/{SrTiO}$_{3}$ heterostructures: Influence of magnetic ordering,
  interface scattering, and dimensionality},}\ }\href {\doibase
  10.1103/PhysRevB.84.075312} {\bibfield  {journal} {\bibinfo  {journal} {Phys.
  Rev. B}\ }\textbf {\bibinfo {volume} {84}},\ \bibinfo {pages} {075312}
  (\bibinfo {year} {2011})}\BibitemShut {NoStop}%
\bibitem [{\citenamefont {R\"odel}\ \emph {et~al.}(2014)\citenamefont
  {R\"odel}, \citenamefont {Bareille}, \citenamefont {Fortuna}, \citenamefont
  {Baumier}, \citenamefont {Bertran}, \citenamefont {Le~F\`evre}, \citenamefont
  {Gabay}, \citenamefont {Hijano~Cubelos}, \citenamefont {Rozenberg},
  \citenamefont {Maroutian}, \citenamefont {Lecoeur},\ and\ \citenamefont
  {Santander-Syro}}]{rodel_orientational_2014}%
  \BibitemOpen
  \bibfield  {author} {\bibinfo {author} {\bibfnamefont {T.~C.}\ \bibnamefont
  {R\"odel}}, \bibinfo {author} {\bibfnamefont {C.}~\bibnamefont {Bareille}},
  \bibinfo {author} {\bibfnamefont {F.}~\bibnamefont {Fortuna}}, \bibinfo
  {author} {\bibfnamefont {C.}~\bibnamefont {Baumier}}, \bibinfo {author}
  {\bibfnamefont {F.}~\bibnamefont {Bertran}}, \bibinfo {author} {\bibfnamefont
  {P.}~\bibnamefont {Le~F\`evre}}, \bibinfo {author} {\bibfnamefont
  {M.}~\bibnamefont {Gabay}}, \bibinfo {author} {\bibfnamefont
  {O.}~\bibnamefont {Hijano~Cubelos}}, \bibinfo {author} {\bibfnamefont
  {M.~J.}\ \bibnamefont {Rozenberg}}, \bibinfo {author} {\bibfnamefont
  {T.}~\bibnamefont {Maroutian}}, \bibinfo {author} {\bibfnamefont
  {P.}~\bibnamefont {Lecoeur}}, \ and\ \bibinfo {author} {\bibfnamefont
  {A.~F.}\ \bibnamefont {Santander-Syro}},\ }\bibfield  {title} {\enquote
  {\bibinfo {title} {{Orientational} {Tuning} of the {Fermi} {Sea} of
  {Confined} {Electrons} at the {SrTiO}$_3$ (110) and (111) {Surfaces}},}\
  }\href {\doibase 10.1103/PhysRevApplied.1.051002} {\bibfield  {journal}
  {\bibinfo  {journal} {Phys. Rev. Applied}\ }\textbf {\bibinfo {volume} {1}},\
  \bibinfo {pages} {051002} (\bibinfo {year} {2014})}\BibitemShut {NoStop}%
\bibitem [{\citenamefont {McKeown~Walker}\ \emph {et~al.}(2014)\citenamefont
  {McKeown~Walker}, \citenamefont {de~la Torre}, \citenamefont {Bruno},
  \citenamefont {Tamai}, \citenamefont {Kim}, \citenamefont {Hoesch},
  \citenamefont {Shi}, \citenamefont {Bahramy}, \citenamefont {King},\ and\
  \citenamefont {Baumberger}}]{mckeown_walker_control_2014}%
  \BibitemOpen
  \bibfield  {author} {\bibinfo {author} {\bibfnamefont {S.}~\bibnamefont
  {McKeown~Walker}}, \bibinfo {author} {\bibfnamefont {A.}~\bibnamefont {de~la
  Torre}}, \bibinfo {author} {\bibfnamefont {F.~Y.}\ \bibnamefont {Bruno}},
  \bibinfo {author} {\bibfnamefont {A.}~\bibnamefont {Tamai}}, \bibinfo
  {author} {\bibfnamefont {T.~K.}\ \bibnamefont {Kim}}, \bibinfo {author}
  {\bibfnamefont {M.}~\bibnamefont {Hoesch}}, \bibinfo {author} {\bibfnamefont
  {M.}~\bibnamefont {Shi}}, \bibinfo {author} {\bibfnamefont {M.~S.}\
  \bibnamefont {Bahramy}}, \bibinfo {author} {\bibfnamefont {P.~D.~C.}\
  \bibnamefont {King}}, \ and\ \bibinfo {author} {\bibfnamefont
  {F.}~\bibnamefont {Baumberger}},\ }\bibfield  {title} {\enquote {\bibinfo
  {title} {Control of a {Two}-{Dimensional} {Electron} {Gas} on
  {SrTiO}$_3(111)$ by {Atomic} {Oxygen}},}\ }\href {\doibase
  10.1103/PhysRevLett.113.177601} {\bibfield  {journal} {\bibinfo  {journal}
  {Phys. Rev. Lett.}\ }\textbf {\bibinfo {volume} {113}},\ \bibinfo {pages}
  {177601} (\bibinfo {year} {2014})}\BibitemShut {NoStop}%
\bibitem [{\citenamefont {Dudy}\ \emph {et~al.}(2016)\citenamefont {Dudy},
  \citenamefont {Sing}, \citenamefont {Scheiderer}, \citenamefont {Denlinger},
  \citenamefont {Sch\"aetz}, \citenamefont {Gabel}, \citenamefont {Buchwald},
  \citenamefont {Schlueter}, \citenamefont {Lee},\ and\ \citenamefont
  {Claessen}}]{Dudy}%
  \BibitemOpen
  \bibfield  {author} {\bibinfo {author} {\bibfnamefont {Lenart}\ \bibnamefont
  {Dudy}}, \bibinfo {author} {\bibfnamefont {Michael}\ \bibnamefont {Sing}},
  \bibinfo {author} {\bibfnamefont {Philipp}\ \bibnamefont {Scheiderer}},
  \bibinfo {author} {\bibfnamefont {Jonathan~D.}\ \bibnamefont {Denlinger}},
  \bibinfo {author} {\bibfnamefont {Philipp}\ \bibnamefont {Sch\"aetz}},
  \bibinfo {author} {\bibfnamefont {Judith}\ \bibnamefont {Gabel}}, \bibinfo
  {author} {\bibfnamefont {Mathias}\ \bibnamefont {Buchwald}}, \bibinfo
  {author} {\bibfnamefont {Christoph}\ \bibnamefont {Schlueter}}, \bibinfo
  {author} {\bibfnamefont {Tien-Lin}\ \bibnamefont {Lee}}, \ and\ \bibinfo
  {author} {\bibfnamefont {Ralph}\ \bibnamefont {Claessen}},\ }\bibfield
  {title} {\enquote {\bibinfo {title} {In situ control of separate electronic
  phases on {SrTiO}$_{3}$ surfaces by oxygen dosing},}\ }\href {\doibase
  10.1002/adma.201600046} {\bibfield  {journal} {\bibinfo  {journal} {Advanced
  Materials}\ }\textbf {\bibinfo {volume} {28}},\ \bibinfo {pages} {7443--7449}
  (\bibinfo {year} {2016})}\BibitemShut {NoStop}%
\bibitem [{\citenamefont {Miao}\ \emph {et~al.}(2016)\citenamefont {Miao},
  \citenamefont {Du}, \citenamefont {Yin},\ and\ \citenamefont
  {Li}}]{miao_anisotropic_2016}%
  \BibitemOpen
  \bibfield  {author} {\bibinfo {author} {\bibfnamefont {Ludi}\ \bibnamefont
  {Miao}}, \bibinfo {author} {\bibfnamefont {Renzhong}\ \bibnamefont {Du}},
  \bibinfo {author} {\bibfnamefont {Yuewei}\ \bibnamefont {Yin}}, \ and\
  \bibinfo {author} {\bibfnamefont {Qi}~\bibnamefont {Li}},\ }\bibfield
  {title} {\enquote {\bibinfo {title} {Anisotropic magneto-transport properties
  of electron gases at {SrTiO}$_3$ (111) and (110) surfaces},}\ }\href
  {\doibase 10.1063/1.4972985} {\bibfield  {journal} {\bibinfo  {journal}
  {Appl. Phys. Lett.}\ }\textbf {\bibinfo {volume} {109}},\ \bibinfo {pages}
  {261604} (\bibinfo {year} {2016})}\BibitemShut {NoStop}%
\bibitem [{\citenamefont {Rout}\ \emph {et~al.}(2017)\citenamefont {Rout},
  \citenamefont {Agireen}, \citenamefont {Maniv}, \citenamefont {Goldstein},\
  and\ \citenamefont {Dagan}}]{rout_six-fold_2017}%
  \BibitemOpen
  \bibfield  {author} {\bibinfo {author} {\bibfnamefont {P.~K.}\ \bibnamefont
  {Rout}}, \bibinfo {author} {\bibfnamefont {I.}~\bibnamefont {Agireen}},
  \bibinfo {author} {\bibfnamefont {E.}~\bibnamefont {Maniv}}, \bibinfo
  {author} {\bibfnamefont {M.}~\bibnamefont {Goldstein}}, \ and\ \bibinfo
  {author} {\bibfnamefont {Y.}~\bibnamefont {Dagan}},\ }\bibfield  {title}
  {\enquote {\bibinfo {title} {Six-fold crystalline anisotropic
  magnetoresistance in (111) {LaAlO}$_3$/{SrTiO}$_3$ oxide interface},}\ }\href
  {\doibase 10.1103/PhysRevB.95.241107} {\bibfield  {journal} {\bibinfo
  {journal} {Phys. Rev. B}\ }\textbf {\bibinfo {volume} {95}},\ \bibinfo
  {pages} {241107} (\bibinfo {year} {2017})}\BibitemShut {NoStop}%
\bibitem [{\citenamefont {Davis}\ \emph
  {et~al.}(2017{\natexlab{a}})\citenamefont {Davis}, \citenamefont {Huang},
  \citenamefont {Han}, \citenamefont {Ariando}, \citenamefont {Venkatesan},\
  and\ \citenamefont {Chandrasekhar}}]{ADMI:ADMI201600830}%
  \BibitemOpen
  \bibfield  {author} {\bibinfo {author} {\bibfnamefont {Samuel~Kenneth}\
  \bibnamefont {Davis}}, \bibinfo {author} {\bibfnamefont {Zhen}\ \bibnamefont
  {Huang}}, \bibinfo {author} {\bibfnamefont {Kun}\ \bibnamefont {Han}},
  \bibinfo {author} {\bibfnamefont {T.}~\bibnamefont {Ariando}}, \bibinfo
  {author} {\bibfnamefont {Thirumalai}\ \bibnamefont {Venkatesan}}, \ and\
  \bibinfo {author} {\bibfnamefont {Venkat}\ \bibnamefont {Chandrasekhar}},\
  }\bibfield  {title} {\enquote {\bibinfo {title} {Electrical transport
  anisotropy controlled by oxygen vacancy concentration in (111)
  {LaAlO}$_3$/{SrTiO}$_3$ interface structures},}\ }\href {\doibase
  10.1002/admi.201600830} {\bibfield  {journal} {\bibinfo  {journal} {Advanced
  Materials Interfaces}\ }\textbf {\bibinfo {volume} {4}},\ \bibinfo {pages}
  {1600830} (\bibinfo {year} {2017}{\natexlab{a}})}\BibitemShut {NoStop}%
\bibitem [{\citenamefont {Monteiro}\ \emph {et~al.}(2017)\citenamefont
  {Monteiro}, \citenamefont {Groenendijk}, \citenamefont {Groen}, \citenamefont
  {de~Bruijckere}, \citenamefont {Gaudenzi}, \citenamefont {van~der Zant},\
  and\ \citenamefont {Caviglia}}]{monteiro_two-dimensional_2017}%
  \BibitemOpen
  \bibfield  {author} {\bibinfo {author} {\bibfnamefont {A.~M. R. V.~L.}\
  \bibnamefont {Monteiro}}, \bibinfo {author} {\bibfnamefont {D.~J.}\
  \bibnamefont {Groenendijk}}, \bibinfo {author} {\bibfnamefont
  {I.}~\bibnamefont {Groen}}, \bibinfo {author} {\bibfnamefont
  {J.}~\bibnamefont {de~Bruijckere}}, \bibinfo {author} {\bibfnamefont
  {R.}~\bibnamefont {Gaudenzi}}, \bibinfo {author} {\bibfnamefont {H.~S.~J.}\
  \bibnamefont {van~der Zant}}, \ and\ \bibinfo {author} {\bibfnamefont
  {A.~D.}\ \bibnamefont {Caviglia}},\ }\bibfield  {title} {\enquote {\bibinfo
  {title} {Two-dimensional superconductivity at the (111)
  {LaAlO}$_3$/{SrTiO}$_3$ interface},}\ }\href {\doibase
  10.1103/PhysRevB.96.020504} {\bibfield  {journal} {\bibinfo  {journal} {Phys.
  Rev. B}\ }\textbf {\bibinfo {volume} {96}},\ \bibinfo {pages} {020504}
  (\bibinfo {year} {2017})}\BibitemShut {NoStop}%
\bibitem [{\citenamefont {Davis}\ \emph
  {et~al.}(2018{\natexlab{a}})\citenamefont {Davis}, \citenamefont {Huang},
  \citenamefont {Han}, \citenamefont {Ariando}, \citenamefont {Venkatesan},\
  and\ \citenamefont {Chandrasekhar}}]{davis_superconductivity_2017}%
  \BibitemOpen
  \bibfield  {author} {\bibinfo {author} {\bibfnamefont {S.}~\bibnamefont
  {Davis}}, \bibinfo {author} {\bibfnamefont {Z.}~\bibnamefont {Huang}},
  \bibinfo {author} {\bibfnamefont {K.}~\bibnamefont {Han}}, \bibinfo {author}
  {\bibnamefont {Ariando}}, \bibinfo {author} {\bibfnamefont {T.}~\bibnamefont
  {Venkatesan}}, \ and\ \bibinfo {author} {\bibfnamefont {V.}~\bibnamefont
  {Chandrasekhar}},\ }\bibfield  {title} {\enquote {\bibinfo {title}
  {Anisotropic superconductivity and frozen electronic states at the (111)
  {LaAlO}$_{3}$/{SrTiO}$_{3}$ interface},}\ }\href {\doibase
  10.1103/PhysRevB.98.024504} {\bibfield  {journal} {\bibinfo  {journal} {Phys.
  Rev. B}\ }\textbf {\bibinfo {volume} {98}},\ \bibinfo {pages} {024504}
  (\bibinfo {year} {2018}{\natexlab{a}})}\BibitemShut {NoStop}%
\bibitem [{\citenamefont {Monteiro}\ \emph {et~al.}(2019)\citenamefont
  {Monteiro}, \citenamefont {Vivek}, \citenamefont {Groenendijk}, \citenamefont
  {Bruneel}, \citenamefont {Leermakers}, \citenamefont {Zeitler}, \citenamefont
  {Gabay},\ and\ \citenamefont {Caviglia}}]{Caviglia2018}%
  \BibitemOpen
  \bibfield  {author} {\bibinfo {author} {\bibfnamefont {A.~M. R. V.~L.}\
  \bibnamefont {Monteiro}}, \bibinfo {author} {\bibfnamefont {M.}~\bibnamefont
  {Vivek}}, \bibinfo {author} {\bibfnamefont {D.~J.}\ \bibnamefont
  {Groenendijk}}, \bibinfo {author} {\bibfnamefont {P.}~\bibnamefont
  {Bruneel}}, \bibinfo {author} {\bibfnamefont {I.}~\bibnamefont {Leermakers}},
  \bibinfo {author} {\bibfnamefont {U.}~\bibnamefont {Zeitler}}, \bibinfo
  {author} {\bibfnamefont {M.}~\bibnamefont {Gabay}}, \ and\ \bibinfo {author}
  {\bibfnamefont {A.~D.}\ \bibnamefont {Caviglia}},\ }\bibfield  {title}
  {\enquote {\bibinfo {title} {Band inversion driven by electronic correlations
  at the (111) {LaAlO}$_{3}$/{SrTiO}$_{3}$ interface},}\ }\href {\doibase
  10.1103/PhysRevB.99.201102} {\bibfield  {journal} {\bibinfo  {journal} {Phys.
  Rev. B}\ }\textbf {\bibinfo {volume} {99}},\ \bibinfo {pages} {201102}
  (\bibinfo {year} {2019})}\BibitemShut {NoStop}%
\bibitem [{\citenamefont {{Khanna}}\ \emph {et~al.}(2019)\citenamefont
  {{Khanna}}, \citenamefont {{Rout}}, \citenamefont {{Mograbi}}, \citenamefont
  {{Tuvia}}, \citenamefont {{Leermakers}}, \citenamefont {{Zeitler}},
  \citenamefont {{Dagan}},\ and\ \citenamefont {{Goldstein}}}]{Goldstein2019}%
  \BibitemOpen
  \bibfield  {author} {\bibinfo {author} {\bibfnamefont {Udit}\ \bibnamefont
  {{Khanna}}}, \bibinfo {author} {\bibfnamefont {Prasanna~K.}\ \bibnamefont
  {{Rout}}}, \bibinfo {author} {\bibfnamefont {Michael}\ \bibnamefont
  {{Mograbi}}}, \bibinfo {author} {\bibfnamefont {Gal}\ \bibnamefont
  {{Tuvia}}}, \bibinfo {author} {\bibfnamefont {Inge}\ \bibnamefont
  {{Leermakers}}}, \bibinfo {author} {\bibfnamefont {Uli}\ \bibnamefont
  {{Zeitler}}}, \bibinfo {author} {\bibfnamefont {Yoram}\ \bibnamefont
  {{Dagan}}}, \ and\ \bibinfo {author} {\bibfnamefont {Moshe}\ \bibnamefont
  {{Goldstein}}},\ }\bibfield  {title} {\enquote {\bibinfo {title} {{Symmetry
  and correlation effects on band structure explain the anomalous transport
  properties of (111) {LaAlO}$_{3}$/{SrTiO}$_{3}$}},}\ }\href@noop {}
  {\bibfield  {journal} {\bibinfo  {journal} {arXiv e-prints}\ ,\ \bibinfo
  {eid} {arXiv:1901.10931}} (\bibinfo {year} {2019})},\ \Eprint
  {http://arxiv.org/abs/1901.10931} {arXiv:1901.10931 [cond-mat.str-el]}
  \BibitemShut {NoStop}%
\bibitem [{\citenamefont {Xiao}\ \emph {et~al.}(2011)\citenamefont {Xiao},
  \citenamefont {Zhu}, \citenamefont {Ran}, \citenamefont {Nagaosa},\ and\
  \citenamefont {Okamoto}}]{Xiao_NComm2011}%
  \BibitemOpen
  \bibfield  {author} {\bibinfo {author} {\bibfnamefont {Di}~\bibnamefont
  {Xiao}}, \bibinfo {author} {\bibfnamefont {Wenguang}\ \bibnamefont {Zhu}},
  \bibinfo {author} {\bibfnamefont {Ying}\ \bibnamefont {Ran}}, \bibinfo
  {author} {\bibfnamefont {Naoto}\ \bibnamefont {Nagaosa}}, \ and\ \bibinfo
  {author} {\bibfnamefont {Satoshi}\ \bibnamefont {Okamoto}},\ }\bibfield
  {title} {\enquote {\bibinfo {title} {Interface engineering of quantum {Hall}
  effects in digital transition metal oxide heterostructures},}\ }\href
  {http://dx.doi.org/10.1038/ncomms1602} {\bibfield  {journal} {\bibinfo
  {journal} {Nat. Comm.}\ }\textbf {\bibinfo {volume} {2}},\ \bibinfo {pages}
  {596 EP --} (\bibinfo {year} {2011})}\BibitemShut {NoStop}%
\bibitem [{\citenamefont {R\"uegg}\ and\ \citenamefont
  {Fiete}(2011)}]{Ruegg_PRB2011}%
  \BibitemOpen
  \bibfield  {author} {\bibinfo {author} {\bibfnamefont {Andreas}\ \bibnamefont
  {R\"uegg}}\ and\ \bibinfo {author} {\bibfnamefont {Gregory~A.}\ \bibnamefont
  {Fiete}},\ }\bibfield  {title} {\enquote {\bibinfo {title} {Topological
  insulators from complex orbital order in transition-metal oxides
  heterostructures},}\ }\href {\doibase 10.1103/PhysRevB.84.201103} {\bibfield
  {journal} {\bibinfo  {journal} {Phys. Rev. B}\ }\textbf {\bibinfo {volume}
  {84}},\ \bibinfo {pages} {201103} (\bibinfo {year} {2011})}\BibitemShut
  {NoStop}%
\bibitem [{\citenamefont {Cook}\ and\ \citenamefont
  {Paramekanti}(2014)}]{Cook_PRL2014}%
  \BibitemOpen
  \bibfield  {author} {\bibinfo {author} {\bibfnamefont {Ashley~M.}\
  \bibnamefont {Cook}}\ and\ \bibinfo {author} {\bibfnamefont {Arun}\
  \bibnamefont {Paramekanti}},\ }\bibfield  {title} {\enquote {\bibinfo {title}
  {Double perovskite heterostructures: Magnetism, {Chern} bands, and {Chern}
  insulators},}\ }\href {\doibase 10.1103/PhysRevLett.113.077203} {\bibfield
  {journal} {\bibinfo  {journal} {Phys. Rev. Lett.}\ }\textbf {\bibinfo
  {volume} {113}},\ \bibinfo {pages} {077203} (\bibinfo {year}
  {2014})}\BibitemShut {NoStop}%
\bibitem [{\citenamefont {Okamoto}\ \emph {et~al.}(2014)\citenamefont
  {Okamoto}, \citenamefont {Zhu}, \citenamefont {Nomura}, \citenamefont
  {Arita}, \citenamefont {Xiao},\ and\ \citenamefont
  {Nagaosa}}]{Okamoto_PRB2014}%
  \BibitemOpen
  \bibfield  {author} {\bibinfo {author} {\bibfnamefont {Satoshi}\ \bibnamefont
  {Okamoto}}, \bibinfo {author} {\bibfnamefont {Wenguang}\ \bibnamefont {Zhu}},
  \bibinfo {author} {\bibfnamefont {Yusuke}\ \bibnamefont {Nomura}}, \bibinfo
  {author} {\bibfnamefont {Ryotaro}\ \bibnamefont {Arita}}, \bibinfo {author}
  {\bibfnamefont {Di}~\bibnamefont {Xiao}}, \ and\ \bibinfo {author}
  {\bibfnamefont {Naoto}\ \bibnamefont {Nagaosa}},\ }\bibfield  {title}
  {\enquote {\bibinfo {title} {Correlation effects in (111) bilayers of
  perovskite transition-metal oxides},}\ }\href {\doibase
  10.1103/PhysRevB.89.195121} {\bibfield  {journal} {\bibinfo  {journal} {Phys.
  Rev. B}\ }\textbf {\bibinfo {volume} {89}},\ \bibinfo {pages} {195121}
  (\bibinfo {year} {2014})}\BibitemShut {NoStop}%
\bibitem [{\citenamefont {Hu}\ \emph {et~al.}(2015)\citenamefont {Hu},
  \citenamefont {Zhong},\ and\ \citenamefont {Fiete}}]{Fiete_SciRep2015}%
  \BibitemOpen
  \bibfield  {author} {\bibinfo {author} {\bibfnamefont {Xiang}\ \bibnamefont
  {Hu}}, \bibinfo {author} {\bibfnamefont {Zhicheng}\ \bibnamefont {Zhong}}, \
  and\ \bibinfo {author} {\bibfnamefont {Gregory~A.}\ \bibnamefont {Fiete}},\
  }\bibfield  {title} {\enquote {\bibinfo {title} {First principles prediction
  of topological phases in thin films of pyrochlore iridates},}\ }\href
  {http://dx.doi.org/10.1038/srep11072} {\bibfield  {journal} {\bibinfo
  {journal} {Sci. Rep.}\ }\textbf {\bibinfo {volume} {5}},\ \bibinfo {pages}
  {11072 EP --} (\bibinfo {year} {2015})}\BibitemShut {NoStop}%
\bibitem [{\citenamefont {Baidya}\ \emph {et~al.}(2016)\citenamefont {Baidya},
  \citenamefont {Waghmare}, \citenamefont {Paramekanti},\ and\ \citenamefont
  {Saha-Dasgupta}}]{Baidya_PRB2016}%
  \BibitemOpen
  \bibfield  {author} {\bibinfo {author} {\bibfnamefont {Santu}\ \bibnamefont
  {Baidya}}, \bibinfo {author} {\bibfnamefont {Umesh~V.}\ \bibnamefont
  {Waghmare}}, \bibinfo {author} {\bibfnamefont {Arun}\ \bibnamefont
  {Paramekanti}}, \ and\ \bibinfo {author} {\bibfnamefont {Tanusri}\
  \bibnamefont {Saha-Dasgupta}},\ }\bibfield  {title} {\enquote {\bibinfo
  {title} {High-temperature large-gap quantum anomalous {Hall} insulating state
  in ultrathin double perovskite films},}\ }\href {\doibase
  10.1103/PhysRevB.94.155405} {\bibfield  {journal} {\bibinfo  {journal} {Phys.
  Rev. B}\ }\textbf {\bibinfo {volume} {94}},\ \bibinfo {pages} {155405}
  (\bibinfo {year} {2016})}\BibitemShut {NoStop}%
\bibitem [{\citenamefont {Si}\ \emph {et~al.}(2017)\citenamefont {Si},
  \citenamefont {Janson}, \citenamefont {Li}, \citenamefont {Zhong},
  \citenamefont {Liao}, \citenamefont {Koster},\ and\ \citenamefont
  {Held}}]{Held_2016}%
  \BibitemOpen
  \bibfield  {author} {\bibinfo {author} {\bibfnamefont {Liang}\ \bibnamefont
  {Si}}, \bibinfo {author} {\bibfnamefont {Oleg}\ \bibnamefont {Janson}},
  \bibinfo {author} {\bibfnamefont {Gang}\ \bibnamefont {Li}}, \bibinfo
  {author} {\bibfnamefont {Zhicheng}\ \bibnamefont {Zhong}}, \bibinfo {author}
  {\bibfnamefont {Zhaoliang}\ \bibnamefont {Liao}}, \bibinfo {author}
  {\bibfnamefont {Gertjan}\ \bibnamefont {Koster}}, \ and\ \bibinfo {author}
  {\bibfnamefont {Karsten}\ \bibnamefont {Held}},\ }\bibfield  {title}
  {\enquote {\bibinfo {title} {Quantum anomalous {Hall} state in ferromagnetic
  {SrRuO}$_3$ (111) bilayers},}\ }\href {\doibase
  10.1103/PhysRevLett.119.026402} {\bibfield  {journal} {\bibinfo  {journal}
  {Phys. Rev. Lett.}\ }\textbf {\bibinfo {volume} {119}},\ \bibinfo {pages}
  {026402} (\bibinfo {year} {2017})}\BibitemShut {NoStop}%
\bibitem [{\citenamefont {Kim}\ and\ \citenamefont
  {Kee}(2017)}]{Kee_NPJQM2017}%
  \BibitemOpen
  \bibfield  {author} {\bibinfo {author} {\bibfnamefont {Heung-Sik}\
  \bibnamefont {Kim}}\ and\ \bibinfo {author} {\bibfnamefont {Hae-Young}\
  \bibnamefont {Kee}},\ }\bibfield  {title} {\enquote {\bibinfo {title}
  {Realizing {Haldane} model in {Fe}-based honeycomb ferromagnetic
  insulators},}\ }\href {\doibase 10.1038/s41535-017-0021-z} {\bibfield
  {journal} {\bibinfo  {journal} {npj Quantum Materials}\ }\textbf {\bibinfo
  {volume} {2}},\ \bibinfo {pages} {20} (\bibinfo {year} {2017})}\BibitemShut
  {NoStop}%
\bibitem [{\citenamefont {Baidya}\ \emph {et~al.}(2015)\citenamefont {Baidya},
  \citenamefont {Waghmare}, \citenamefont {Paramekanti},\ and\ \citenamefont
  {Saha-Dasgupta}}]{baidya2015}%
  \BibitemOpen
  \bibfield  {author} {\bibinfo {author} {\bibfnamefont {Santu}\ \bibnamefont
  {Baidya}}, \bibinfo {author} {\bibfnamefont {Umesh~V.}\ \bibnamefont
  {Waghmare}}, \bibinfo {author} {\bibfnamefont {Arun}\ \bibnamefont
  {Paramekanti}}, \ and\ \bibinfo {author} {\bibfnamefont {Tanusri}\
  \bibnamefont {Saha-Dasgupta}},\ }\bibfield  {title} {\enquote {\bibinfo
  {title} {Controlled confinement of half-metallic two-dimensional electron gas
  in {BaTiO}$_{3}$/{Ba}$_{2}${FeReO}$_{6}$/{BaTiO}$_{3}$ heterostructures: A
  first-principles study},}\ }\href {\doibase 10.1103/PhysRevB.92.161106}
  {\bibfield  {journal} {\bibinfo  {journal} {Phys. Rev. B}\ }\textbf {\bibinfo
  {volume} {92}},\ \bibinfo {pages} {161106} (\bibinfo {year}
  {2015})}\BibitemShut {NoStop}%
\bibitem [{\citenamefont {Davis}\ \emph
  {et~al.}(2017{\natexlab{b}})\citenamefont {Davis}, \citenamefont
  {Chandrasekhar}, \citenamefont {Huang}, \citenamefont {Han}, \citenamefont
  {Ariando},\ and\ \citenamefont {Venkatesan}}]{Venkatesan2017}%
  \BibitemOpen
  \bibfield  {author} {\bibinfo {author} {\bibfnamefont {S.}~\bibnamefont
  {Davis}}, \bibinfo {author} {\bibfnamefont {V.}~\bibnamefont
  {Chandrasekhar}}, \bibinfo {author} {\bibfnamefont {Z.}~\bibnamefont
  {Huang}}, \bibinfo {author} {\bibfnamefont {K.}~\bibnamefont {Han}}, \bibinfo
  {author} {\bibfnamefont {T.}~\bibnamefont {Ariando}}, \ and\ \bibinfo
  {author} {\bibfnamefont {T.}~\bibnamefont {Venkatesan}},\ }\bibfield  {title}
  {\enquote {\bibinfo {title} {Anisotropic multicarrier transport at the (111)
  {LaAlO}$_3$/{SrTiO}$_3$ interface},}\ }\href {\doibase
  10.1103/PhysRevB.95.035127} {\bibfield  {journal} {\bibinfo  {journal} {Phys.
  Rev. B}\ }\textbf {\bibinfo {volume} {95}},\ \bibinfo {pages} {035127}
  (\bibinfo {year} {2017}{\natexlab{b}})}\BibitemShut {NoStop}%
\bibitem [{\citenamefont {{Burgos}}\ \emph {et~al.}(2018)\citenamefont
  {{Burgos}}, \citenamefont {{Warnes}},\ and\ \citenamefont {{De La
  Espriella}}}]{AMR2Band}%
  \BibitemOpen
  \bibfield  {author} {\bibinfo {author} {\bibfnamefont {R.}~\bibnamefont
  {{Burgos}}}, \bibinfo {author} {\bibfnamefont {J.~H.}\ \bibnamefont
  {{Warnes}}}, \ and\ \bibinfo {author} {\bibfnamefont {N.}~\bibnamefont {{De
  La Espriella}}},\ }\bibfield  {title} {\enquote {\bibinfo {title}
  {{Anisotropic magnetoresistance in 2DEG with {Rashba} spin-orbit
  coupling}},}\ }\href {\doibase 10.1016/j.jmmm.2018.07.013} {\bibfield
  {journal} {\bibinfo  {journal} {Journal of Magnetism and Magnetic Materials}\
  }\textbf {\bibinfo {volume} {466}},\ \bibinfo {pages} {234--237} (\bibinfo
  {year} {2018})}\BibitemShut {NoStop}%
\bibitem [{\citenamefont {Trushin}\ \emph {et~al.}(2009)\citenamefont
  {Trushin}, \citenamefont {V\'yborn\'y}, \citenamefont {Moraczewski},
  \citenamefont {Kovalev}, \citenamefont {Schliemann},\ and\ \citenamefont
  {Jungwirth}}]{AMRMagnetic}%
  \BibitemOpen
  \bibfield  {author} {\bibinfo {author} {\bibfnamefont {Maxim}\ \bibnamefont
  {Trushin}}, \bibinfo {author} {\bibfnamefont {Karel}\ \bibnamefont
  {V\'yborn\'y}}, \bibinfo {author} {\bibfnamefont {Peter}\ \bibnamefont
  {Moraczewski}}, \bibinfo {author} {\bibfnamefont {Alexey~A.}\ \bibnamefont
  {Kovalev}}, \bibinfo {author} {\bibfnamefont {John}\ \bibnamefont
  {Schliemann}}, \ and\ \bibinfo {author} {\bibfnamefont {T.}~\bibnamefont
  {Jungwirth}},\ }\bibfield  {title} {\enquote {\bibinfo {title} {Anisotropic
  magnetoresistance of spin-orbit coupled carriers scattered from polarized
  magnetic impurities},}\ }\href {\doibase 10.1103/PhysRevB.80.134405}
  {\bibfield  {journal} {\bibinfo  {journal} {Phys. Rev. B}\ }\textbf {\bibinfo
  {volume} {80}},\ \bibinfo {pages} {134405} (\bibinfo {year}
  {2009})}\BibitemShut {NoStop}%
\bibitem [{\citenamefont {McGuire}\ and\ \citenamefont
  {Potter}(1975)}]{mcguire1975anisotropic}%
  \BibitemOpen
  \bibfield  {author} {\bibinfo {author} {\bibfnamefont {T}~\bibnamefont
  {McGuire}}\ and\ \bibinfo {author} {\bibfnamefont {RL}~\bibnamefont
  {Potter}},\ }\bibfield  {title} {\enquote {\bibinfo {title} {Anisotropic
  magnetoresistance in ferromagnetic 3{D} alloys},}\ }\href@noop {} {\bibfield
  {journal} {\bibinfo  {journal} {IEEE Transactions on Magnetics}\ }\textbf
  {\bibinfo {volume} {11}},\ \bibinfo {pages} {1018--1038} (\bibinfo {year}
  {1975})}\BibitemShut {NoStop}%
\bibitem [{\citenamefont {Wang}\ \emph {et~al.}(2005)\citenamefont {Wang},
  \citenamefont {Edmonds}, \citenamefont {Campion}, \citenamefont {Zhao},
  \citenamefont {Foxon},\ and\ \citenamefont {Gallagher}}]{AMRgaas}%
  \BibitemOpen
  \bibfield  {author} {\bibinfo {author} {\bibfnamefont {K.~Y.}\ \bibnamefont
  {Wang}}, \bibinfo {author} {\bibfnamefont {K.~W.}\ \bibnamefont {Edmonds}},
  \bibinfo {author} {\bibfnamefont {R.~P.}\ \bibnamefont {Campion}}, \bibinfo
  {author} {\bibfnamefont {L.~X.}\ \bibnamefont {Zhao}}, \bibinfo {author}
  {\bibfnamefont {C.~T.}\ \bibnamefont {Foxon}}, \ and\ \bibinfo {author}
  {\bibfnamefont {B.~L.}\ \bibnamefont {Gallagher}},\ }\bibfield  {title}
  {\enquote {\bibinfo {title} {Anisotropic magnetoresistance and magnetic
  anisotropy in high-quality ({Ga},{Mn}){As} films},}\ }\href {\doibase
  10.1103/PhysRevB.72.085201} {\bibfield  {journal} {\bibinfo  {journal} {Phys.
  Rev. B}\ }\textbf {\bibinfo {volume} {72}},\ \bibinfo {pages} {085201}
  (\bibinfo {year} {2005})}\BibitemShut {NoStop}%
\bibitem [{\citenamefont {Ranieri}\ \emph {et~al.}(2008)\citenamefont
  {Ranieri}, \citenamefont {Rushforth}, \citenamefont {V{\'{y}}born{\'{y}}},
  \citenamefont {Rana}, \citenamefont {Ahmad}, \citenamefont {Campion},
  \citenamefont {Foxon}, \citenamefont {Gallagher}, \citenamefont {Irvine},
  \citenamefont {Wunderlich},\ and\ \citenamefont
  {Jungwirth}}]{De_Ranieri_2008}%
  \BibitemOpen
  \bibfield  {author} {\bibinfo {author} {\bibfnamefont {E~De}\ \bibnamefont
  {Ranieri}}, \bibinfo {author} {\bibfnamefont {A~W}\ \bibnamefont
  {Rushforth}}, \bibinfo {author} {\bibfnamefont {K}~\bibnamefont
  {V{\'{y}}born{\'{y}}}}, \bibinfo {author} {\bibfnamefont {U}~\bibnamefont
  {Rana}}, \bibinfo {author} {\bibfnamefont {E}~\bibnamefont {Ahmad}}, \bibinfo
  {author} {\bibfnamefont {R~P}\ \bibnamefont {Campion}}, \bibinfo {author}
  {\bibfnamefont {C~T}\ \bibnamefont {Foxon}}, \bibinfo {author} {\bibfnamefont
  {B~L}\ \bibnamefont {Gallagher}}, \bibinfo {author} {\bibfnamefont {A~C}\
  \bibnamefont {Irvine}}, \bibinfo {author} {\bibfnamefont {J}~\bibnamefont
  {Wunderlich}}, \ and\ \bibinfo {author} {\bibfnamefont {T}~\bibnamefont
  {Jungwirth}},\ }\bibfield  {title} {\enquote {\bibinfo {title}
  {Lithographically and electrically controlled strain effects on anisotropic
  magnetoresistance in ({Ga},{Mn}){As}},}\ }\href {\doibase
  10.1088/1367-2630/10/6/065003} {\bibfield  {journal} {\bibinfo  {journal}
  {New Journal of Physics}\ }\textbf {\bibinfo {volume} {10}},\ \bibinfo
  {pages} {065003} (\bibinfo {year} {2008})}\BibitemShut {NoStop}%
\bibitem [{\citenamefont {Bovenzi}\ and\ \citenamefont
  {Diez}(2017)}]{diez2016}%
  \BibitemOpen
  \bibfield  {author} {\bibinfo {author} {\bibfnamefont {N.}~\bibnamefont
  {Bovenzi}}\ and\ \bibinfo {author} {\bibfnamefont {M.}~\bibnamefont {Diez}},\
  }\bibfield  {title} {\enquote {\bibinfo {title} {Semiclassical theory of
  anisotropic transport at {LaAlO}$_{3}$/{SrTiO}$_{3}$ interfaces under an
  in-plane magnetic field},}\ }\href {\doibase 10.1103/PhysRevB.95.205430}
  {\bibfield  {journal} {\bibinfo  {journal} {Phys. Rev. B}\ }\textbf {\bibinfo
  {volume} {95}},\ \bibinfo {pages} {205430} (\bibinfo {year}
  {2017})}\BibitemShut {NoStop}%
\bibitem [{\citenamefont {Davis}\ \emph
  {et~al.}(2018{\natexlab{b}})\citenamefont {Davis}, \citenamefont {Huang},
  \citenamefont {Han}, \citenamefont {Ariando}, \citenamefont {Venkatesan},\
  and\ \citenamefont {Chandrasekhar}}]{Chandrasekhar2018}%
  \BibitemOpen
  \bibfield  {author} {\bibinfo {author} {\bibfnamefont {S.}~\bibnamefont
  {Davis}}, \bibinfo {author} {\bibfnamefont {Z.}~\bibnamefont {Huang}},
  \bibinfo {author} {\bibfnamefont {K.}~\bibnamefont {Han}}, \bibinfo {author}
  {\bibnamefont {Ariando}}, \bibinfo {author} {\bibfnamefont {T.}~\bibnamefont
  {Venkatesan}}, \ and\ \bibinfo {author} {\bibfnamefont {V.}~\bibnamefont
  {Chandrasekhar}},\ }\bibfield  {title} {\enquote {\bibinfo {title}
  {Signatures of electronic nematicity in (111) {LaAlO}$_3$/{SrTiO}$_3$
  interfaces},}\ }\href {\doibase 10.1103/PhysRevB.97.041408} {\bibfield
  {journal} {\bibinfo  {journal} {Phys. Rev. B}\ }\textbf {\bibinfo {volume}
  {97}},\ \bibinfo {pages} {041408} (\bibinfo {year}
  {2018}{\natexlab{b}})}\BibitemShut {NoStop}%
\bibitem [{\citenamefont {Reyes-Lillo}\ \emph {et~al.}(2019)\citenamefont
  {Reyes-Lillo}, \citenamefont {Rabe},\ and\ \citenamefont
  {Neaton}}]{Neaton2019}%
  \BibitemOpen
  \bibfield  {author} {\bibinfo {author} {\bibfnamefont {Sebastian~E.}\
  \bibnamefont {Reyes-Lillo}}, \bibinfo {author} {\bibfnamefont {Karin~M.}\
  \bibnamefont {Rabe}}, \ and\ \bibinfo {author} {\bibfnamefont {Jeffrey~B.}\
  \bibnamefont {Neaton}},\ }\bibfield  {title} {\enquote {\bibinfo {title}
  {Ferroelectricity in [111]-oriented epitaxially strained
  ${\mathrm{srtio}}_{3}$ from first principles},}\ }\href {\doibase
  10.1103/PhysRevMaterials.3.030601} {\bibfield  {journal} {\bibinfo  {journal}
  {Phys. Rev. Materials}\ }\textbf {\bibinfo {volume} {3}},\ \bibinfo {pages}
  {030601} (\bibinfo {year} {2019})}\BibitemShut {NoStop}%
\bibitem [{\citenamefont {Aharony}\ \emph {et~al.}(1977)\citenamefont
  {Aharony}, \citenamefont {M\"uller},\ and\ \citenamefont
  {Berlinger}}]{aharony_trigonal--tetragonal_1977}%
  \BibitemOpen
  \bibfield  {author} {\bibinfo {author} {\bibfnamefont {Amnon}\ \bibnamefont
  {Aharony}}, \bibinfo {author} {\bibfnamefont {K.~A.}\ \bibnamefont
  {M\"uller}}, \ and\ \bibinfo {author} {\bibfnamefont {W.}~\bibnamefont
  {Berlinger}},\ }\bibfield  {title} {\enquote {\bibinfo {title}
  {Trigonal-to-{Tetragonal} {Transition} in {Stressed} {SrTiO}$_3$: {A}
  {Realization} of the {Three}-{State} {Potts} {Model}},}\ }\href {\doibase
  10.1103/PhysRevLett.38.33} {\bibfield  {journal} {\bibinfo  {journal} {Phys.
  Rev. Lett.}\ }\textbf {\bibinfo {volume} {38}},\ \bibinfo {pages} {33--36}
  (\bibinfo {year} {1977})}\BibitemShut {NoStop}%
\bibitem [{\citenamefont {Boudjada}\ \emph {et~al.}(2018)\citenamefont
  {Boudjada}, \citenamefont {Wachtel},\ and\ \citenamefont
  {Paramekanti}}]{paramekanti2018}%
  \BibitemOpen
  \bibfield  {author} {\bibinfo {author} {\bibfnamefont {Nazim}\ \bibnamefont
  {Boudjada}}, \bibinfo {author} {\bibfnamefont {Gideon}\ \bibnamefont
  {Wachtel}}, \ and\ \bibinfo {author} {\bibfnamefont {Arun}\ \bibnamefont
  {Paramekanti}},\ }\bibfield  {title} {\enquote {\bibinfo {title} {Magnetic
  and {Nematic} {Orders} of the {Two}-{Dimensional} {Electron} {Gas} at {Oxide}
  (111) {Surfaces} and {Interfaces}},}\ }\href {\doibase
  10.1103/PhysRevLett.120.086802} {\bibfield  {journal} {\bibinfo  {journal}
  {Phys. Rev. Lett.}\ }\textbf {\bibinfo {volume} {120}},\ \bibinfo {pages}
  {086802} (\bibinfo {year} {2018})}\BibitemShut {NoStop}%
\bibitem [{\citenamefont {M\"uller}\ \emph {et~al.}(1991)\citenamefont
  {M\"uller}, \citenamefont {Berlinger},\ and\ \citenamefont
  {Tosatti}}]{muller_indication_1991}%
  \BibitemOpen
  \bibfield  {author} {\bibinfo {author} {\bibfnamefont {K.~Alex}\ \bibnamefont
  {M\"uller}}, \bibinfo {author} {\bibfnamefont {W.}~\bibnamefont {Berlinger}},
  \ and\ \bibinfo {author} {\bibfnamefont {E.}~\bibnamefont {Tosatti}},\
  }\bibfield  {title} {\enquote {\bibinfo {title} {Indication for a novel phase
  in the quantum paraelectric regime of {SrTiO}$_3$},}\ }\href {\doibase
  10.1007/BF01313549} {\bibfield  {journal} {\bibinfo  {journal} {Z. Physik B -
  Condensed Matter}\ }\textbf {\bibinfo {volume} {84}},\ \bibinfo {pages}
  {277--283} (\bibinfo {year} {1991})}\BibitemShut {NoStop}%
\bibitem [{\citenamefont {Dugaev}\ \emph {et~al.}(2012)\citenamefont {Dugaev},
  \citenamefont {Inglot}, \citenamefont {Sherman}, \citenamefont {Berakdar},\
  and\ \citenamefont {Barna\ifmmode~\acute{s}\else \'{s}\fi{}}}]{Sherman2012}%
  \BibitemOpen
  \bibfield  {author} {\bibinfo {author} {\bibfnamefont {V.~K.}\ \bibnamefont
  {Dugaev}}, \bibinfo {author} {\bibfnamefont {M.}~\bibnamefont {Inglot}},
  \bibinfo {author} {\bibfnamefont {E.~Ya.}\ \bibnamefont {Sherman}}, \bibinfo
  {author} {\bibfnamefont {J.}~\bibnamefont {Berakdar}}, \ and\ \bibinfo
  {author} {\bibfnamefont {J.}~\bibnamefont {Barna\ifmmode~\acute{s}\else
  \'{s}\fi{}}},\ }\bibfield  {title} {\enquote {\bibinfo {title} {Nonlinear
  anomalous {Hall} effect and negative magnetoresistance in a system with
  random {Rashba} field},}\ }\href {\doibase 10.1103/PhysRevLett.109.206601}
  {\bibfield  {journal} {\bibinfo  {journal} {Phys. Rev. Lett.}\ }\textbf
  {\bibinfo {volume} {109}},\ \bibinfo {pages} {206601} (\bibinfo {year}
  {2012})}\BibitemShut {NoStop}%
\bibitem [{\citenamefont {Bovenzi}\ \emph {et~al.}(2017)\citenamefont
  {Bovenzi}, \citenamefont {Caprara}, \citenamefont {Grilli}, \citenamefont
  {Raimondi}, \citenamefont {Scopigno},\ and\ \citenamefont
  {Seibold}}]{Bovenzi2017}%
  \BibitemOpen
  \bibfield  {author} {\bibinfo {author} {\bibfnamefont {N.}~\bibnamefont
  {Bovenzi}}, \bibinfo {author} {\bibfnamefont {S.}~\bibnamefont {Caprara}},
  \bibinfo {author} {\bibfnamefont {M.}~\bibnamefont {Grilli}}, \bibinfo
  {author} {\bibfnamefont {R.}~\bibnamefont {Raimondi}}, \bibinfo {author}
  {\bibfnamefont {N.}~\bibnamefont {Scopigno}}, \ and\ \bibinfo {author}
  {\bibfnamefont {G.}~\bibnamefont {Seibold}},\ }\bibfield  {title} {\enquote
  {\bibinfo {title} {Density inhomogeneities and {Rashba} spin-orbit coupling
  interplay in oxide interfaces},}\ }\href
  {http://www.sciencedirect.com/science/article/pii/S0022369717305991}
  {\bibfield  {journal} {\bibinfo  {journal} {Journal of Physics and Chemistry
  of Solids}\ } (\bibinfo {year} {2017})}\BibitemShut {NoStop}%
\end{thebibliography}%

\appendix

\section{Numerical solution of the Boltzmann equation}
\label{boltz}
It is helpful to rewrite Eq. \eqref{eqn:boltzmann} with discretized momenta as
\begin{widetext}
\begin{eqnarray}
- \frac{\partial f_{n,\bk}}{\partial \varepsilon_{n,\bk}}   E^i v^i_{n,\bk} = \!\!\! \sum_{m,l,\bk',\bk''} \!
(\delta_{nl}\delta_{\bk,\bk''} -\delta_{ml}\delta_{\bk',\bk''})~g_{l,\bk''}~|\langle n\bk|\hat{U}|m\bk'\rangle|^2~\delta(\varepsilon_{n,\bk}-\varepsilon_{m,\bk'}),
\label{eqn:boltzmann2}
\end{eqnarray}	
\end{widetext}
where we have dropped factors of $({\cal N}, e)$ since we will only be interested in {\it ratios} of transport coefficients where these will cancel.
Taking a derivative with respect to $E^i$, and lumping together band and momentum indices via $\mu\equiv(n,\bk)$, $\nu\equiv(m,\bk')$ and $\alpha\equiv(l,\bk'')$, we get
\begin{eqnarray}
	\left(- 
	\frac{\partial f}{\partial \varepsilon} v^i\right)_{\!\!\!\mu}
	\!\! &=& \! \sum_{\nu,\alpha}(\delta_{\alpha\mu} \! -\! \delta_{\alpha\nu})|U_{\mu\nu}|^2\delta(\varepsilon_\mu \! -\! \varepsilon_\nu)
	\left(\frac{\partial g}{\partial E^i}\right)_{\!\!\!\alpha}\nonumber\\
	&\equiv& \sum_{\alpha}(A_{\mu\alpha}-B_{\mu\alpha})\left(\frac{\partial g}{\partial E^i}\right)_{\!\!\!\alpha},
	\label{eqn:matboltzmann}
\end{eqnarray}
where $U_{\mu\nu} \equiv \la \mu | \hat{U} | \nu \ra$,
${B_{\mu\alpha}\equiv|\langle \mu|\hat{U}|\alpha\rangle|^2\delta(\varepsilon_\mu-\varepsilon_\alpha)}$ and ${A_{\mu\alpha}\equiv
\sum_{\nu}B_{\mu\nu}\delta_{\mu\alpha}}$. Equation \eqref{eqn:matboltzmann} is a matrix inversion problem for the `vector' $(\partial g/\partial E^i)$;
however $(A-B)^{-1}$ is not well defined since it has a zero eigenvalue associated with a constant (band and momentum independent) vector 
$(\partial g/\partial E^i) \propto \mathds{1}$.

Rather than using SVD algorithms and pseudoinverse techniques, we found that an efficient way to solve Eq.~\eqref{eqn:matboltzmann} is to write it as an iterative equation
\begin{eqnarray}
	\left[\frac{\partial g}{\partial E^i}\right]_{p+1} = A^{-1} \left(B \cdot \left[\frac{\partial g}{\partial E^i}\right]_p -
	\frac{\partial f}{\partial \varepsilon} v^i\right).
	\label{eqn:matboltzmann2}
\end{eqnarray}
We begin at step $p=0$ by guessing a random initial vector $[\partial g/\partial E^i]_{p=0}$, removing its projection in the constant eigenvector subspace,
and evaluating the right-hand side of the above equation to obtain a new $[\partial g/\partial E^i]_{p=1}$. We then repeat this process, taking care at each step
to remove the projection of $[\partial g/\partial E^i]_{p}$ to the constant eigenvector (to avoid errors that can creep in from numerical precision).
Convergence is reached when the $L_2$ norm $||\frac{\partial f}{\partial \varepsilon} v^i + (A-B)\frac{\partial g}{\partial E^i}||_2<\text{tol}$, 
where tol is typically chosen to be $10^{-10}$. 

We find that for proper convergence one must choose momentum mesh sizes of the order $2500\times2500$ for temperatures of $T \sim 5$K. Since the $(A,B)$ matrix dimensions scales as $(6\cdot2500^2\times6\cdot2500^2)$ for a 6-band problem, the memory requirements far exceed what can be handled by clusters. To reduce the dimensionality of the problem, we use the fact that $g$ and $\partial g/\partial E^i$ must go to zero far from the Fermi momenta and work with momenta within a certain temperature $T$ window of the Fermi momenta. 
It is found that a temperature window of $\pm 6 \times T$ is reasonable for low densities ($n\lesssim0.1$) while one needs up to $\pm12 \times T$ at high densities $n\gtrsim0.4$.

\section{Landau theory of nematic/polar 2DEG}
\label{landau}
We start with the Landau theory for ferroelectric order in a bulk 3D cubic crystal in terms of the vector order parameter $\vec\varphi=(\varphi_x,\varphi_y,\varphi_z)$
representing the electric dipole moment vector.
In a displacive ferroelectric, this arises due to displacements of the ions away from high symmetry positions; for instance in BaTiO$_3$ or $^{18}$O isotope-substituted 
SrTiO$_3$, it would involve off-center displacements of the
Ba$^{2+}$ or Sr$^{2+}$ ions from the cube center, and the Ti$^{4+}$ ions within the oxygen octahedra. 
The $(x,y,z)$ components refer to the cubic axes of crystal. The symmetry allowed bulk terms are
\begin{equation}
    \mathcal{F}_{\rm bulk}=r_B \vec{\varphi}^2+u_B \vec{\varphi}^4+w_B (\varphi_x^4+\varphi_y^4+\varphi_z^4),
       \label{eqn:FEbulk}
\end{equation}
with subscript $B$ on the coefficients denoting bulk.
Explicit spatio-temporal gradients of the order parameter, stemming from thermal or quantum fluctuation effects are ignored here; they are 
only taken into account to the extent that they renormalize the coefficients of this effective Landau theory.
Here, ${r_B \propto (T - T_c)}$ where $T_c$ is the mean field ordering temperature into the 3D ferroelectric state. Since SrTiO$_3$ remains a paraelectric,  it 
has $r_B > 0$ down to the lowest temperature, but proximity to a quantum critical point can lead to small $r_B (T=0)$.
Even in such cases, where the bulk remains a paraelectric, a spontaneous symmetry breaking state might still arise at the surface.

With $\hat{n} \parallel [111]$, additional terms are allowed at the surface,
\bea
        \!\! \Delta \mathcal{F}_{111}&\!=\!& \alpha \vec{\varphi}\cdot\hat{n}\! + \! g (\vec{\varphi}\cdot\hat{n})^2 \!+\! \lambda_1 (\vec \varphi \cdot \hat{n})^3 \! + \! \lambda_2 (\vec\varphi\cdot\hat{n}) \vec \varphi^2 \nonumber\\
        &+& \lambda_3 \varphi_x \varphi_y \varphi_z + w_1 (\vec\varphi\cdot \hat{n})^4 + w_2 (\vec\varphi\cdot\hat{n})^2 \vec\varphi^2 \nonumber\\
        &+& w_3 (\varphi_x^2 \varphi_y^2+\varphi_y^2 \varphi_z^2+\varphi_z^2 \varphi_x^2).
    \label{eqn:FEsur}
\eea

From the bulk free energy, $w_B <0$ favors states in which the dipole moment points along $[100]$ or symmetry related axes for a total of six symmetry-related ground
states. At a $(111)$ surface, the inversion breaking term $\alpha$ splits this sixfold degeneracy into two triplets $(+\hat{x},+\hat{y},+\hat{z})$ and 
$(-\hat{x},-\hat{y},-\hat{z})$. Any symmetry-breaking surface phase transitions will involve breaking the residual  $C_3$ and mirror symmetries of these
triplets.

Similarly, $w_B > 0$ in the bulk free energy favours the eight degenerate states where the dipole moment points along $[111]$ and its symmetry equivalents. 
At the $(111)$ surface, $g>0$ breaks this degeneracy into a high energy doublet (with 
dipoles along $\pm\hat{n}$) and a low energy sextet with dipoles along the other 
directions: $(11\bar{1})$, $(1\bar{1}1)$, $(\bar{1}11)$, $(1\bar{1}\bar{1})$, $(\bar{1}1\bar{1})$, 
and $(\bar{1}\bar{1}1)$. The term $\alpha$ splits this low energy sextet into two triplets $[(11\bar{1}),(1\bar{1}1),(\bar{1}11)]$ and
$[(1\bar{1}\bar{1}),(\bar{1}1\bar{1}),(\bar{1}\bar{1}1)]$. As for the case with $w_B <0$, the residual surface symmetry breaking will involve breaking
$C_3$ and mirror symmetries.

\begin{figure}[h]
	\begin{overpic}[width=\linewidth]{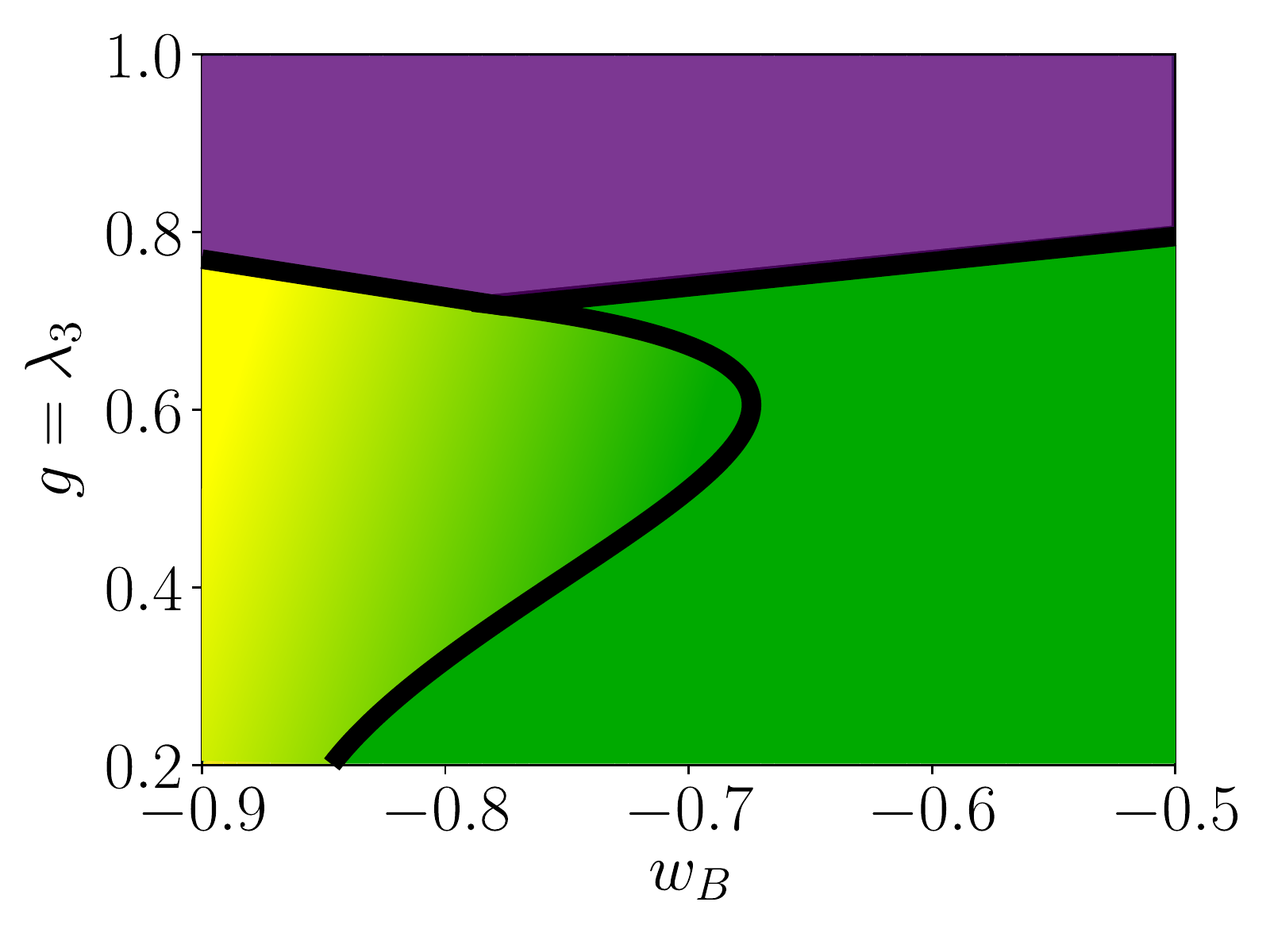}
	\put(55,20){\textcolor{white}{\Large$\varphi_\perp\neq0,\psi=0$}}
	\put(30,60){\textcolor{white}{\Large$\varphi_\perp=0,\psi\neq0$}}
	\put(18,40){\Large$\varphi_\perp\neq0,\psi\neq0$}
	\end{overpic}
	\caption{Phase diagram of the Landau theory, given by Eq.~\eqref{eqn:FEbulk} and Eq.~\eqref{eqn:FEsur}, 
	as a function of $w_B$ and $g=\lambda_3$, keeping all other coefficients fixed as indicated in the text, and
	where $\varphi_\perp$ and $\psi$ are defined in Eq.~\eqref{eq:defpsi}.
	The  phase with $\psi=0$ is a ``paraelectric'' phase, and it is associated with a moment pointing along $\hat{n}$ (which is
	always symmetry allowed at a surface or interface), while the phases with $\psi\neq0$ are symmetry-broken phases, having a purely in-plane ($\varphi_\perp=0$) or partially in-plane ($\varphi_\perp\neq0$) electric dipole moments. Conduction electrons will convert these symmetry broken phases with $\psi \neq 0$ into 2D polar metals. }
	\label{fig:PD}
\end{figure}

As an illustrative example, we plot the phase diagram of such a Landau theory in Fig. \ref{fig:PD}. Here, we choose to work in units where the bulk Landau theory
coefficient $u_B=1$, we set $r_B = 0.01$, and we vary $w_B$. For the surface coefficients, we set $\alpha= - 0.2$,
$\lambda_1 = \lambda_2=0$, and we drop quartic invariants $w_1=w_2=w_3=0$ while varying $g = \lambda_3$.
Since the surface breaks $\hat{n} \!\! \to \!\! -\hat{n}$ inversion, it is useful to parametrize 
\be
\vec{\varphi} \equiv \varphi_\perp\hat{n}+\psi \vec{\gamma}+\psi^* \vec{\gamma}^*,
\label{eq:defpsi}
\ee
where ${\hat{n} \equiv (1,1,1)/\sqrt{3},\vec\gamma \equiv (1,\omega,\omega^2)/\sqrt{3}}$, and ${\omega=\mathrm{e}^{\i2\pi/3}}$.
Here, $\varphi_\perp$ is a non-symmetry-breaking polarization, while the complex $\psi \neq 0$ reflects spontaneous symmetry breaking of the surface symmetries.

We can simplify the Landau theory to focus only on the in-plane spontaneous symmetry breaking order parameter $\psi$.
Substituting the above expression in the
free energy ${\cal F} + \Delta {\cal F}_{111}$, we arrive at the simplified free energy at the (111) surface
\bea
{\cal F}_{111} = r |\psi|^2 + w (\psi^3 + \psi^{*3}) + u |\psi|^4 + \ldots,
\eea
used in the main text. In the absence of conduction electrons, the state with $\psi \neq 0$ is a nematic which breaks rotational symmetry. Due to the symmetry-allowed
cubic term $w$ which breaks $\psi \to -\psi$ symmetry, it is also a surface ferroelectric with an in-plane ferroelectric moment. The conducting
2DEG at the surface will convert this into a ``polar metal'' phase which exhibits nematic transport.

\section{Two-band model}
\label{2band}
In order to get some insight about the sign of the $C_2$ Fourier coefficient, we consider the following Hamiltonian in the $\{|\uparrow\rangle,|\downarrow\rangle\}$ basis, with a continuum Rashba term and a Zeeman field:

\begin{equation}
	H(\bk)=\varepsilon^0_\bk \sigma_0 + \lambda (\vec{\sigma}\times\bk)\cdot\hat{z} + \tilde{g} \mathcal{\vec{B}}\cdot\vec{\sigma}.
\end{equation}

The eigenvalues and eigenvectors are given by:
\begin{equation}
\resizebox{0.85\linewidth}{!}{%
	$\varepsilon_{\bk,\pm}=\varepsilon^0_\bk\pm\sqrt{\lambda^2|\bk|^2+g^2|\mathcal{\vec{B}}|^2+2\lambda \tilde{g}(k_y\mathcal{B}_x-k_x\mathcal{B}_y)},$%
}
\end{equation}
\begin{equation}
	|u_\pm(\bk)\rangle=\frac{1}{\sqrt{2}}(\pm \mathrm{e}^{\i\phi_\bk},1)^T,
\end{equation}
where $\tan(\phi_\bk)\equiv\left(\frac{\lambda k_x-g\mathcal{B}_y}{\lambda k_y+\tilde{g}\mathcal{B}_x}\right)$. At a given electronic density, one can calculate the overlaps between points $\bk$ and $\bk+\delta\bk$ both sitting at the Fermi level of two bands with different chirality. For a scalar scattering potential we can only focus on the eigenstates overlaps:
\begin{eqnarray}
	|\langle u_\pm(\bk+\delta\bk)|u_\mp(\bk)\rangle|^2&=&\sin^2\left(\frac{\phi_{\bk+\delta\bk}-\phi_\bk}{2}\right),\\
	|\langle u_\pm(\bk+\delta\bk)|u_\pm(\bk)\rangle|^2&=&\cos^2\left(\frac{\phi_{\bk+\delta\bk}-\phi_\bk}{2}\right).
\end{eqnarray}

Parameterizing the in-plane components in polar coordinates via ${\mathcal{\vec{B}}=|\mathcal{\vec{B}}|(\cos\vartheta,\sin\vartheta)}$,${\bk=|\bk|(\cos\alpha,\sin\alpha)}$, ${\delta\bk=|\delta\bk|(\cos\beta,\sin\beta)}$ and rescaling ${g = \tilde{g}|\mathcal{\vec{B}}|}$, we can expand to leading order in $\lambda/g$ (large magnetic field limit) and $g/\lambda$ (large Rashba limit):

\begin{eqnarray}
|\langle u_\pm(\bk+\delta\bk)|u_\mp(\bk)\rangle|^2&\stackrel{\frac{\lambda}{g}\rightarrow0}{=}&\frac{|\delta\bk|^2\lambda^2}{4g^2}\cos^2(\beta-\vartheta)\nonumber\\&+&\mathcal{O}\left(\left(\frac{\lambda}{g}\right)^3\right)\label{eqn:B},\\
|\langle u_\pm(\bk+\delta\bk)|u_\mp(\bk)\rangle|^2&\stackrel{\frac{g}{\lambda}\rightarrow0}{=}&\sin^2\left(\frac{\alpha-\beta}{2}\right)\nonumber\\&+&\mathcal{O}\left(\frac{g}{\lambda}\right).\label{eqn:R}
\end{eqnarray}

From this, it is clear that in the large field limit the overlap is minimal for $\beta=\vartheta\pm\frac{\pi}{2}$ and its maximum is when $\beta=\{\vartheta,\vartheta\pm\pi\}$, independently of the position of $\bk$. In other words, the spins on each band will either align or anti-align with the magnetic field. In the large Rashba limit we distinguish two cases: (i) For $|\delta\bk|>|\bk|$ the overlap is maximized when $\beta=\alpha\pm\pi$, and (ii) For $|\delta\bk|<|\bk|$ it is maximized for $\beta=\alpha\pm\cos^{-1}(-|\delta\bk|/|\bk|)$. The details of this will be determined by the explicit shape of the scattering potential and the scattering lengthscale. For a point at $k_y=0$, i.e. $\alpha=0$ like those considered in Figs. \ref{fig:FS001B20} and \ref{fig:FS111B20}, the two overlaps \eqref{eqn:B} and \eqref{eqn:R} can be simultaneously maximized for a field in the $x$-direction by choosing $\beta=\pm\pi$ leading to an overall higher resistivity. On the other hand, for a field pointing in the $y$ direction no choice of $\beta$ can maximize the two overlaps, giving rise to a competition between the Rashba and magnetic field energy scales and thus reducing the total overlap and resistivity. 
\end{document}